# Extreme solar events

Edward W. Cliver[1,2] · Carolus J. Schrijver[1] · Kazunari Shibata[3,4] · Ilya G. Usoskin[5]

**Abstract**
We trace the evolution of research on extreme solar and solar-terrestrial events from the 1859 Carrington event to the rapid development of the last twenty years. Our focus is on the largest observed/inferred/theoretical cases of sunspot groups, flares on the Sun and Sun-like stars, coronal mass ejections, solar proton events, and geomagnetic storms. The reviewed studies are based on modern observations, historical or long-term data including the auroral and cosmogenic radionuclide record, and *Kepler* observations of Sun-like stars. We compile a table of 100- and 1000-year events based on occurrence frequency distributions for the space weather phenomena listed above. Questions considered include the Sun-like nature of superflare stars and the existence of impactful but unpredictable solar "black swans" and extreme "dragon king" solar phenomena that can involve different physics from that operating in events which are merely large.

## Contents

✉ Edward W. Cliver
ecliver@nso.edu

[1] National Solar Observatory, Boulder, CO, USA
[2] School of Physics and Astronomy, University of Glasgow, Glasgow, UK
[3] Kwasan Observatory, Kyoto University, Yamashina, Kyoto, Japan
[4] Doshisha University, Kyotanabe, Kyoto, Japan
[5] Space Physics and Astronomy Research Unit and Sodankylä Geophysical Observatory, University of Oulu, Oulu, Finland

# 1 Introduction

Research on extreme solar and solar-terrestrial activity dates to the notable event of 1859 (Carrington 1859; Hodgson 1859; Stewart 1861), but it is only within the last twenty years that extreme events, as a separate class, have been examined in detail. The threat posed by extreme space weather events to Earth's technological infrastructure provided much of the impetus for this development. Detailed studies of the impact of a severe space weather event on modern society have been conducted by the US National Research Council (NRC; 2008), Lloyds of London (2010), JASON (2011), and the UK Royal Academy of Engineering (2013), among others.[1] The NRC report contains an estimate for the economic costs of such a storm of "$1 trillion to $2 trillion during the first year alone … with recovery times of 4–10 years". From the Lloyds' report: "Sustained loss of power could mean that society reverts to nineteenth century practices. Severe space weather events that could cause such a major impact may be rare, but they are nonetheless a risk and cannot be completely discounted."

The investigation of extreme space weather has broadened as new windows—historical cosmogenic nuclide events (Miyake et al. 2012) and *Kepler* observations of flares on Sun-like stars (Maehara et al. 2012)—were opened. In this review, we trace the evolution of research on extreme solar activity and review work on the limits of the various types of extreme space weather and their occurrence probabilities.

## 1.1 1859: in the beginning

Richard Carrington, the accomplished nineteenth century English astronomer (Cliver and Keer 2012), described his discovery of the first solar flare—on 1 September 1859—in a brief *Monthly Notices* paper that is a mixture of scientific rigor (e.g., avoiding any correspondence with Hodgson, who also observed the flare, to maintain the independence of their accounts), excitement ("being somewhat flurried by the surprise, I hastily ran to call someone to witness the exhibition with me, and on returning within 60 s, was mortified to find that it [the flare] was already much changed and enfeebled"), and caution ("While the contemporary occurrence [of solar flare and geomagnetic disturbance] may deserve noting, he [Carrington] would not have it supposed that he even leans toward hastily connecting them. 'One swallow does not make a summer.'"). From his detailed account of the "singular appearance seen in the Sun", it seems clear that Carrington knew that the transient bright emission patches he observed in the large spot group near central meridian (Fig. 1) represented a new and important solar phenomenon. What he could not know was that the "sudden conflagration" he observed on the Sun and the ensuing geomagnetic storm were historically large—possibly the largest that have been directly observed (Cliver and Dietrich 2013). It was recognized at the time that the 1859 magnetic storm was strong, but just how strong was hard to say. It was accompanied by a widespread aurora (Loomis 1859, 1860, 1861) and the magnetometers at Greenwich (east of London) and Kew (west), both near to Carrington's observatory in Reigate (south of London) and Hodgson's in Highgate (north), were driven off-scale, but systematic magnetic records only dated from the 1830s (Chapman and Bartels 1940) and there was little basis for comparison.

---

[1] NRC (http://lasp.colorado.edu/home/wp-content/uploads/2011/07/lowres-Severe-Space-Weather-FINAL.pdf); Lloyds: (https://assets.lloyds.com/media/ec9c7308-7420-4f1a-83c3-9653b1f00a4c/7311_Lloyds_360_Space%20Weather_03.pdf); JASON: (https://fas.org/irp/agency/dod/jason/spaceweather.pdf); Royal Academy: (http://www.raeng.org.uk/publications/reports/space-weather-full-report); RAL Technical Report RAL-TR-2020-005 (Second revised edition). https://epubs.stfc.ac.uk/work/46642513.

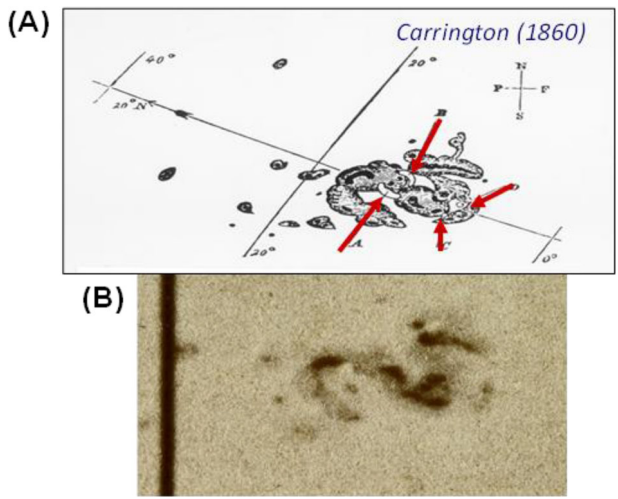

**Fig. 1** **a** Carrington's (1859) carefully executed drawing of his sunspot group 520 on 1 September 1859, the first visual record of a solar flare. The initial (A and B) and final (C and D) positions of the white-light emission are indicated. Solar east is to the right. **b** Enlargement of region 520 from an early solar photograph made at Kew Observatory by Warren De la Rue on 31 August 1859. Image reproduced with permission from Cliver and Keer (2012), copyright by Springer; the enlarged portion of RGO 67/266 in (B) is reproduced by kind permission of the Syndics of Cambridge University Library

### 1.2 2003–2004: surge in interest in extreme events

The establishment of the extreme strength of the Carrington storm awaited the study by Tsurutani et al. (2003) who analyzed long-buried geomagnetic observations of the event from Colaba Observatory where the automatic recording that led to off-scale readings at Greenwich and Kew had not yet been instituted. The manual observations at ∼ 10-min intervals at Bombay indicated a storm three times as intense as any that has been observed since. While the three-times assessment is likely an overestimate (Akasofu and Kamide 2005; Siscoe et al. 2006; Cliver and Dietrich 2013), the 1859 storm remains among the strongest events ever observed (Hayakawa et al. 2020c, 2022). The year 2003 was ripe for a renewed appreciation of the intensity of the Carrington storm. The discipline of space weather was becoming established and interest in long-term space weather, sparked by Eddy's (1976) rediscovery of the Maunder Minimum, was fueled by the concern of global warming. The journal *Space Weather* began publication in 2003 (Lanzerotti 2003) and a series of Space Climate Symposia (Mursula et al. 2004) was inaugurated in 2004. In this context, the Tsurutani et al. (2003) paper became a seminal paper for research on extreme solar activity.

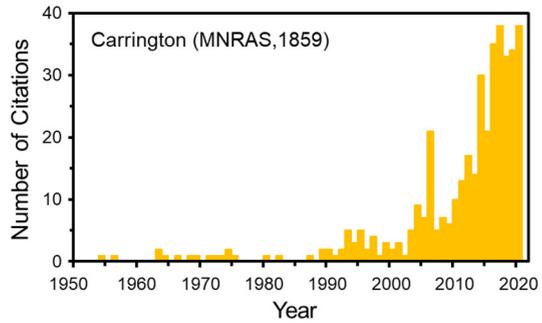

Fig. 2 Recent growth (through 2020) in citations to Carrington's (1859) *Monthly Notices* paper on the discovery of the 1859 flare

The NASA Astrophysics Data System (ADS) annual citation rate for Carrington's *Monthly Notices* paper on the 1859 event (Fig. 2) shows the recent growth of interest in the first recognized extreme solar event and in extreme events generally. The citation spike in 2006 reflects the publication in *Advances in Space Research* of the proceedings of a workshop on the Carrington event at the University of Michigan in late 2003 (Clauer and Siscoe 2006).

In 2004, Cliver and Svalgaard presented a paper at the first Space Climate Symposium that listed the observed/inferred extreme values of: the peak intensity of solar flares (based on soft X-ray observations and magnetic crochet amplitudes); Sun-Earth disturbance transit time (a proxy for the average speed of coronal mass ejections (CMEs)); solar proton event amplitude, inferred—erroneously, it now appears (Wolff et al. 2012; Sukhodolov et al. 2017)—from nitrate concentrations in ice cores; geomagnetic storm intensity; and low-latitude auroral extent. The limiting cases of the various types of events they considered are analogous to the 100-year floods or hurricanes of terrestrial climate. Note the inclusion of the CME (Webb and Howard 2012), intermediary to flare and storm, in this list of space weather phenomena. CMEs are a relatively new aspect of solar activity first directly observed in the early 1970s (Koomen et al. 1974, and references therein; Gopalswamy 2016) that were pointedly brought to the forefront in solar-terrestrial physics by Gosling (1993) with an important precursor in Kahler (1992).

### 1.3 2012: new windows

The year 2012 witnessed significant advances, based on disparate data sets, in the study of extreme solar activity. Miyake et al. (2012) obtained high-time-resolution measurements of the $^{14}$C concentration in the tree rings of two Japanese cedar trees that showed a transient increase of $\sim 12$‰ from 774 to 775 AD (Fig. 3). The origin of the 774–775 AD $^{14}$C enhancement was immediately a matter of debate. Was it caused by a very large solar energetic proton (SEP) event (or a closely-spaced cluster of high-energy proton events such as occurred from August to October 1989)

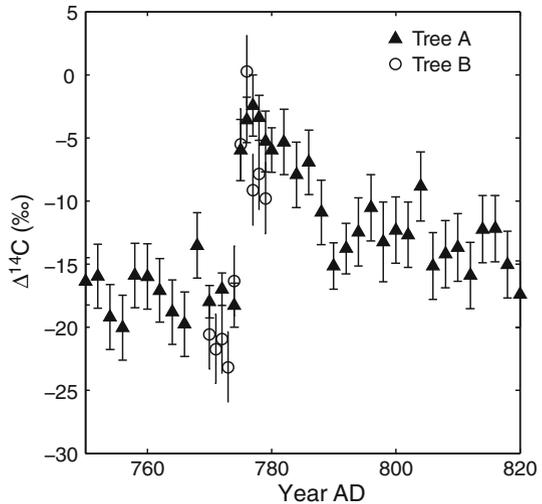

**Fig. 3** Measured $^{14}$C content at 1–2-year time resolution for two Japanese cedar trees showing the cosmogenic nuclide event of 774–775 AD. Image reproduced with permission from Miyake et al. (2012), copyright by Macmillan

(Usoskin et al. 2013) or by a galactic gamma-ray burst (GRB; Hambaryan and Neuhäuser 2013; Pavlov et al. 2013a, b)? Since 2012, numerous studies, including a multi-isotope investigation (Mekhaldi et al. 2015) of the 774–775 AD cosmogenic-nuclide event and a similar event in 993–994 AD discovered by Miyake et al. (2013), have provided strong evidence for the solar scenario (Miyake et al. 2020a).

Potential evidence far from the Sun bearing on the limits of the strength of solar flares was provided by the precise, long-term, and continuous photometry of stars by the *Kepler* satellite (Koch et al. 2010) that had the search for extra-solar planets as its primary task. For the first 120 days of the *Kepler* mission, Maehara et al. (2012) reported the observations of 14 superflares (with energy $> 10^{34}$ erg) on 14,000 Sun-like stars slowly-rotating G-type main-sequence stars with surface temperatures of 5600–6000 K). They calculated that a flare with bolometric energy $> 10^{34}$ erg could occur on a Sun-like star once every 800 years. The question of whether the Sun in its present state is capable of producing a flare of this size remains unsettled (Aulanier et al. 2013; Shibata et al. 2013; Tschernitz et al. 2018; Schmieder 2018; Notsu et al. 2019; Okamoto et al. 2021). In a detailed survey of 38 large eruptive (i.e., CME-associated) solar flares from 2002 to 2006, Emslie et al. (2012) found that the X28 soft X-ray flare on 4 November 2003 had the largest radiative energy ($4.3 \times 10^{32}$ erg), while the X17 flare on 28 October 2003 had the largest total (radiative plus CME kinetic energy) energy ($1.6 \times 10^{33}$ erg). The corresponding inferred values for the Carrington flare are essentially identical to these estimates: $\sim 5 \times 10^{32}$ erg (bolometric) and $\sim 2 \times 10^{33}$ erg (total) (Cliver and Dietrich 2013).

There was a further important development in 2012. On 23 July of that year a major backside eruption on the Sun was observed both remotely and in situ by the *STEREO* spacecraft (Kaiser et al. 2008). Work by several teams of investigators (Russell et al. 2013; Baker et al. 2013; Liu et al. 2014a, b; Riley et al. 2016; cf.

Ngwira et al. 2013) indicated that had the eruption occurred on the front side of the Sun, it might have produced a magnetic storm greater than that inferred for the Carrington event.

### 1.4 Subsequent developments

Hayakawa et al. (2017) presented evidence for a space weather event in 1770 that rivaled or exceeded aspects of the 1859 event. Aurorae from 16 to 18 September 1770 were observed at geomagnetic latitudes as low as $\sim 20°$ in both the southern and northern hemispheres, comparable to those following the 1859 event, and the estimated area of the likely associated sunspot region (from a contemporary drawing) was $\sim 6000$ millionths of a solar hemisphere, approximately three times that of the mean area for the source region of the Carrington flare (Newton 1943; Jones 1955). Similarly, Love et al. (2019c) deduced a minimum Dst intensity for the 15 May 1921 geomagnetic storm that equaled (within uncertainties) that inferred for the Carrington storm. Other developments include the Knipp et al. (2016) review of the notable May 1967 space weather event which drew attention to an aspect of extreme space weather that deserves increased attention—extreme radio bursts that pose a threat to radar operations and radio communications (e.g., Cerruti et al. 2008)—and the discovery and verification of a third historical cosmogenic nuclide event in $\sim 660$ BC (Park et al. 2017; O'Hare et al. 2019) that was comparable to the 774–775 AD event. More recently, Cliver et al. (2020b) inferred a bolometric energy of $\sim 2 \times 10^{33}$ erg for the flare associated with the 774 AD proton event and Reinhold et al. (2020) presented evidence suggesting that the Sun is currently in a state of subdued activity relative to its stellar peers.

### 1.5 Related work

Previous reviews on extreme events have been published by Riley (2012), Schrijver et al. (2012), Hudson (2015, 2021), Riley et al. (2018), Gopalswamy (2018) and Hapgood et al. (2021). Here, in addition to the phenomena of solar flares, CMEs, geomagnetic storms, and low-energy proton events, we consider sunspot groups, flares on Sun-like stars, solar radio bursts, fast transit interplanetary coronal mass ejections (ICMEs), low-latitude aurorae, and high-energy proton events that give rise to cosmogenic nuclide enhancements—topics that were not included or were more lightly treated in the reviews of extreme events listed above. We do not, however, consider the effects of extreme solar events on the ionosphere (e.g., sudden ionospheric disturbances and polar cap absorption events), atmosphere (e.g., ozone depletion) and lithosphere (e.g., geomagnetically induced currents), all of which are addressed by Riley et al. (2018). The emphasis in Hapgood et al. (2021) is on the terrestrial impacts of extreme solar activity. Recent reviews by Tsurutani et al. (2020) and Temmer (2021) focuses on space weather generally. Although the rarity of extreme events makes their footing less certain, there is evidence for certain of the phenomena we consider that the physics of extreme space weather events can differ from that in events which are merely large.

Our focus will be on the largest directly observed, inferred, and theoretically derived values of size/intensity measures of the various types of solar emissions and of geomagnetic storms. For solar flares, geomagnetic storms, and solar proton events, long-term indirect observations are provided by magnetograms (via magnetic crochets), auroral records, and cosmogenic nuclide data, respectively. We compile lists of the largest observed events in each category for comparison with future events and tabulate estimates of 100- and 1000-year events based on occurrence probability distribution functions.

In Sects. 2–7 we consider the different types of extreme activity, in turn, and in Sect. 8 we present and discuss a summary table of extreme events.

## 2 Sunspot groups

### 2.1 100- and 1000-year spot groups

The primary data base used to estimate the areas of 100-year and 1000-year solar spot groups is that compiled at the Royal Greenwich Observatory from 1874 to 1976 (RGO; Willis et al. 2013a, b; Erwin et al. 2013) and extended to the present using data from the US Air Force's Solar Observing Optical Network (SOON) and other observatories (Muñoz-Jaramillo et al. 2015; Giersch et al. 2018; Mandal et al. 2020).[2] Such estimates begin with histograms of projection-corrected spot group areas (measured in millionths of a solar hemisphere (μsh; 1 μsh = $3.0 \times 10^6$ km$^2$)). Various functional forms have been used to fit the size distribution of spot groups, dependent in part on the data sets and time intervals considered.

Bogdan et al. (1988), Baumann and Solanki (2005), and Leuzzi et al. (2018) used lognormal functions to fit distributions of group spot areas. Baumann and Solanki considered the RGO data set from 1874 to 1976, restricting their analysis to groups within ± 30° from central meridian, to minimize the errors resulting from visibility corrections. They only considered groups with umbral areas > 15 μsh and total areas > 60 μsh in their fits because of intrinsic measurement errors and distortion due to seeing for small groups. Their lognormal number density functions for the maximum (peak value observed during disk passage) and instantaneous (all daily observations), umbral and total (umbral plus penumbral), group areas are shown in Fig. 4. In Fig. 4b, the curves for maximum and snapshot total spot areas are essentially identical.

Double or bi-lognormal functions have also been used to fit distributions of spot group areas (Kuklin 1980; Nagovitsyn et al. 2012, 2017; Nagovitsyn and Pevtsov et al. 2021). Nagovitsyn and Pevtsov considered the maximum total areas of groups observed by Greenwich from 1874 to 1976 and from Kislovodsk Mountain Astronomical Station (KMAS) from 1977 to 2018. The crossover point for the two lognormal distributions of Nagovitsyn and Pevtsov corresponds to the 60 μsh lower limit of the area range that Baumann and Solanki considered for their single lognormal distribution. From analysis of instantaneous group areas from multiple

---

[2] A listing of sunspot areas (1874–2016) is given at https://solarscience.msfc.nasa.gov/greenwch.shtml.

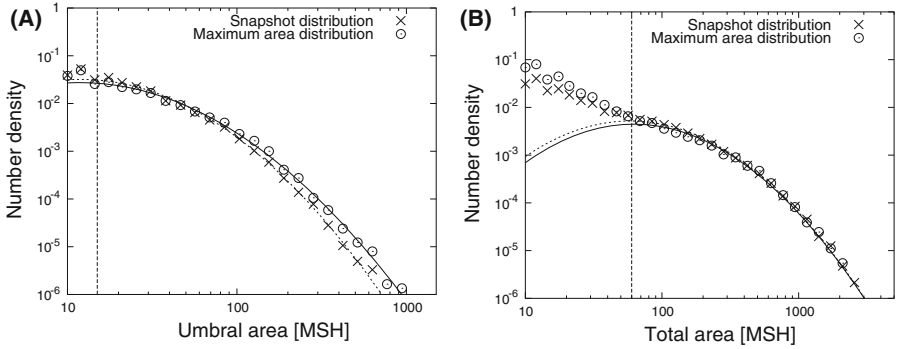

**Fig. 4 a** Size distribution of maximum (circles) and instantaneous (crosses) sunspot group umbral areas. **b** Same as **a** for total group areas. The log-normal fits are over-plotted. The vertical lines indicate the smallest areas considered for the fits. Image reproduced with permission from Baumann and Solanki (2005), copyright by ESO data sets including RGO and KMAS, Muñoz-Jaramillo et al. (2015) concluded that while the larger spot groups had a lognormal distribution, the smaller groups were better represented by a Weibull (1951) function (cf. Nagovitsyn and Pevtsov 2021).

Figure 5 shows a comparison made by Muñoz-Jaramillo et al. (2015) of fits to the instantaneous total group sunspot number area from RGO for four different functions: lognormal, power law, exponential, and Weibull. For the five data sets Muñoz-Jaramillo et al. considered (RGO, KMAS, SOON, Pulkovo Observatory (Nagovitsyn et al. 2008), and the Heliospheric and Magnetic Imager (HMI; Scherrer et al. 2012) on the *Solar Dynamics Observatory* (SDO; Pesnell et al. 2012), the best fit was provided by the Weibull function distribution in each case although it only passed the Kolomogorov-Smirnov test for the HMI data set.

Figure 6 shows a downward cumulative distribution of spot group areas (left hand axis) from Gopalswamy et al. (2018) for a data set consisting of observed daily whole sunspot areas for spot groups from 1874 to 2016 (based on RGO data from 1874 to 1976 and SOON data after 1976). The blue curve is a modified exponential function to the annualized distribution on the right hand axis

$$Y = a\left(1 - \exp\left[-\left(\frac{-X+b}{c}\right)\right]\right) \quad (1)$$

with a normalization factor (a), in addition to location (b) and scale (c) parameters, that gives the occurrence frequency distribution (OFD).[3] The scale factor (c) reflects

---

[3] Gopalswamy (2018) refers to Eq. (1) as a Weibull function (https://en.wikipedia.org/wiki/Weibull_distribution; https://www.weibull.com/hotwire/issue14/relbasics14.htm) but the choice of "1" as exponent for the $((-X+b)/c)$ term reduces it to an exponential function. The modification to the standard Weibull (exponential) function is the introduction of an additional factor (a). The minus sign before the X-variable is required for the downward cumulative distribution.

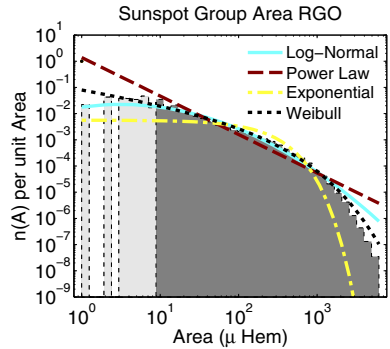

**Fig. 5** Distribution of the instantaneous total group sunspot area from RGO, fitted by four different functions. The dark-gray shaded area indicates the range over which the fits were made. Image reproduced with permission from Muñoz-Jaramillo et al. (2015), copyright by AAS

the spread of the distribution in the X-parameter. The data are fitted to this function by making an initial guess of the three parameters and using an IDL routine called MPFIT (e.g., https://pages.physics.wisc.edu/~craigm/idl/fitting.

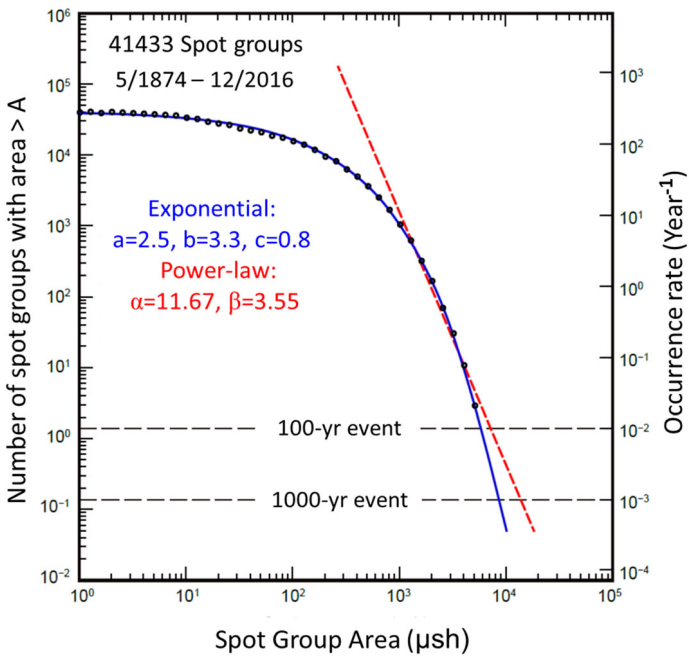

**Fig. 6** Downward cumulative distribution (left hand axis) of the number of solar spot groups from 1874 to 2016 with instantaneously-measured total areas greater than a given value A (black circle data points). This annualized distribution (right hand axis) is fitted with a modified exponential function (solid blue line) and power-law (dashed red line; for the tail of the distribution) to give the occurrence frequency distribution (OFD). The fit parameters to Eq. (1; exponential) and Eq. (2; power law) are given in the figure. The intersections of the dashed horizontal lines with the fitted curves give the areas of the 100-year and 1000-year spot groups. Adapted from Gopalswamy (2018)

html) (Gopalswamy, personal communication, 2021) to determine the best-fit values.

Gopalswamy (2018) also obtained a power-law fit (Newman 2005) for the tail of the distribution (dashed red line),

$$Y = \alpha - \beta X \quad (2)$$

with the minimum $X$-value determined by the maximum-likelihood estimator (MLE; Clauset et al. 2009). The parameters for both the exponential and power-law fits are given in the figure. In both Eqs. (1) and (2), $X$ is the log of the spot group area and $Y$ is the log of the number of number of groups per year for this area.

Equations (1) and (2) define occurrence frequency distributions (OFDs) that give the probability of a spot group with area A $\geq$ a given value occurring during one year. The intersection of the exponential and power-law fits with the dashed horizontal lines drawn from $10^{-2}$ and $10^{-3}$ on the right-hand y-axis indicate the spot areas of 100-year and 1000-year spot groups. The exponential fit indicates a maximum 100-year spot area of 5780 μsh sunspot that is comparable to the largest (corrected for foreshortening) whole sunspot area of 6132 μsh for Greenwich sunspot group 14886 on 8 April 1947 (Newton 1955). The corresponding estimate for a 1000-year sunspot is 8200 μsh. The power-law fit yields 100-year and 1000-year estimates for maximum group areas of 7100 and 13,600 μsh.

From the above it is clear that the choice of a function to fit a size distribution is not straightforward and the form used will affect the 100-year and 1000-year estimates obtained. For this reason we adopt a conservative empirical approach that favors the modified exponential function of Gopalswamy (2018) for several of the phenomena considered below; with its three free parameters, this function generally fits distributions well over their full parameter range, as in Fig. 6. Because power-law functions are commonly used for flare X-ray and radio distributions (e.g., Aschwanden 2014), we consider them as well, comparing exponential and power-law 100-year and 1000-year event estimates when both functions are available.

### 2.2 Sunspot groups versus active regions

To date, sunspot group area is the parameter most commonly used to make estimates of the largest possible solar flare (as discussed in Sect. 3.1.7 below). This is likely due to the ready availability of the digital RGO record but there is no guarantee that group spot area is the optimum parameter from which to determine the limiting energy of extreme flares. The entire magnetic active region (AR) determines field configuration, although spots dominate where the (free) energy can be stored. Harvey and Zwaan (1993) considered AR size at maximum development and found a power law with slope close to $-2$ for the distribution, without a clear indication of turnover at the largest sizes. It may well be significant that sunspot groups (the strong core fields of active regions) have a lognormal distribution while AR sizes that include more dispersed peripheries have a power law.

# 3 Flares on the sun and sun-like stars

## 3.1 Solar flares

### 3.1.1 Solar flare soft X-ray burst classification

The current standard measure for solar flare intensity is the widely-used *Geostationary Environmental Satellite System* (GOES) ABCMX SXR classification system which is defined as follows: SXR classes A1-9 through X1-9 correspond to flare peak 1–8 Å fluxes of $1-9 \times 10^{-n}$ W m$^{-2}$ where n = 8, 7, 6, 5, and 4, for classes A, B, C, M, and X, respectively. Occasionally, flares are observed which peak intensities $\geq 10^{-3}$ W m$^{-2}$. Rather than being assigned a separate letter designation, such flares are referred to as X10 events and above.[4]

Approximately 20 flares of class X10 or higher have been observed during the last $\sim$ 40 years.[5] The largest GOES SXR flare yet recorded occurred on 4 November 2003. The 1–8 Å detector saturated at a level of X18.4, with an estimated SXR class of X35 $\pm$ 5 (Cliver and Dietrich 2013) based on consideration of values given in Kiplinger and Garcia (2004), Thomson et al. (2004, 2005), Brodrick et al. (2005), and Tranquille et al. (2009). The bulk of the SXR class estimates for this flare were obtained via comparisons of flare SXR intensities and the amplitudes of flare-associated sudden ionospheric disturbances (SIDs; Mitra 1974; Prölls 2004; Tsurutani et al. 2009) caused by energetic flare photons leading to increased electrical conductivity in the day-side ionosphere, e.g., the magnetic crochet observed for the 1859 event. Cliver and Dietrich (2013) obtained an estimate for the Carrington flare of X45 $\pm$ 5 based on the magnetic crochet (a type of SID recorded by ground-based magnetometers) observed for this event (Cliver and Svalgaard 2004; Boteler 2006; Clarke et al. 2010). The November 2003 and September 1859 flares provide the current benchmarks for extreme flare activity. Less certain is an estimate of X285 $\pm$ 140 (Cliver et al. 2020b: Sect. 7.8 below) for the flare associated with the inferred SEP event of 774 AD (Miyake et al. 2012; Usoskin et al. 2013). Because flares such as those observed/inferred in 1859 and 774 AD are rare, we need to look at the ensemble of lesser flares observed by the GOES system since 1976 to estimate the occurrence frequency of extreme flares of this size and larger.

### 3.1.2 Solar flare frequency distributions

The GOES 1−8 Å soft X-ray measurements provide the longest, uninterrupted, and uniformly calibrated data set available for solar flares. Thus these records provide the basis for statistical studies that seek to establish the frequency of occurrence of flares as a function of flare magnitude. The most often used magnitude metric is the peak intensity of the flare as measured in that passband (expressed in W m$^{-2}$), i.e., the ABCMX GOES classification.

---

[4] The longest record of flare observations is that in Hα which dates to 1934 following the invention of the spectrohelioscope by Hale (1929) and the institution of the worldwide flare patrol (Hale 1931; Cliver 2006a). While the Hα data base covers approximately twice as many years as that for GOES SXRs (1976–present), the Hα flare classification system (Švestka 1976, p 24), consisting of a numeric flare area indicator (1–3) and a letter intensity indicator (F = faint, N = normal, B = brilliant), lacks the resolution and precision needed for detailed quantitative analysis.

[5] A list of $\geq$ X10 SXR flares since 1976 is given at https://www.sws.bom.gov.au/Educational/2/3/9.

Flares of classes A and B are under-reported in the GOES records because during active phases of the solar cycle such faint flares are often hard to separate from (or detect above) the supposedly quiescent background and are judged to be of low interest. As a further complication, flare magnitude is often not corrected for background emission, so that the weak flares that are reported are intrinsically overestimated in their strength. The least observationally biased records are those for flares of classes C, M, and X.

Many studies over the past decades have established that, at least for the larger flares, the frequency distribution for peak brightness of flares is well approximated by a power law (a thorough review in the literature of power-law fits to flare strength parameters can be found in Aschwanden et al. (2016) who also discuss waiting time distributions). The first such distribution listed by Aschwanden is that of Akabane (1956) who presented a distribution for burst peak radio intensities at 3.0 GHz. Subsequently, Hudson et al. (1969) showed a frequency histogram based on solar hard X-rays as observed by the *Orbiting Solar Observatory* (OSO-3) satellite that, with hindsight, hints at a power-law distribution for the more intense flares, although it lacks an explicit power-law fit to the data. Such an approximation was made by Drake (1971) based on soft X-ray measurements by the Explorer 33 and 35 spacecraft. Similar power-law distributions were first reported a few years earlier for UV Ceti type flare stars (very cool main sequence stars that exhibit frequent flaring; e.g., Kunkel 1968; Gershberg 1972; Lacy et al. 1976; Kowalski et al. 2013).

Rosner and Vaiana (1978) suggested that one way to create such frequency distributions would be to have a system in which exponential growth in stored energy is interrupted at random times by a flare-like event in which a substantial amount of the stored energy is removed from the system. But if such an energy build-up would occur in regions on the solar surface, a correlation between flare brightness and waiting time would be expected. The absence of such a correlation in both solar observations (Aschwanden et al. 1998) and stellar data (as already noted by, e.g., Lacy et al. 1976) led to other ideas, including that of self-organized criticality (SOC; Bak et al. 1987, 1988; applied to solar flares by Lu and Hamilton 1991). The SOC model for solar flares is based on the assumption that flares occur as a time-varying, or non-stationary, Poisson process. We point the reader to Aschwanden et al. (1998, 2016) for descriptions of the developments of such interpretations, focusing here on the distribution functions rather than the processes behind them.

In an extensive study of $\sim$ 50,000 GOES SXR bursts observed over a 25-year period from 1976 through 2000–38,000 of which were in classes C-X, Veronig et al. (2002a) obtained well-defined fluence and peak flux distributions that could be fitted by power laws with similar exponents: $-2.03 \pm 0.09$ and $-2.11 \pm 0.13$, respectively. These power-law approximations to the frequency distribution hold for over 2.5 orders of magnitude, with no significant indication of a change in behavior up to the largest flares observed. At a flare magnitude bin at $\sim$ X15 (based on only two flares), the histogram simply aborts at the end of a straight power law (Fig. 7).

The work by Gopalswamy (2018) includes a comparable analysis of flare magnitudes over a 47-year period from 1969 through 2016, including 55,285 flares of class C1 or larger (it did not consider the $\sim$ 11,500 B class flares included in the Veronig et al. 2002a, sample). This larger sample was obtained over a time base that almost doubles that of Veronig et al. (2002a) by including more recent data and also extending the GOES records with *Solar Radiation* (SOLRAD) satellite data for the period 1969–1975. The downward-cumulative representation of the frequency distribution for that data set (Fig. 8) exhibits a deviation from a power-law distribution (see also, e.g., Riley 2012). In a log–log downward cumulative representation, an approximation by power laws suggests a break towards less-frequent larger flares somewhere around X4, defined by a total of about 100 flares at that magnitude or larger. Such a downward break is consistent with the conclusions reached by Schrijver et al. (2012).

The century- and millennium-level flares based on the modified exponential function fit to the SXR peak values in Fig. 8 are X44 and X101, respectively. The corresponding bolometric energy values from Gopalswamy (2018) are $\sim 4 \times 10^{32}$ erg and $\sim 10^{33}$ erg. The 100- and 1000-year estimates based on power laws are comparable (X42 and X115; $\sim 4 \times 10^{32}$ erg and $\sim 1.2 \times 10^{33}$ erg). Both functional forms fit the tail of the distribution well. For a 10,000-year flare, the exponential function yields $\sim$ X200 versus $\sim$ X310 for the power-law fit.

While the radiative energy of the strongest observed solar flare is often given variously as $10^{32}$ erg, about $10^{32}$ erg, or $\sim 10^{32}$ erg (e.g., Candelaresi et al. 2014; Peter et al. 2014; Osten and Wolk 2015; Hudson 2016; Maehara et al. 2017; Notsu et al. 2019; Brasseur et al. 2019), the measured value is a half-decade larger. Emslie et al. (2012) reported that the inferred/observed radiative energies of 12 events from 2002 to 2006 were above the $10^{32}$ erg threshold. For the three largest such events (4 November 2003, $4.3 \times 10^{32}$ erg, X35 $\pm$ 5; 28 October 2003, $3.6 \times 10^{32}$ erg, X17;

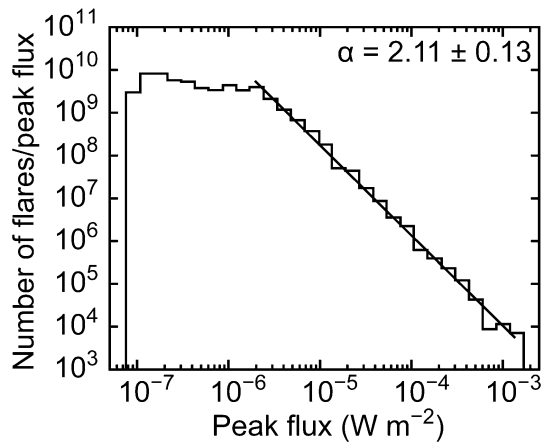

**Fig. 7** Frequency distribution of flare 1–8 Å peak flux with least squares fit for GOES SXR flares from 1976 to 2000. Image reproduced with permission from Veronig et al. (2002a), copyright by ESO

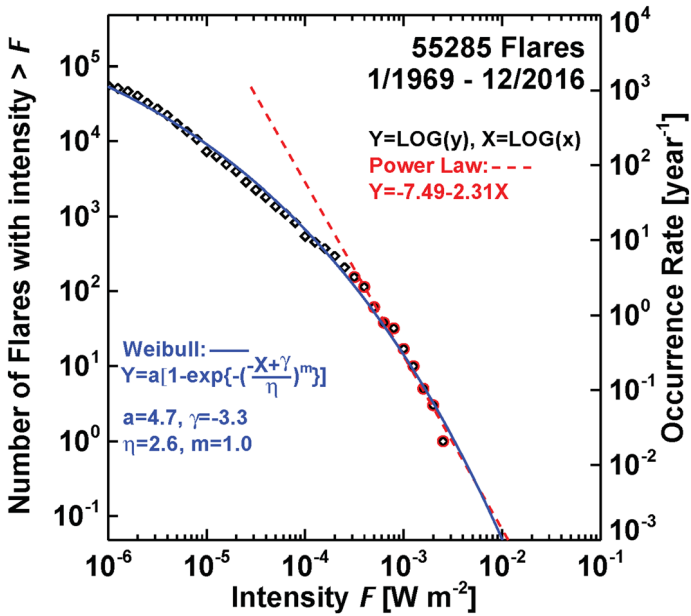

**Fig. 8** Downward cumulative distribution (left hand axis) of the number of flares from 1969 to 2016 with peak SXR fluxes greater than a given value F (blue diamond and red circle data points). This annualized distribution (right hand axis) is fitted with a modified exponential function (solid blue line) and power law (dashed red line; for the tail of the distribution (red circle data points)) to give the occurrence frequency distribution. The fit equations and parameters are given in the figure. Adapted from Gopalswamy (2018)

7 September 2005, $3.2 \times 10^{32}$ erg, X17), the bolometric energy was measured directly by the Total Irradiance Monitor (TIM; Kopp and Lawrence 2005) on the *Solar Radiation and Climate Experiment* (SORCE; Rottman 2005). Thus, the largest observed (November 2003) and inferred (September 1859; $\sim 5 \times 10^{32}$ erg, ensemble estimate of X45 ± 5; Cliver and Dietrich 2013; based on electromagnetic emissions) flares are comparable to the 100-year flare given by Fig. 8.

### 3.1.3 White-light flares

The flare Carrington and Hodgson independently recorded on 1 September 1859 was observed in integrated light. Thus their Monthly Notices papers were the first reports of what came to be known as "white light flares" (WLFs). A compilation of reports of such events from (1859–1982) is given in Neidig and Cliver (1983a). At that time, Neidig and Cliver (1983b) reckoned that such flares were relatively rare, with an occurrence frequency of $\sim 15$ per year based on a $\sim 2.5$-year period from June 1980 to December 1982 following the maximum of solar cycle 21. They determined that a flare with SXR class $\geq$ X2 originating in a large ($\geq$ 500 µsh), magnetically complex, sunspot group was a sufficient condition for a white light flare. Two decades later, Hudson et al. (2006) used *Transition Region and Coronal Explorer* (TRACE; Handy et al. 1999) images, which encompassed a wavelength range from 1500 to 4000 Å, and identified 11 white-light flares that had SXR classes ranging from C1.6 to M9.1 (median = M1.2) to "support the conclusion of Neidig (1989) that white-light continuum occurs in essentially all flares." Subsequently—based on observations of many flares with the Variability of Solar Irradiance and Gravity Oscillations (VIRGO: Fröhlich et al. 1995) experiment on the *Solar and Heliospheric Observatory* (SOHO; Domingo et al. 1995)—Kretzschmar (2011) estimated that white-light emission might typically account for $\sim 70\%$ of the total radiated flare energy, following an earlier estimate by Neidig (1989) of $\lesssim 90\%$ based on analysis of the 24 April 1981 white-light flare (Neidig 1983).

White-light flares are of particular importance in studies of extreme flares on Sun-like stars because of the observations of such flares by the high-precision (better than 0.01% for moderately bright stars) visible wavelength (i.e., white-light, 4000–9000 Å) photometer on the *Kepler* spacecraft (Koch et al. 2010), which operated from 2009 to 2018. The durations of stellar white-light flares can exceed several hours (Maehara et al. 2012) versus typical durations of $\lesssim 10$ min (Neidig and Cliver 1983a) for their largest solar counterparts, although some of this discrepancy may be due to the lower sensitivity of early WLF observations (see Fig. 13 below). Stellar flares detected by *Kepler* have bolometric energies exceeding $10^{33}$ erg (Maehara et al. 2012; Shibayama et al. 2013), exceeding those of the largest solar flares ($\sim 4 \times 10^{32}$ erg; Emslie et al. 2012; cf. Kane et al. 2005) by about a factor of three. Whereas the *Kepler* photometer detects essentially all superflares with amplitudes $> 1\%$ of the average stellar flux on solar-type (G-type dwarfs, 5100 K $< T_{\rm eff} <$ 6000 K and log $g > 4.0$) stars, corresponding to a flare bolometric energy of $\sim 5 \times 10^{34}$ erg (Shibayama et al. 2013), the estimated detection completeness for $\sim 10^{34}$ and $10^{33}$ erg flares are $\sim 0.1$ and 0.001, respectively (Maehara et al. 2012).

It is well-accepted that the paradigm of the solar flare also holds also for the large ($10^{34-35}$ erg) stellar flares (Gershberg 2005). In the standard picture of solar flares, a rapid conversion occurs, via magnetic reconnection, of energy stored in the magnetic field to energies distributed over bulk kinetic energy and thermal and non-thermal particle distributions. A significant fraction of this converted energy is eventually released as a pulse in the visible radiation, the observable diagnostic of the *Kepler* superflares on Sun-like stars. Like solar flares, stellar flares: (1) occur on single "solar-type" stars, viz., G-type main sequence stars with $T_{\rm eff}$ and gravity similar to that of the Sun and rotation periods $> 10$ days and without hot Jupiters (Maehara et al. 2012; Shibayama et al. 2013; Okamoto et al. 2021; Sect. 3.2.2); (2) show evidence of non-thermal particle populations and of temperatures $\gtrsim 10^7$ K (e.g., Osten et al. 2007; Benz and Güdel 2010); (3) exhibit a characteristic fast rise and slower, near-exponential, decay in X-rays (Kahler et al. 1982; Haisch et al. 1983); and (4) have lower-energy emissions that often scale with the time integral of their high-energy emissions (equivalent to the Neupert effect for solar flares; Hawley et al. 1995; Güdel 2002; Osten et al. 2004; see Sect. 3.1.4). The above evidence for the commonality of stellar flare characteristics and processes with those that occur during smaller solar flares implies that the statistics of energetic flares on a multitude of Sun-like stars (see Sect. 3.2) can provide information on extreme solar flares that may occur only once per millennium or even less frequently (Notsu et al. 2019; Okamoto et al. 2021).

### 3.1.4 Impulsive and gradual phases of flares: The Neupert effect

The separation of flare emission into a fast-rise impulsive phase followed by a slowly decaying gradual phase has long been noted in multiple wavelengths (Fletcher et al. 2011). The impulsive phase of flares (Dennis and Schwartz 1989; Benz 2017), characterized by non-thermal hard X-ray and radio bursts, marks the time of the early principal energy release in a flare. See Fletcher et al. (2011) and Benz (2017) for detailed reviews of flare observations.

In a prescient six-page paper based on early flare soft X-ray data for three flares, Neupert (1968) found that the integral of impulsive phase microwave emission in each flare resembled the rise of the SXR light curve. He hypothesized that collisional losses by the energetic electrons responsible for the microwave burst heated the chromosphere to sufficient temperatures to eject plasma into the low corona where its cooling was manifested by the slow decay of thermal SXR emission during the flare gradual phase.

The support structure for Neupert's conjecture on the conversion of non-thermal energy to thermal energy in a solar flare lay in the future. It consisted of: the thick target model of hard X-ray emission (Brown 1971; Kane and Donnelly 1971), the CSHKP model for eruptive solar flares (Carmichael 1964; Sturrock 1968; Hirayama 1974; Kopp and Pneuman 1976; Hudson 2021), and the establishment of chromospheric evaporation (Antonucci et al. 1984; Fisher et al. 1985). In addition, evidence that white light emission is powered by energetic electrons accelerated during the flare impulsive phase accumulated (Hudson 1972; Machado and Rust 1974; Rust and Hegwer 1975; Neidig 1989; Hudson et al. 1992; Neidig and Kane, 1993). Figure 9 (taken from Hayes et al. 2016) shows the inverse aspect of the Neupert effect—hard X-ray emission is the derivative of the SXR time profile (Dennis and Zarro 1993)—using modern data.

Figure 10 shows a CSHKP schematic for an eruptive flare that illustrates various elements of the Neupert effect: acceleration of electrons and protons at a neutral current sheet, propagation of electrons to the low atmosphere giving rise to hard X-ray and microwave emission and heating of the chromosphere via particle bombardment, evaporation of the heated plasma to fill SXR emitting loops, and subsequent retraction and cooling of these loops as manifested by the appearance of post-flare Hα loops (Švestka et al. 1987).

According to the Neupert effect, the impulsive phase of flares can be interpreted as the time interval during which magnetic reconnection occurs. In the two-dimensional schematic in Fig. 10, this phase would end when the last closed loop of the CME is disconnected from the pinched off lower loop system. In theory, the SXR intensity at this time would be at it or near its maximum. In reality, Veronig et al. (2002b) found that this coincidence or near-coincidence of the end of the impulsive phase with SXR maximum was observed for only about 50% of their sample of ∼ 1100 C2-X4 flares. In ∼ 25% of the cases, the SXR emission peak

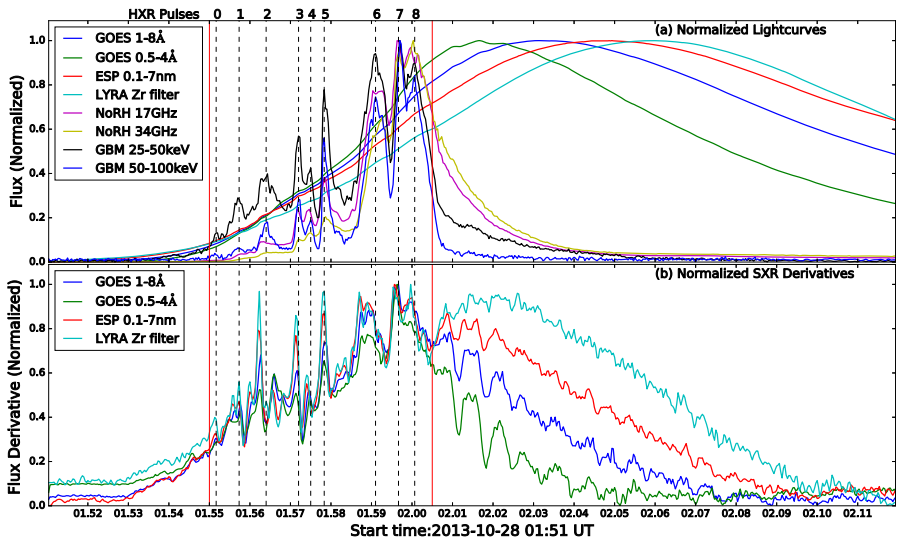

**Fig. 9** Illustration of the Neupert (1968) effect in which the time derivative of flare soft X-ray (1–8 Å) emission during the impulsive phase of a flare mimics the flare hard X-ray time profile (Dennis and Zarro 1993). (Top panel) Normalized light curves for different wavelengths/radio frequencies/energies for a flare on 28 October 2013 flare. (Bottom panel) Normalized derivatives of the four SXR channels in the top panel. The red vertical lines bracket the flare impulsive phase. Image reproduced with permission from Hayes et al. (2016), copyright by AAS

occurred after the impulsive phase hard X-ray emission, and another ∼ 25% of cases were unclear. For a much smaller subset of 66 large impulsive hard X-ray flares, Dennis and Zarro (1993) found that 80% of the events were consistent with the Neupert effect, suggesting that a relationship between SXR emission and white light emission can be valid for extreme events.

The significance of the Neupert effect for this review is two-fold: (1) it provides a link between the widely used (and more readily available) SXR intensity as a measure of flare size and the physically more significant measure of flare white-light energy which dominates the radiative energy budget of flares; and (2) as will be seen in Sect. 3.2, it allows flares on Sun-like stars to be directly compared with solar flares in terms of the standard CMX SXR classification.

The classic Neupert effect suggests that magnetic reconnection, particle acceleration, and post-flare loop formation, are confined to the flare impulsive phase, a useful generalization. Subsequently, Zhang et al. (2001) demonstrated that the SXR rise phase corresponded to the interval of CME acceleration. That said, magnetic reconnection, particle acceleration, and loop formation can continue for up to ≳ 10 h (e.g., Bruzek 1964; Kahler 1977; Akimov et al. 1996; Gallagher et al. 2002; Tripathi et al. 2004), well beyond the impulsive phase, as can particle acceleration by CME-driven shocks. Moreover, these late phase phenomena can give rise to strong, and even extreme (e.g., Frost and Dennis 1971; Chupp et al.

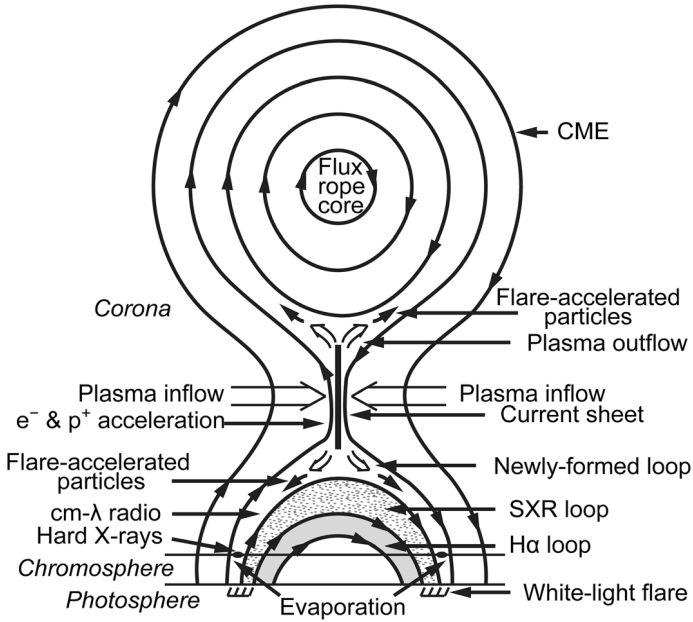

**Fig. 10** CSHKP schematic showing the global structure of an eruptive solar flare and the major energy conversion via reconnection viewed in cross section along the magnetic neutral line. Adapted from Martens and Kuin (1989); with input from Anzer and Pneuman (1982), Howard et al. (2017), and Veronig et al. (2018)

1987; Gary 2008; Hudson 2018) solar phenomena. Time profiles of flare emission at various frequencies/energies are given in Fig. 11. The hard and soft X-ray traces during the impulsive phase in panels (b) and (c) display the classic Neupert effect, which applies for the majority of flares, particularly the smaller confined events, but also for eruptive events with late phase reconnection too weak (or too high in the corona) to significantly affect the SXR time profile in panel (c)—consistent with the radiative energy domination of the impulsive phase. As shown in panels (b), (d), and (e), and discussed in Sect. 4, electrons accelerated in certain of these eruptive events can result in gradual hard X-ray and microwave bursts as well as intense decimetric bursts at $\sim$ 1 GHz, with peak flux densities as large as $10^5$–$10^6$ solar flux units.

Late phase particle acceleration in eruptive flares can also manifest itself in remotely-sensed high-energy solar $\gamma$-ray emission. When first observed in the 1980s and early 1990s, prolonged 100 MeV $\gamma$-ray emission, attributed to the decay of neutral pions which require acceleration of protons to $\gtrsim$ 300 MeV energies—the highest that can be inferred from $\gamma$-ray observations—for their production, contrasted sharply with the Neupert effect in which flare energy degrades over time from non-thermal X-ray and radio emissions to thermal soft X-rays. Rather than flare electromagnetic emission becoming progressively less energetic after the impulsive phase, 100 MeV γ-ray emission characteristically occurs during the late phase of flares (Share et al. 2018). The most economical (Occam's razor) explanation for acceleration of the energetic protons responsible for such emission is that they are accelerated by the same coronal shocks (manifested by the slow-drift metric type II burst in panel (a)) that give rise to the solar energetic particles detected by spacecraft near Earth (Sect. 7).

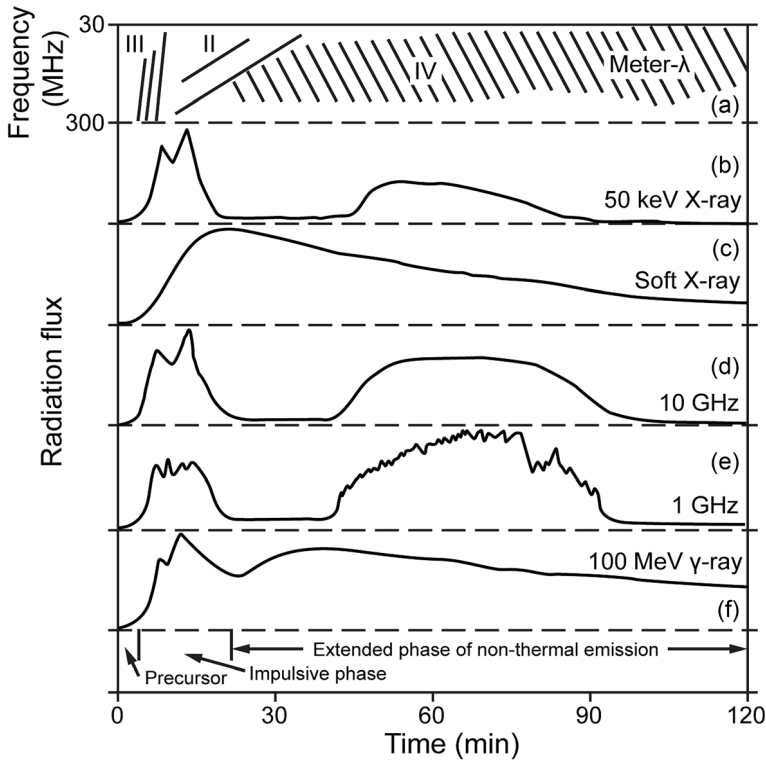

**Fig. 11** Schematic of time profiles of flare emissions at various wavelengths and energy ranges. The meter wavelength range (m–λ; 30–300 MHz in panel **a**) exhibits various spectral types of which fast drift type III bursts are a characteristic emission of the impulsive phase and slow-drift type II bursts are the defining emission of the second phase of fully developed radio events (McLean and Labrum 1985). The various time profiles shown have been simplified to emphasize the separation of the delayed non-thermal emissions both from the impulsive phase and from each other. The delayed electron-generated emissions in panels **b**, **d**, and **e** are attributed to late-phase reconnection in post flare loop systems while the delayed > 100 MeV γ-ray emission in panel **f** is attributed to shock-accelerated high-energy protons that precipitate to the photosphere. These extended phase emissions have little effect on the flare SXR time profile (panel **c**)

### 3.1.5 Energetics of flares and CMEs

1−8 Å soft X-ray and broad-band visible light emissions are but two of the channels into which the processes related to solar flares deposit energy. Magnetic energy is converted not only into photon emission in other channels, but also into kinetic energy of non-thermal particle populations, and the bulk kinetic energy associated with the motion of material ejected from the flaring active region. Specific electromagnetic, plasma, and particle aspects of solar eruptions will be considered in following sections of this review.

Figure 12 (taken from Emslie et al. 2012) shows the various pathways by which energy is converted and released during eruptive solar flares. Emslie et al. analyzed wide-ranging data for a sample of 38 M- and, mostly, X-class flares. Figure 12 summarizes their findings for six well-studied X-class flares. All of the listed phenomena in the figure derive their energy ultimately from the free (or excess) magnetic energy of an active region ($E_{fm}$), defined to be the non-potential energy beyond (or in excess of) that obtained from a potential field model. Emslie et al. (2012) estimated that, on average, the amount of $E_{fm}$ of an active region was equal to 30% of the modelled potential energy determined from line-of-sight magnetograms. For the six events upon which Fig. 12 is based, with an average flare magnitude of X6, the median $E_{fm}$ was $\sim 15 \times 10^{32}$ erg. Emslie et al. estimated that $\sim 30\%$ of the free energy in an active region was released in an eruptive event. This, in turn, implies that the energy released in large flares amounts to $\sim 10\%$ of the magnetic energy in an active region (see also Shibata et al. 2013). Thus, it is not surprising that magnetic field changes between before and after a flare are hard to spot unambiguously in the overall, always-evolving, multi-thermal coronal configuration. In contrast, rapid changes associated with flares are detected in the photospheric field, most readily in association with SXR flares of class M or X (e.g., Wang et al. 2002; Liu et al. 2005; Sudol and Harvey 2005; Castellanos Durán et al. 2018). The search for such changes was lengthy, however, beginning with Carrington (1859) who looked unsuccessfully following the 1 September 1859 flare (Fig. 1) for any changes in the sunspot group based on the sketch he had made prior to the flare.

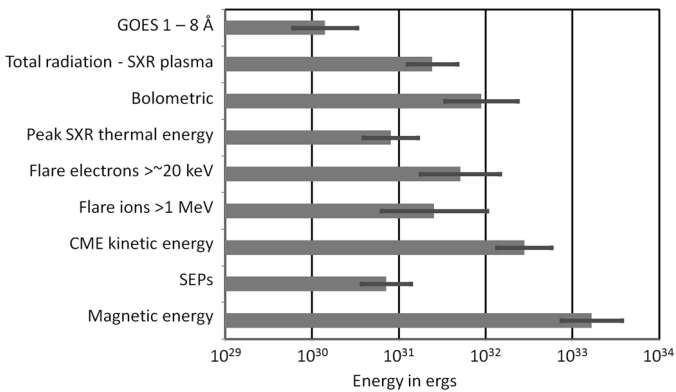

**Fig. 12** Bar chart showing the free magnetic energy (denoted by the bottom bar) contained in various energy sinks (with their standard deviations in logarithmic values) for six major eruptive flares (X2.5, X3.0, X3.8, X7.1, X8.3, X10; averaging to X5.8). "Magnetic energy" is the assumed free energy, taken to be 30% of the estimated energy in a potential field over the active region based on observed line-of-sight magnetograms. Energy release is dominated by CMEs and flare electromagnetic radiation. Image reproduced with permission from Emslie et al. (2012), copyright by AAS

In keeping with the increased emphasis on CMEs versus flares per se for space weather during the last $\sim 25$ years, Emslie et al. (2012) find that, on average—for the two dominant terms of energy release (Fig. 12)—CME energy (kinetic plus gravitational potential) is $\sim 3$ times larger than the flare bolometric energy. Aschwanden et al. (2017) performed an analogous analysis to that of Emslie et al. (2012) for a sample of 157 M- and X-class eruptive flares observed from 2002–2006 and found that the ratio of CME to bolometric energy was $\sim 1:1$ versus $\sim 3:1$ determined by Emslie et al. (2012). These different samples preclude a straightforward direct comparison. X-class flares constituted 76% of the Emslie et al. (2012) sample with a median of X2 versus < 10% of that for Aschwanden et al. (2017; median < M2). As this review focuses on the largest, most energetic events, we use the Emslie et al. result.

This $\sim 3:1$ ratio CME to flare energy obtained by Emslie et al. is presumably an underestimate because the magnetic energy of the CME (Webb et al. 1980) is not taken into account in the determination. In addition, the CME masses and speeds in the SOHO Large Angle Spectroscopic Coronagraph (LASCO; Brueckner et al. 1995) CME data base (https://cdaw.gsfc.nasa.gov/CME_list/; Yashiro et al. 2004; Gopalswamy et al. 2009) used by Emslie et al. are underestimated because of projection effects (Vourlidas et al. 2002, 2010; Vršnak et al. 2007; Paouris et al. 2021). Emslie et al. (2012) write that the mass is underestimated by about a factor of two for CMEs that are associated with flares $\lesssim 40°$ from the limb—approximately half of their sample—with greater underestimates for those closer to disk center. Vršnak et al. find that average velocities of non-halo limb-CMEs are 1.5–2 times higher than for such CMEs originating near disk center. This underestimation of CME parameters impacts estimates of the largest possible solar flare (Sect. 3.1.7) that are based on active region energy and flare energy released by reducing the radiative energy budget. For a plausible 9:1 apportionment (vs. 3:1), only 10% of the released energy would go into the flare.

In reference to extreme solar flares, we can safely conclude that they are eruptive. Schrijver (2009) reviewed studies on flares with and without CMEs, including Andrews (2003), Yashiro et al. (2005), and Wang and Zhang (2007), that showed whereas $\sim 20\%$ of low to mid C-class flares have associated CMEs, $\sim 90\%$ of X-class flares are eruptive. The largest reported flares that lacked CMEs were an X3.1 flare on 24 October 2014 from NOAA spot group 12192 that produced several confined X-class flares (e.g., Thalmann et al. 2015; Liu et al. 2016; Green et al. 2018; Gopalswamy 2018), an X3.6 flare on 16 July 2004 (Wang and Zhang 2007), and an X4.0 flare on 9 March 1989 (Feynman and Hundhausen (1994). For flares on active stars, it has been argued (Drake et al. 2016; Alvarado-Gómez et al. 2018; Moschou et al. 2019; Li et al. 2020, 2021) that much stronger magnetic fields than observed on the Sun would not allow free energy to be released via eruption below some threshold. Based on observations of solar CMEs, Li et al. (2021) speculate that for an active region with unsigned flux of $10^{24}$ Mx on a solar-type star, the CME association rate for X100 flares would be < 50%. To date, the most intense solar flares yet inferred (viz., 1 September 1859 and 774 AD (Sect. 7.8)) are eruptive by virtue of their terrestrial effects—geomagnetic storm and proton event, respectively—which both imply CME association (Kahler 1992; Gosling 1993).

### 3.1.6 Empirical relationship between flare total solar irradiance and SXR class

Even though most of the solar flare electromagnetic radiation, or bolometric output, occurs in the visible range, the contrast with the quiescent photosphere is low. Only the larger white-light flares stand out clearly against the photospheric background in high-resolution images while only the most extreme flares such as 28 October or 4 November 2003 can be recognized as spikes in the record of total solar irradiance (TSI; Woods et al. 2004, 2006; Kretzschmar 2011). As shown in Fig. 13 (from Kopp 2016), the X17 flare on 28 October in the Halloween sequence in 2003 reached a peak brightness in the visible wavelength range of only 0.028% of TSI.

However, when averaging observations over multiple flares, the signal-to-noise ratio increases. Kretzschmar et al. (2010) and Kretzschmar (2011) used this ensemble technique in a superposed-epoch analysis of observations made with the VIRGO experiment on SOHO to correlate the GOES measured soft X-ray energy ($\mathcal{F}_{GOES}$) and bolometric energy of over 2100 solar flares (with GOES classes from C4 up to X17) over an 11-year period. Kretzschmar (2011) presented his results averaged over fairly wide energy ranges in order to have sufficient signal-to-noise. In their analysis of solar and stellar flares, Schrijver et al. (2012) fitted a power-law relationship to these SOHO and GOES measurements to obtain

$$\mathcal{F}_{\text{TSI}} = 2.4 \times 10^{12} \mathcal{F}_{\text{GOES}}^{0.65 \pm 0.05}, \quad (3)$$

for the conversion from 1 to 8 Å GOES radiated energy to bolometric energy (both expressed in erg) based on flares with bolometric energies ranging from $3.6 \times 10^{30}$ erg to $5.9 \times 10^{31}$ erg. The GOES 1−8 Å channel, commonly used to characterize flares and their frequency distribution, captures < 2% of the total radiation emitted by large flares (Woods et al. 2006; Kretzschmar 2011; Emslie et al. 2012).

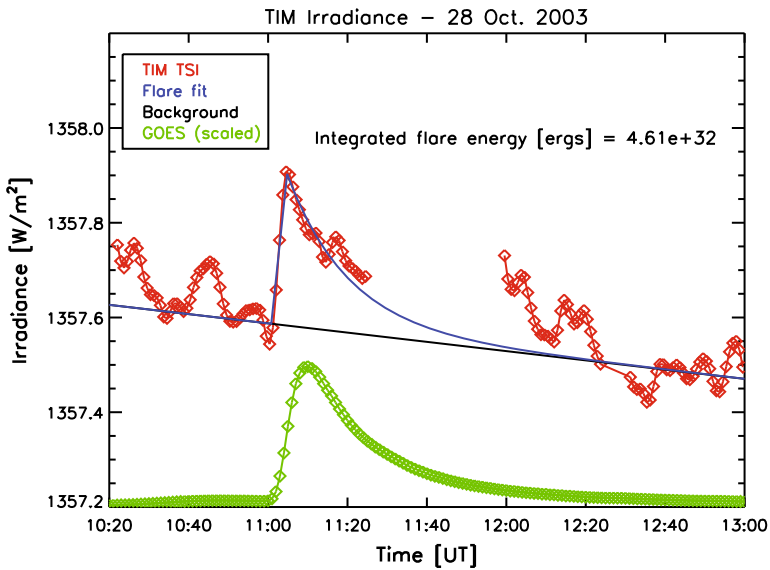

**Fig. 13** Comparison of the total solar irradiance (TSI) signal from the X17 flare on 28 October 2003 with a bolometric energy of $3.6 \times 10^{32}$ erg, and the corresponding scaled GOES 1−8 Å signal. Image reproduced with permission from Kopp (2016), copyright by the author

To go from soft X-ray radiated energy to peak brightness (which sets the GOES flare class) we can use observations of the almost 50,000 flares that were analyzed by Veronig et al. (2002a). They find that the GOES 1−8 Å radiated energy,

$$\mathcal{F}_{\text{GOES}} = 2.8 \times 10^{29} \left( \frac{\mathcal{C}_{\text{GOES}}}{\mathcal{C}_{\text{GOES,X1}}} \right)^{1.10}, \quad (4)$$

with their SXR fluence measurement converted to energy by multiplying by $4\pi \times (1\text{ AU})^2$ and peak SXR flux scaled to equal unity for an X1 class flare ($10^{-4}$ W m$^{-2}$), where 1 AU = 1 astronomical unit ($1.5 \times 10^8$ km).

Combining Eqs. (3) and (4) leads to a relationship between total radiated energy, $\mathcal{F}_{\text{TSI}}$ (in erg), and the scaled GOES flare class:

$$\mathcal{F}_{\text{TSI}} = 0.33 \times 10^{32} \left( \frac{\mathcal{C}_{\text{GOES}}}{\mathcal{C}_{\text{GOES,X1}}} \right)^{0.72}. \quad (5)$$

Thus values of $10^{32}$ ($10^{33}$) erg correspond to a ∼ X5 (∼ X115) flare, with uncertainties in the mean relationships as well as flare-to-flare differences resulting in overall uncertainties of easily half an order of magnitude (e.g., Emslie et al. 2012; Benz 2017; Gopalswamy 2018). Because TSI is dominated by white-light emission in the impulsive phase of flares, Eq. (5) can be considered a corollary of the Neupert effect.

### 3.1.7 Estimates of the largest possible solar flare based on the largest observed sunspot group

*(a) Estimates based on reconnection flux*

Large flares require large active regions with substantial amounts of magnetic flux near a high-gradient polarity separation line (Schrijver 2007). As noted in Sect. 2.1, the largest sunspot group recorded since systematic area measurements began at RGO in 1874 (Greenwich group 14886) occurred on 8 April 1947 (Fig. 14), with a corrected total spot area of 6132 μsh. Estimates of the total unsigned magnetic flux of such an active region range from 2 to $6 \times 10^{23}$ Mx (Toriumi et al. 2017; Schrijver et al. 2012). Figure 15 (adapted from Toriumi et al.)

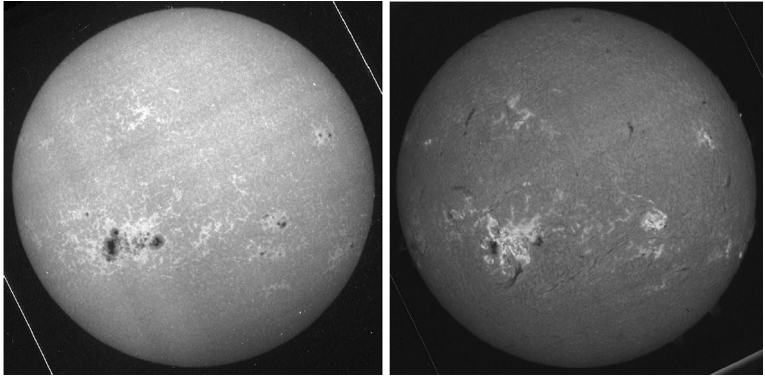

**Fig. 14** The largest sunspot group reported since 1874, as observed on 5 April 1947 in Ca II K1v (left) and Hα (right) by the Meudon spectroheliograph. The largest area of this group (Greenwich 14886), as reported by the Royal Greenwich Observatory (RGO) was 6132 μsh on 8 April. Image reproduced with permission from Aulanier et al. (2013), copyright by ESO

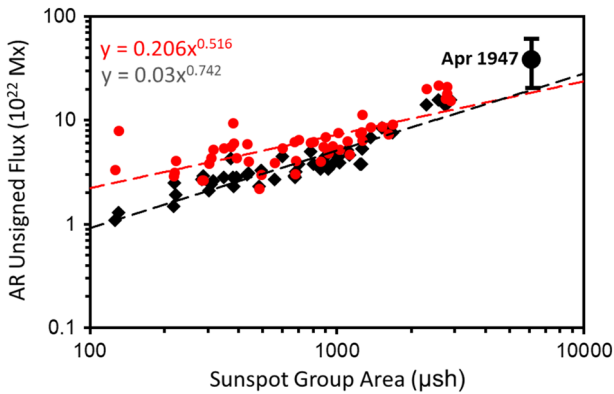

**Fig. 15** Scatter plot of total unsigned active region (AR) magnetic flux versus sunspot area (black diamonds) for $51 \geq M5.0$ flares (from 29 active regions) located within $45°$ of disk center from 2011 to 2015 (Toriumi et al. 2017). The location of the black circle point for the April 1947 spot group (Greenwich 14886) is based on an average of the estimated total unsigned flux from Schrijver et al. (2012; $6 \times 10^{23}$ Mx) and Toriumi et al. (2017; $2 \times 10^{23}$ Mx). The black straight line is the result of a linear fit to data in the log–log plot. The red circle data points with fitted dashed line are based on the unsigned flux values for these events from Kazachenko et al. (2017). Adapted from Toriumi et al. (2017)

shows a plot of unsigned magnetic flux (from spots and plage) for active regions associated with 51 flares located within $45°$ of solar central meridian from 2011–2015 (black diamonds). The red circle data points with fitted dashed line are based on unsigned flux values from Kazachenko et al. (2017). Linear fits to both data sets give a value of $\sim 2 \times 10^{23}$ Mx for the April 1947 spot group. At the same time, the largest spot groups in this figure (all but one from spot group 12192 in October 2014) hint at a non-linear fit line in the log–log plot that could reach the $6 \times 10^{23}$ Mx value used by Tschernitz et al. The large black circle in Fig. 15 represents a compromise average flux value of $4 \times 10^{23}$ Mx.

The Schrijver et al. estimate of $\sim 6 \times 10^{23}$ Mx assumes a uniform field strength of $\sim 3000$ G for the spotted area of group. While $\sim 3000$ G is not unreasonable for the peak field strength of a large single spot, such a value is reached $< 10\%$ of the time in sunspot umbrae (Pevtsov et al. 2014). Moreover, the corrected umbral area of Greenwich 14886, is only 739 μsh. Schrijver et al. (2012) used an overall sunspot group area of 6000 μsh but did not consider plage. This neglect of the plage contribution offsets the assumption of the high uniform field strength in the sunspots.

Cycle 18 (1944–1954), which included the great spot group of April 1947 is known as the cycle of "giant" sunspots (Dodson et al. 1974). The five spot groups since 1874 with observed areas $> 4500$ μsh all occurred in this cycle. (The next largest group, with an area of 3716 μsh, occurred in January 1926.) While large spot areas are indicative of the potential for extreme events, they are not sufficient for their occurrence. Of these five spot groups, two (February 1946 ($-$ 220 nT; http://dcx.oulu.fi/dldatadefinite.php) and July 1946 ($-$ 268 nT) were associated with "great", but not "outstanding" magnetic storms, the designation given the May 1921 event (Jones 1955). As we shall see in Sects. 6 and 7, the intensities of geomagnetic storms and solar energetic proton events, respectively, can be affected by other factors than the size of the parent sunspot group.

Tschernitz et al. (2018) find a strong correlation between the GOES class of a flare and the rate of magnetic reconnection for a sample of 51 flares (SXR class B3-X17). They estimate the reconnection rate by mapping the Hα ribbon evolution in time onto the region's magnetogram (Fig. 16a). Among the most pronounced correlations ($r = 0.92$) they find a scaling between GOES SXR class and the total reconnected flux $\Phi_r$ (in Mx; 1 Mx = $10^{-8}$ Wb) (Fig. 16b). Tschernitz et al. (2018) find that the largest events involve about half of the magnetic flux contained in the active region. For the April 1947 spot group with a total magnetic flux of $6 \times 10^{23}$ Mx (after Schrijver et al. 2012), Tschernitz et al. inferred a largest possible SXR flare of class $\sim$ X500 (with confidence bounds from X200-X1000) as shown in Fig. 16b, with a corresponding bolometric energy of $\sim 3 \times 10^{33}$ erg. However, because the correlation of Tschernitz et al. used the average of the positive and negative magnetic flux rather than the total unsigned flux as estimated by Schrijver et al. (2012), the reconnection flux for the 1947 region should be reduced by another factor of two from $3.0 \times 10^{23}$ Mx to $1.5 \times 10^{23}$ Mx (A. Veronig, personal communication, 2021), reducing the SXR class estimate to X180 ($-$ 100, $+$ 300) (1.4 ($-$ 0.6, $+$ 1.4) $\times 10^{33}$ erg) via Eq. (5). For the compromise total (average) unsigned flux of 4(2) $\times 10^{23}$ Mx for the April 1947 active region in Fig. 15 (with 50% flux involvement), Fig. 16b yields a SXR class of X80 ($-$ 40, $+$ 120) (with bolometric energy of 0.8 ($-$ 0.3, $+$ 0.7) $\times 10^{33}$ erg). For reasons given in 3.1.7(c) below, our preferred estimate based on Fig. 16b is the X180($-$ 100, $+$ 300) value for an unsigned flux of $6 \times 10^{23}$ Mx. Figure 17 gives an impression of the areas of sunspot groups that appear to be required to power flares of different magnitudes. The inset shows the large 1947 active region for comparison with a modelled spot group in the upper right of the figure that, based on the above calculation, could produce a $10^{33}$ erg flare.

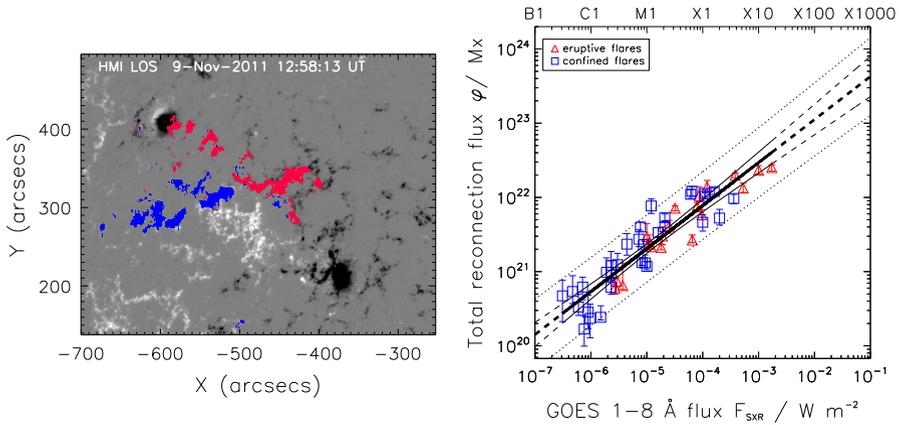

**Fig. 16** **a** Determination of reconnection flux $\varphi$ for an eruptive M1.1 flare on 2011 November 9. The cumulated pixels swept over by the flare ribbons are superimposed on the HMI line of sight magnetogram (scaled to $\pm$ 500 G). Red (blue) areas indicate negative (positive) magnetic polarity. **b** Total flare reconnection flux $\Phi_r$ (defined as the mean of the absolute values of the reconnection fluxes in both polarity regions) versus GOES 1–8 Å SXR peak flux $F_{SXR}$. Blue squares indicate confined events, and red triangles indicate eruptive events. The linear regression line derived in log–log space for all events (thick line) is plotted along with the 95% confidence intervals (thin lines). Image reproduced with permission from Tschernitz et al. (2018), copyright by AAS

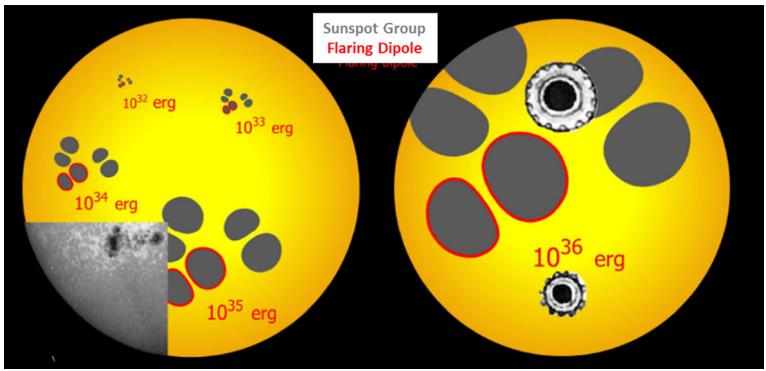

**Fig. 17** A cartoon depicting estimated areas of sunspot groups (exclusive of any surrounding facular areas) needed to power flares with different energy budgets. These estimates were originally developed by, and depicted in, Aulanier et al. (2013) for the parts of the sunspot groups involved in the flaring, which is assumed to be up to a third of the overall group size. In this modified version, Schmieder (2018) tripled the areas of the groups for the bipole involved in flaring (with red outlines) to show the sizes of spot groups required to reach up to a maximum value of a large stellar flare of around $10^{36}$ erg. In the Sun and stars, the flux in two thirds of the region may not be fully contained in spots, of course, but could also be distributed over extended facular regions. In a modification of the figure from Schmieder (2018), we overlaid, in the solar image on the left, the largest observed sunspot group since 1874 (shown in Ca II K1v; full disk image in Fig. 14) for comparison with the group in the upper right quadrant which has an estimated peak flare energy of $10^{33}$ erg. The two symbols in the right-hand image that resemble tires are sunspot drawings of uncertain scale by John of Worcester from 1128 AD December 8 (https://sunearthday.nasa.gov/2006/locations/firstdrawing.php)

Kazachenko et al. (2017) performed a similar study to that of Tschernitz et al. based on over 3000 flare events (C1.0-X5.4) by measuring the ribbons observed in the 1600 Å channel of SDO/AIA (Atmospheric Imaging Assembly; Lemen et al. 2012) instead of Hα spectroheliograms, and mapping these to SDO/HMI magnetograms. They find

$$\Phi_r = 10^{22.13} \left( \frac{\mathcal{C}_{GOES}}{\mathcal{C}_{GOES,X1}} \right)^{0.67} \quad (6)$$

Equation (6) was derived as a linear reduced major axis fit to logarithmic variables (Isobe et al. 1990). The slope of 0.58 derived by Tschernitz et al. (2018) for the fit line in Fig. 16b and that of Kazachenko et al. in Eq. (6) are statistically consistent. Toriumi et al. (2017) obtain a slope of 0.28 ± 0.10 for a sample of 51 events but for a small (M5.0-X5.4) range. For an active region with total unsigned flux of $4 \times 10^{23}$ Mx and 50% flux involvement, Eq. (6) yields an ∼ X55 SXR flare with radiative energy of ∼ $6 \times 10^{32}$ erg from Eq. (5), comparable to the ∼ X45 and $5 \times 10^{32}$ erg values inferred for the 1 September 1859 event.

*(b) Estimates based on active region energy and released flare energy*

As noted in Sect. 3.1.5, solar flares and eruptions are powered by some fraction of the excess or free (i.e., non-potential) magnetic energy ($E_{fm}$) of an active region. $E_{fm}$ is distributed throughout the magnetic field of an active region. Sometimes, authors define a "free energy density" as if that would provide information on where the primary contributions to free energy would be located. However, despite the fact that the difference of the integral of the field energy for non-potential and potential fields can be mathematically written as the integral over the difference of what looks like a local quantity, namely

$$E_{fm} = \int_V \frac{B^2_{non-pot}}{8\pi} dV - \int_V \frac{B^2_{pot}}{8\pi} dV = \int_V \left( \frac{B^2_{non-pot}}{8\pi} - \frac{B^2_{pot}}{8\pi} \right) dV \quad (7)$$

this should not be interpreted to mean that the physical quantity of the resulting integrand maps to, say, the electrical currents that most contribute to the free energy. $E_{fm}$ is intrinsically a large-scale quantity.

Estimation of $E_{fm}$ is complicated by the facts that we cannot directly measure the coronal magnetic field and that models of that field based on the measurements of polarized photospheric light are, at best, uncertain at present. Nevertheless, as summarized by Emslie et al. (2012): "Numerous efforts have been undertaken to estimate nonpotential magnetic energies in active regions near disk center. The methods include: (1) using the magnetic virial theorem estimates from chromospheric vector magnetograms (Metcalf et al. 1995, 2005), (2) semi-empirical flux-rope modelling using Hα and EUV images with MDI LOS magnetograms (Bobra et al. 2008), and (3) MHD modelling (Metcalf et al. 1995; Jiao et al. 1997) and non-potential field extrapolation based upon photospheric vector magnetograms (Guo et al. 2008; Schrijver et al. 2008; Thalmann and Wiegelmann 2008; Thalmann et al. 2008). These methods are labor intensive, and uncertainties in their energy estimates are large. For example, error bars on virial free-energy estimates can exceed the potential magnetic energy. Also, there is considerable scatter in estimates from studies that employ several methods to analyze the same data (e.g., Schrijver et al. 2008). A couple of generalizations, however, can be made. Free energies determined by virial methods matched or exceeded the potential field energy, while free energies estimated using other techniques typically amounted to a few tens of percent of the potential field energy. Published values for free energies in analytic (Schrijver et al. 2006) and semi-empirical (Metcalf et al. 2008) fields meant to model solar fields also hover around a few tens of percent of the potential field energy." Thus, as noted above, Emslie et al. (2012) adopted 30% of the modelled potential energy of an active region as their estimate of the free energy available for eruptive flares.

It may well be that a value of $\sim 30\%$ of the total magnetic energy of an active region represents a maximum value of the free energy beyond which active-region coronal fields cannot be stable. A model example of this can be found in the magnetohydrodynamic experiment performed by Aulanier et al. (2010, 2012) and analyzed in light of extreme flaring by Aulanier et al. (2013). In their bipolar region, one polarity is subjected to a rotational shear, which builds up until an instability in the field develops, manifested in the eruption of a flux rope mimicking a CME. Aulanier et al. (2013) note that the "CME itself was triggered by the ideal loss-of-equilibrium of a weakly twisted coronal flux rope …, corresponding to the torus instability (Kliem & Török 2006; Démoulin and Aulanier 2010)". During the eruption, the field above their modelled bipolar region released 19% of the total magnetic energy, with most of that energy available for the thermal evolution of the corona (i.e., a flare) and the remaining 5% going into the bulk kinetic energy of the CME. While the fraction of total energy released in this numerical experiment approaches 30%, the $\sim 20{:}1$ apportionment between flare and CME is opposite to that of Emslie et al. (2012).

Aulanier et al. (2013) obtained relationships between active region length scale, typical field strength, flux, and energy content. Reformulation of their results yields the following scaling between the total energy $E(AR)$ (erg) contained in the active region magnetic field, the region's total unsigned flux $\Phi$ (Mx), and the core field strength of the simulated spot pair, $B_C$ (Mx cm$^{-2}$):

$$E(AR) = 0.14 \left(\frac{B_C}{1000\text{G}}\right)^{1/2} \Phi^{3/2}. \tag{8}$$

This equation permits an independent determination of the largest possible flare from that determined in Sect. 3.1.7(a). With the scaling of Eq. (8), a maximum field strength of $B_C = 3.5$ kG (Aulanier et al. 2013) and an active region flux of $\Phi_{\text{max}} = 4 \times 10^{23}$ Mx yields a total active region energy $E$ of $6.6 \times 10^{34}$ erg. Apportioning per Sect. 3.1.5, the released energy (approximately one-tenth of active region magnetic potential energy) is $6.6 \times 10^{33}$ erg, with bolometric energy ($\sim$ one-fourth of released energy) of $\sim 1.7 \times 10^{33}$ erg and a SXR class of $\sim 240$ via Eq. (5)[6].

In another estimate of this type, Toriumi et al. (2017) calculated the total energy released in a flare based on the integral of the flux contained in the flare ribbons over time. For an estimated total magnetic flux of $2 \times 10^{23}$ Mx for the April 1947 active region (Fig. 15) and under the assumptions that the ratio of ribbon area to spot area was 0.85 and that two-thirds of the computed energy would be released in a flare, they obtained a released energy of $\sim 1 \times 10^{34}$ erg, which after Emslie et al. (2012) would translate to a bolometric energy $\sim 2.5 \times 10^{33}$ erg, and a $\sim$ X410 SXR flare from Eq. (5).

The likely underestimation of CME energies noted in Sect. 3.1.5 will reduce the SXR class estimates based on the work of Aulanier et al. (2013) and Toriumi et al. (2017). Assuming a 6:1 (vs. 3:1) apportionment of released free energy between CMEs and flares, respectively, results in a SXR class estimate of X105 (X185) for Aulanier et al. (Toriumi et al.).

(c) *Composite estimate*

Altogether, the above estimates have a surprisingly low range of values, all lying within the X80 ($-$ 40, $+$ 120) range for the estimate from Fig. 16b, based on a total unsigned flux of $4 \times 10^{23}$ Mx from Fig. 15. The lower limit SXR class and radiative energy estimates of Kazachenko et al. (X55; $6 \times 10^{32}$ erg) and Tschernitz et al. (X40; $5 \times 10^{32}$ erg) are comparable to the parameters for the 1 September 1859 and 4 November 2003 flares, which originated in spot groups less than half as large as April 1947. Because of this and the apparent upward curvature of magnetic flux with increasing group spot area in Fig. 15, we adopt: (1) the reconnection flux based estimate of X180 ($-$ 100, $+$ 300) given by Fig. 16b from Tschernitz et al. as our preferred estimate of the largest possible flare, and (2) the total unsigned flux value of $6 \times 10^{23}$ Mx they used as the best estimate of the unsigned flux for the April 1947 active region. An unsigned flux of $6 \times 10^{23}$ Mx would increase the SXR flare class (radiative energy) estimate of Kazachenko et al. to X105 ($9.4 \times 10^{32}$ erg), and those of Aulanier et al. and Toriumi et al. to $\sim$ X240 ($1.7 \times 10^{33}$ erg) and $\sim$ X4000 ($1.3 \times 10^{34}$ erg), respectively. We reject the $\sim$ X4000 estimate of Toriumi et al. as a clear outlier—an overestimate apparently due to assumptions regarding spot:ribbon area ratio and the fraction of the computed energy released. From Fig. 8, a 10,000-year flare would have a bolometric energy of $\sim 2 \times 10^{33}$ erg versus the $1.3 \times 10^{34}$ erg value calculated from Eq. (5) for the $\sim$ X4000 estimate from Toriumi et al. Nominal SXR class and energy values of X180 and $1.4 \times 10^{33}$ erg are approximately five and three times as large as the respective values ($\sim$ X40; $\sim 5 \times 10^{33}$ erg) for the 1859 and 2003 flares. Our preferred estimate of X180 ($-$ 100, $+$ 300), indicated by bold/italic font in Table 1, can be

---

[6] The Aulanier et al. (2013) simulations do not have a chromosphere and may be a poor basis for discussing energetics (e.g., no WLF, no H-alpha, no Ly-alpha etc.), a limitation of MHD modelling acknowledged by Aulanier et al. (2012; their Sect. 4.4). But the energy contained in the non-potential field can be estimated. The limitations of the MHD modelling then led us to use the 'apportioning' as per Emslie et al. (2012) from observations rather than from an MHD model.

**Table 1** Estimates of the largest possible present-era solar flare based on an unsigned magnetic flux of $6 \times 10^{23}$ Mx for the active region of April 1947 (peak daily area = 6132 μsh), and a 6:1 CME:flare energy apportionment

| Nos. | Bolom. energy (erg) | SXR class | April 1947 unsigned mag flux (Mx) | Assumptions | Based on |
|---|---|---|---|---|---|
| (1) | $9.4 \times 10^{32}$ | ∼ X105 | $6 \times 10^{23}$ | 50% flux involvement | Kazachenko et al. (2017) (reconnection flux) |
| (2) | ***$1.4 \times 10^{33}$*** | ***X180 (− 100, + 300)*** | ***$6 \times 10^{23}$*** | 50% flux involvement | Tschernitz et al. (2018) (reconnection flux) |
| (3) | $1.7 \times 10^{33}$ | ∼ X240 | $6 \times 10^{23}$ | See Footnote 6 | Aulanier et al. (2013) (active region energy) |

The uncertainty range of the preferred estimate (bold/italics) based on Tschernitz et al. encompasses the other two estimates considered a composite because it encompasses the ∼ X105 and ∼ X240 estimates of Kazachenko et al. and Aulanier et al., respectively, with all three estimates based on an unsigned magnetic flux of $6 \times 10^{23}$ Mx. The lower end of the X180 (− 100, + 300) estimate for the largest possible flare based on the April 1947 active region exceeds that of the largest flares observed since 1859 (from smaller spot groups) and the upper limit encompasses the X285 (± 140) SXR class estimated for the inferred flare for the 774–775 AD SEP event (Sect. 7.8).

### 3.2 Flares on Sun-like stars

#### 3.2.1 Stellar flare research before *Kepler*

The first generally recognized observations of stellar flares (on the faint dwarf star L 726-8; also designated UV Ceti) occurred on 25 September 1948 (Joy and Humason 1949) and 7 December 1948 (Luyten 1949a, b). Luyten (1949a) noted that Edwin Carpenter's photographic plates, taken on 7 December with the 36-in. telescope at Steward Observatory in Tucson (Fig. 18), give "the first observation of the extreme rapidity of the [flare brightness] change—to twelve times the original luminosity [of the star] in less than 3 min …" Precursor observations of stellar flares were made by Hertzsprung in 1924 (Bastian 1990; Gershberg 2005), Luyten (1926) and Van Maanen (1940, 1945). Subsequently, radio (Lovell 1963; Slee et al. 1963a, b; Orchiston 2004) and X- ray emissions (Heise et al. 1975; Güdel 2004) were found to accompany optical stellar flares.

Joy and Humason (1949) and Luyten (1949a, b) used the terms "flare-up" or "flare" when describing the stellar phenomenon but neither paper makes an analogy to solar flares. Luyten (1949b) suggested that the cause might be "the same as that which produces a nova, or an SS Cygni star." In the preface of his monograph on solar activity in main sequence stars, Gershberg (2005) places the establishment of the similarity of solar and stellar flares in the mid-1960s, following the first high-time resolution spectra of flares on UV Ceti variables. As noted in Sect. 3.1.3, several lines of evidence support the close relationship of physical processes observed in stellar and solar flares.

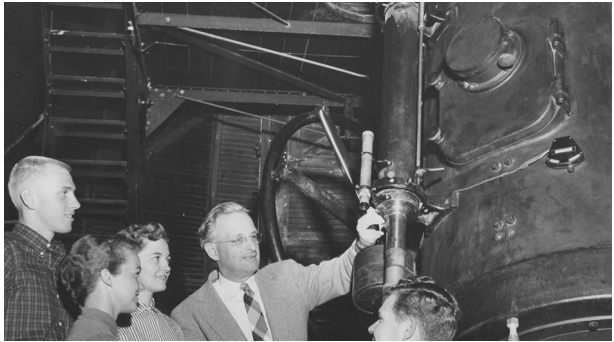

**Fig. 18** Edwin Carpenter (center) shows University of Arizona students the 36-in. telescope of the Steward Observatory in 1957. Carpenter was the first to record the rapidity with which some stellar flares can unfold. Image reproduced with permission from https://speccoll.library.arizona.edu/, copyright by University of Arizona

### 3.2.2 Discovery of superflares on solar-type (vs. Sun-like) stars

X-ray observations show that young stars can have very large flares, termed "superflares" (Schaefer et al. 2000), defined as flares with radiative energy more than $10^{33}$ erg ($\sim$ 2–3 times more energetic than the largest directly observed solar flares; Emslie et al. 2012). These young stars rotate very fast with rotational periods of a few days, much less than solar sidereal rotation period of $\sim$ 25 days at the equator. Generally, such fast rotating stars are strong X-ray sources even in their quiescent phase, indicating the presence of strong magnetic field and often with evidence that large fractions of the stellar surface are covered by starspots (Pevtsov et al. 2003). Consequently, it has been assumed that superflares would never occur on the present Sun, since the Sun is old and is slowly rotating.

However, by analyzing previously existing astronomical data, Schaefer et al. (2000) identified nine superflares with energies $\sim 10^{33-38}$ erg in ordinary solar-type stars (i.e., G type main sequence stars with rotational velocities less than 10 km s$^{-1}$; rotation period > 5 days). Concurrently, Rubenstein and Schaefer (2000) argued from analogy with RS CVn binary star systems that superflares on solar-type stars were caused by a "hot Jupiter" (Mayor and Queloz 1995; Marcy and Butler 1998; Schilling 1996), a Jovian-type exoplanet orbiting close-in to these stars. Rubenstein and Schaefer thus implied that—because of the lack of a hot Jupiter—the Sun is not a candidate to produce superflares. Cuntz et al. (2000) laid the framework for how a close-in giant planet could cause stellar activity via gravitational and magnetic interaction. Ip et al. (2004) used a numerical MHD simulation to show that a star-planet magnetic interconnection could lead to energy release comparable to that of a typical solar flare and Lanza (2008) explained the phase relation between stellar hot spots and planetary location in terms of such a linkage. Direct observational evidence between stars with close-in Jovian planets and superflares was lacking, however (e.g., Saar et al. 2004).

One of the authors of this review, Kazunari Shibata, questioned the hot-Jupiter hypothesis of Rubenstein and Schaefer (2000) because the RS CVn stars on which the hypothesis is based have short rotation periods due to tidal locking. Hence in 2010 Shibata started to encourage young researchers and students to search for superflares on solar-type stars (G type dwarfs) observed by *Kepler*. As a result, Maehara et al. (2012) reported 365 superflares (with radiative energy $> 10^{33}$ erg) on 148 "solar-type" stars (defined as G-type main sequence stars with effective temperature of 5100–6000 K and log (surface gravity $(cm/s^2)) \geq 4.0$) during 120 days of *Kepler* observations. For more restrictive criteria, they identified [9] superflares on 5 "Sun-like" stars in the sample with effective temperatures of 5600–6000 K and rotational periods between 11.0 d and 17.1 d (Maehara et al.; Supplementary Information). As a reflection of Shibata's encouragement, the Nature paper reporting these results included five undergraduate researchers as co-authors (T. Shibayama, S. Notsu, Y. Notsu, T. Nagao, and S. Kusaba).

Later, Shibayama et al. (2013) extended the survey and confirmed the work of Maehara et al. by finding 1547 superflares (including those reported by Maehara et al. 2012) on 279 G-type dwarf stars during 500 days (from 2009 May to 2010 September) of *Kepler* observations. In all, they found 44 superflares on 19 Sun-like stars. Of these 19 Sun-like stars, three had rotational periods longer than the solar rotation period of $\sim 25$ days, suggesting that slowly rotating stars like the Sun could exhibit superflares.

Maehara et al. (2012) and Shibayama et al. (2013) argued that the superflares they discovered were not due to hot Jupiters orbiting the superflare stars. Quoting from Shibayama et al., "According to the *Kepler* candidate planet data explorer (Batalha et al. 2013), 2321 planets have been found in 1790 stars among 156,453 stars. Hence, the probability of finding exoplanets orbiting stars is about 1%. Howard et al. (2012) showed that the probability of finding a hot Jupiter was 0.5%. However, none of our superflare stars (279 G-type dwarfs [of which 69 had rotation periods $> 10$ d]) have a hot Jupiter according to the data explorer. For a G-type dwarf with a hot Jupiter, the probability of [detecting] a transit of the planet across the star is about 10% averaged over all possible orbital inclinations (Kane and von Braun 2008). If all of our 279 [69] superflare stars are caused by a hot Jupiter as suggested by Rubenstein and Schaefer (2000), *Kepler* should detect 28 [7] of them from transits" versus the zero that were found. Recently, for a sample of 265 solar-type super flare stars (G-type main-sequence; effective temperature of 5100–6000 K; 139 with rotation periods $> 10$ days), with 43 in common with the 279 G-type dwarfs of Shibayama et al. 2013), Okamoto et al. (2021) found only three stars with candidate exoplanets in the NASA Exoplanet Archive (https://exoplanetarchive.ipac.caltech.edu/).

Figure 19a shows a typical example of a superflare observed by *Kepler*, with a spike-like increase peaking at 1.5% of the stellar brightness for a few hours. Flares on Sun-like (solar-type) stars have brightness variations in the range of $\sim 0.1$

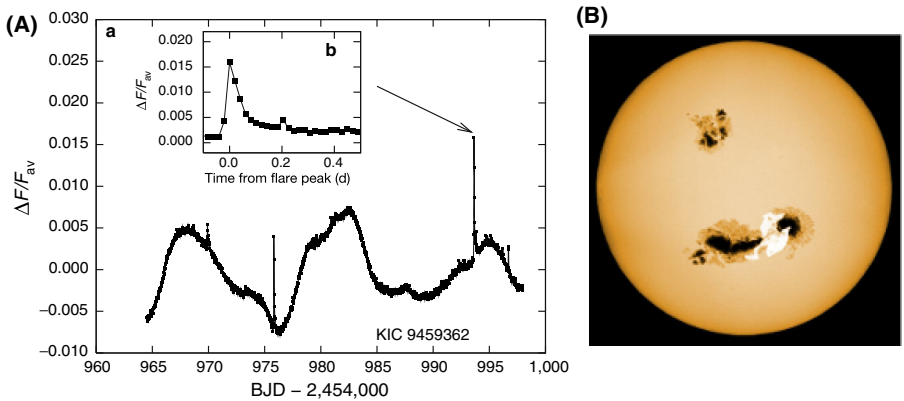

**Fig. 19** **a** (left) Typical example of a superflare (shown on an expanded time scale in the inset) on a solar-type star. **b** (right) An artist's conception of a superflare and big starspots on a solar-type star based on *Kepler* observations. Images reproduced with permission from [left] Maehara et al. (2012), copyright by Macmillan; and [right] courtesy of H. Magara

to ∼ 20% of the stellar luminosity (Notsu et al. 2013a). In contrast, one of the largest solar flares in the past 20 years (X17 SXR flare on 28 October 2003 shown in Fig. 13; Woods et al. 2004; Kopp 2016) showed only a ∼ 0.03% solar brightness increase for ∼ 20 min (FWHM). The estimated bolometric energy of the superflare in Fig. 19a is ∼ $10^{35}$ erg, ∼ 200 times larger than that for the largest solar flare at ∼ $5 \times 10^{32}$ erg.

In Fig. 19a the stellar brightness itself shows significant rotational variation with an amplitude of ∼ 1% at a characteristic time of 10–15 days. Almost all superflare stars show such a variation ranging from 0.1 to 1% (Maehara et al. 2012) versus ∼ 0.1% for the 11-year solar cycle TSI variation (Willson and Hudson 1991; Kopp 2016). This stellar brightness variation can be interpreted in terms of the rotation of a star with a substantial coverage by unevenly distributed starspots (Strassmeier et al. 2002, and references therein; Strassmeier 2009; Notsu et al. 2013a; Namekata et al. 2019), such as depicted in Fig. 19b. The interpretation of *Kepler* brightness variations in terms of rotation of large area starspot groups can be used to measure the rotation period of stars and to infer the fractional coverage of starspots (technically, one can only infer the maximum difference between opposing hemispheres of total spot coverage; see below), or total magnetic flux, assuming the average magnetic flux density is the same as that of a large sunspot. The derivation of superflare star, and superflare, parameters from *Kepler* light curves will be discussed in Sect. 3.2.4 below.

Before proceeding further, we need to add a caveat. Up to this point, we have been using the terms radiative energy, bolometric energy, and TSI interchangeably, taking it to mean flare-radiated energy time-integrated over all wavelengths. For *Kepler* white light flare observations, the term "bolometric flare energy" has a different, more limited, definition. The "bolometric energy" as referenced for *Kepler* here is the flare-radiated energy in the blackbody continuum, assuming the flare to have an effective temperature of 10,000 K to convert the emission that occurs in the *Kepler* 4000–9000 Å bandpass into a bolometric energy (e.g., Notsu et al. 2013a) based on a solar study by Kretzschmar (2011). We further note that there is no consideration of line versus continuum emission in the photosphere and chromosphere or allowance for coronal emission. Thus far such computations have not been done for solar-like stars other than the Sun. Consequently, the *Kepler* flare bolometric energies could be underestimated by up to a factor of ten in comparison with solar values based on actual TSI measurements (Osten and Wolk 2015).

### 3.2.3 Verifying the Sun-like nature of superflare stars

*(a) Initial verification*

It is long known that close binary systems such as the RS CVn-type (Hall 1976) can produce superflares with energies $\sim 10^{35}$ erg or more (e.g., Doyle et al. 1991). Doppler observations can be used to eliminate such binary systems (except for systems seen (nearly) pole-on), as well as single stars rotating much faster than the Sun, from the *Kepler* sample of nominally Sun-like superflare stars. Other spectroscopic observations can be used to substantiate the Sun-like character of superflare stars. These include measurements of: Hα and Ca II 8542 fluxes (gauges of chromospheric activity; Linsky et al. 1979; Herbig 1985; Foing et al. 1989; Soderblom et al. 1993), Ca II K flux (indicator of the stellar mean magnetic field strength; Skumanich et al. 1975; Schrijver et al. 1989); Fe/H ratio (metallicity) and Li abundance (both indicators of stellar age; Edvardsson et al. 1993; Skumanich 1972; Soderblom 2010), and equivalent widths of Fe I and Fe II lines (effective temperature and surface gravity; Takeda et al. 2002, 2005; Valenti and Fischer 2005). Early spectroscopic studies of superflare stars were conducted by Notsu et al. (2013b; one star), Nogami et al. (2014; two stars), and Wichmann et al. (2014; 11 stars). The detailed discussion of the 11 stars examined in Wichmann et al. illustrates the complexity of the task, with all indicators of "Sun-likeness" seldom pointing in the same direction.

Notsu et al. (2015a, b) reported on their spectroscopic investigation of 34 non-binary solar-type (G-type main sequence) superflare stars observed with the High Dispersion Spectrograph (HDS; Noguchi et al. 2002) on the Japanese 8.2-m Subaru telescope in Hawaii. As a check on light-curve based stellar rotation periods (and inferred rotation rates), Notsu et al. (2015a), used HDS data to determine the projected rotation rates ($v \sin i$; where $i$ is the inclination angle between the stellar rotational axis and the observer's line of sight, measured from the pole; e.g., for $i = 0°$, the stellar rotation axis points directly along the line-of-sight) of these 34 stars[7] and parameters such as metallicity, effective temperature, and surface gravity (all via Fe I and Fe II line widths) to isolate a sample of seven Sun-like stars that met the following criteria: $5600 \leq T_{\mathrm{eff}} \leq 6000$ K, $\log g \geq 4.0$, and rotation periods (based on brightness variation) > 10 days.

The seven Sun-like stars identified by Notsu et al. (2015a) are listed in the first seven rows of Table 2 along with their effective temperature ($T_{\mathrm{eff}}$), surface gravity

---

[7] Following Takeda et al. (2008) and Gray (1929, 2005), Notsu et al. (2013a; 2015b) determined the rotational turbulence ($v_{\mathrm{rt}}$) from its relationship with instrumental ($v_{\mathrm{ip}}$) and macroturbulence ($v_{\mathrm{mt}}$) line broadening parameters, with $v \sin i$ (and $i$) deduced from an empirical relationship with $v_{\mathrm{rt}}$. The inclination angle $i = \arcsin (v \sin \mathrm{i} / v_{\mathrm{lc}})$.

Table 2  Sun-like superflare stars identified by *Kepler* (from Nogami et al. 2014; Notsu et al. 2015a, b, 2019)

| KIC number | $T_{eff}$ (°K) | log (g) (cm/s$^2$) | Fe/H (%) | $BV^a$ (%) | $fB^b$ (G) | Aspot (μsh) | $v_{lc}$ (km/s) | $v \sin(i)$ (km/s) | $P_{rot}$ (d) | Max flare ($10^{33}$ erg) |
|---|---|---|---|---|---|---|---|---|---|---|
| KIC 6865484  | 5800 (± 28) | 4.52 (± 0.07) | − 0.12 (± 0.03) | 0.72 | 52 (± 41) | 12,000 | 4.5 | 2.7 | 10.3 | 99 |
| KIC 7354508  | 5620 (± 25) | 4.09 (± 0.06) | − 0.10 (± 0.03) | 0.50 | 16 (± 24) | 9000   | 4.5 | 3.2 | 16.8 | 7.7 |
| KIC 9766237  | 5606 (± 40) | 4.25 (± 0.11) | − 0.16 (± 0.04) | 0.09 | 1 (± 6)   | ≤1000  | 4.2 | 2.1 | 14.2 | 13.7 |
| KIC 9944137  | 5666 (± 35) | 4.46 (± 0.09) | − 0.10 (± 0.03) | 0.07 | 1 (± 6)   | ≤1000  | 3.7 | 1.9 | 12.6 | 9.9 |
| KIC 10471412 | 5776 (± 20) | 4.53 (± 0.05) | 0.15 (± 0.03)   | 0.24 | 7 (± 24)  | 2400   | 3.2 | 2.7 | 15.2 | 520 |
| KIC 11390058 | 5921 (± 25) | 4.43 (± 0.07) | − 0.15 (± 0.03) | 0.33 | 4 (± 24)  | 3300   | 4.2 | 3.5 | 12.0 | 17 |
| KIC 11494048 | 5907 (± 25) | 4.49 (± 0.07) | 0.01 (± 0.03)   | 0.30 | 4 (± 24)  | 3000   | 3.4 | 3.4 | 14.8 | 28 |
| KIC 11253827 | 5686 (± 23) | 4.76 (± 0.06) | 0.19 (± 0.05)   | 1.59 | –         | 16,000 | 3.4 | <4  | 13.4 | 1.6 |
| Sun          | 5780        | 4.44          | 0.01            | ∼ 0.06 | ∼ 10$^c$ | –      | 2.0 |     | ∼ 25 | 0.4 |

This list of Sun-like stars originally consisted of nine events (Notsu et al. 2015a). One of these nine stars (KIC11764567), commented on in Notsu et al. (2015b), did not satisfy the $v \sin i \lesssim v_{lc}$ condition and is not included here. Also, KIC8547383 was removed because its *Gaia*-DR2 radius value is in the sub-giant range (Notsu et al. 2019)

$^a$Average amplitude of brightness variability over a rotation period (McQuillan et al. 2014; Notsu et al. 2019; Witzke et al. (2020) for the Sun)

$^b f$ is a filling factor (Notsu et al. 2015a, b); 1 G = $10^{-4}$ T

$^c$Fig. 3 in Schrijver and Harvey (1994) suggests $6 \times 10^{23}$ Mx as a typical total unsigned flux value for the entire Sun, which translates to a globally-averaged value of ∼ 10 G

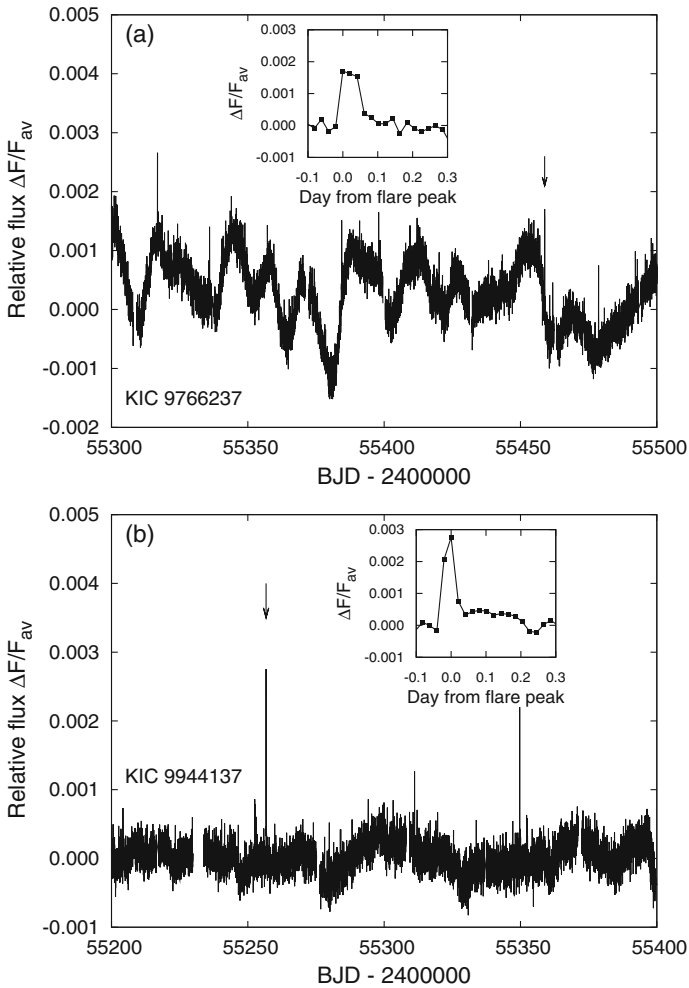

**Fig. 20 a** Typical light curve of KIC 9766237 from the long cadence *Kepler* data. The y-axis represents the relative flux normalized by the average flux: $(F - F_{av})/F_{av}$. Quasi-periodic modulations with a timescale of about 14 days (see text) are seen. An arrow points out the superflare. The inset figure shows an enlarged light curve around the superflare. The amplitude is about 0.17%, and the duration is about 0.1 days. Though many "spikes" other than that of the superflare are seen, they consist of only one point. **b** Same as **a**, but for KIC 9944137. The amplitude of the superflare in the inset figure is about 0.28%, and the duration exceeds 0.2 days. Image reproduced with permission from Nogami et al. (2014), copyright by the authors

(g), amplitude of brightness variation (*BV*), metallicity (Fe/H), mean magnetic flux density (*fB*), spot area ($A_{spot}$), rotational velocity (determined both photometrically ($v_{lc}$ (lc = light curve)) and spectroscopically ($v \sin(i)$), rotation period ($P_{rot}$), and bolometric energy of the largest observed flare (Max flare). Recently an additional Sun-like star (KIC 11253827) has been identified by Notsu et al. (2019). Its parameters are given in row 8. Corresponding parameters for the Sun are given in the bottom row.

Light curves for the two stars in Table 2 that were analyzed in detail by Nogami et al. (2014) are shown in Fig. 20. Those authors reported photometrically-based rotation periods of 21.8 days and 25.3 days for KIC 9766237 and KIC 9944137, respectively. Subsequently, these two periods were reduced in Notsu et al. (2015a, see their Appendix A1 (Supplementary data)) to the 14.2 day and 12.6 day values shown in Table 2. The Nogami et al. (2014) periods were calculated by Shibayama et al. (2013) from *Kepler* Quarters 0–6 data while those from Notsu et al. (2015b) were based on updated Quarter 2–16 data.

Witzke et al. (2020) document the difficulty of obtaining photometry-based light curves for "solar-like" stars by analyzing a set of synthesized light curves for stars that are identical to the Sun except for their metallicity. They find a minimum in the detection rate of the rotation period for stars having a metallicity value close to that (0.01) of the Sun because of the close balance of facular and spot contributions in our star. Witzke et al. substantiate this result by considering a sample of *Kepler* solar-like stars with a range of metallicities from − 0.35 to 0.35 and near-solar effective temperatures and photometric variabilities. The underlying assumption of long-term coherence, i.e., of long-lived spots comparable to the stellar rotation period, is not tenable for the Sun (in its present state). As of 2019, KIC 9766237 and KIC 9944137 appeared to be the most Sun-like superflare stars in terms of their brightness variation ($BV$) and rotation velocities (Y. Notsu, personal communication, 2019), but as noted below, even the reduced rotation periods from Notsu et al. (2015a) are now in question. The two associated superflares (one per star) had energies of $\sim 10^{34}$ erg, at the lower end of the *Kepler* superflare range and had associated Sun-like spot areas of $\lesssim 1000$. Such spot areas are at or below the limit of *Kepler* detection (Nogami et al. 2014), but see Sect. 3.2.4 below for a different interpretation of these inferred values.

Maehara et al. (2017) identified two other candidates for Table 2, both with photometrically-determined rotation periods > 20 days. KIC 6347656 and KIC 10011070 have spot area brightness-variation-based rotation periods (spot areas, $T_{\text{eff}}$) of 28.4 days (3600 μsh, 5623 K) and 24.0 days (3400 μsh, 5669 K), respectively. Each of these stars had one superflare with bolometric energy of $3.1 \times 10^{34}$ erg for KIC 6347656 and $2.6 \times 10^{34}$ erg for KIC 10011070. The two stars have yet to be examined spectroscopically (Y. Notsu, personal communication, 2019).

In Table 2, the mean magnetic flux densities (fB) for the Sun-like stars cannot be distinguished from that of the Sun. The eight listed stars in Table 2 have a median flux density of 4($\pm$ 24) G versus $\sim 10$ G for the Sun. The values of $<fB>$ are derived from a calibration against the intensity of the stellar Ca II infrared triplet relative to its local continuum (Notsu et al. 2015b). However, this calibration assumes observations of an area around a single active region. The authors do not discuss the consequences of using such observations to calibrate the signal against the disk-integrated signal. Furthermore, the uncertainties that they list are characteristic of the spread of points around the mean relationship, not the uncertainty in the mean itself. Finally, we note that that relationship when applied to the Sun yields a mean magnetic flux density of 0.2G, where observations show values of $\sim$ 10 G, i.e., $\sim$ 1.5 orders of magnitude larger. In addition, Notsu et al. (2015a, b) make no mention of center-to-limb effects for the transformation from their single spot group measurement to a disk-integrated signals for stars (cf. Namekata et al. 2017).

### (b) A subsequent challenge

Subsequent development and refinement of the *Kepler* data base challenged the case for superflares on Sun-like stars. Quoting from Notsu et al. (2019): "The recent *Gaia*-DR2 stellar radius data (Berger et al. 2018; Lindegren et al. 2007, 2018) have suggested the possibility of severe contaminations of subgiant stars in the classification of *Kepler* solar type (G-type main-sequence) stars used for the statistical studies of superflares … The classification of solar-type stars in our previous studies … was based on $T_{\text{eff}}$ and log $g$ values from the *Kepler* Input Catalog … and there can be large differences between the real and catalog values. For example, Brown et al. (2011) reported that uncertainties of $T_{\text{eff}}$ and log $g$ in the initial KIC are $\pm$ 200 K and 0.4 dex [decimal exponent], respectively. … In a strict sense, we cannot even deny an extreme possibility that all of the slowly rotating Sun-like superflare stars [we considered] were the results of contaminations of subgiant stars. In addition, the Kepler solar-type superflare stars discussed in our previous studies can include some number of binary stars [excepting the events in Table 2 that were confirmed to be single, and Sun-like, by spectroscopic examination]. This is a problem, especially for investigating whether even truly Sun-like stars can have large super flares or not."

To address the above concern, Notsu et al. (2019) revisited earlier analyses (e.g., Maehara et al. 2012, 2015, 2017; Shibayama et al. 2013) by using the *Gaia*-DR2 stellar radius estimates (Berger et al. 2018) and $T_{\text{eff}}$ values updated in DR25-KSPC (Mathur et al. 2017; Pinsonneault et al. 2012). In so doing, they found that of the 245 (of 279) stars in the Shibayama et al. (2013) sample of solar-type (G-type dwarfs) superflare stars (from 2009 May to 2010 September) for which Berger et al. (2018) determined a radius, 108 ($\sim$ 45%) were subgiants (with one red giant) and, of the remaining 136 identified as main-sequence stars, only 106 had $T_{\text{eff,DR25}}$ from 5100–6000 K. Spectroscopic analysis (3200–10,000 Å) was then conducted using the Apache Point Observatory 3.5 m telescope for 18 main-sequence superflare stars observed by *Kepler* at 1-min cadence for which *Gaia*-DR2 (Release 2) stellar radius and $T_{\text{eff,DR25}}$ values were available; 13 of these 18 stars were found to be single.

### (c) Further development

In a further development based on the entire *Kepler* primary mission data set (*Kepler* Data Release 25 (Thompson et al. 2016; Mathur et al. 2017); 2009 May to 2013 May) and the Gaia-DR2 catalog, Okamoto et al. (2021) used a flare detection method that employed high-pass filtering to remove low-frequency rotational variations due to starspots to identify superflares on 15 Sun-like stars (based on $T_{\text{eff}}$ from 5600 to 6000 K, a more stringent $P_{\text{rot}}$ requirement of $>$ 20 d (with $P_{\text{rot}}$ taken from McQuillan, et al. 2014), and age of 4.6 Gyr). The parameters for these 15 Sun-like superflare stars are given in Table 3 (for 15 vs. the 16 in the table, see Flag 2). This list includes the two Sun-like stars with rotation periods > 20 days identified by Maehara et al. (2017) noted above. The range of spot areas from ∼ 1500 to 13,000 μsh for the superflare stars in Table 3 are comparable to those in Table 2 while, as noted by Okamoto et al. (2021), the faster rotating stars in Table 2 are capable of producing stronger superflares (by up to an order of magnitude) for the listed events. At present the case for superflares on Sun-like stars is based largely on Table 3.

In a reappraisal of KIC 9766237 and KIC 9944137 (from Table 2) for which light curves are shown in Fig. 20 above, Okamoto et al. (2021) assess them to be "Sun-like star candidates", rather than Sun-like stars, because "the amplitude of the rotational brightness variation is so small that [its] value is not reported in McQuillan et al. (2014), and the accuracy of the $P_{\rm rot}$ value is considered to be low." In contrast, as discussed in Sect. 8.3.1 below, the inability to photometrically detect a rotation period may be a signature of a Sun-like star.

### 3.2.4 Inferring superflare star properties from brightness variations

From the *Kepler* photometric observations, it is possible to estimate the stellar rotation period, starspot area, and flare energy. We will consider each of these in turn.

Maehara et al. (2012) and Shibayama et al. (2013) calculated the rotational periods of superflare stars by discrete Fourier analysis of the stellar brightness variations for the first 500 days of the *Kepler* mission. They chose the largest peak that exceeded the error level as the stellar rotation period. Alternatively, McQuillan et al. (2014) used an automated autocorrelation function (ACF) approach to determine rotation periods based on the entire *Kepler* data set.

The lower limit of the spot area of superflare stars ($A_{\rm spot}$) can be inferred from the $\Delta F_{\rm rot}$ (defined as the full amplitude of the rotational brightness variation normalized by the average brightness) as follows:

$$\Delta F_{\rm rot} \approx \left[1 - \left(\frac{T_{\rm spot}}{T_{\rm star}}\right)^4\right] \frac{A_{\rm spot}}{A_{\rm star}} \qquad (9)$$

where $T_{\rm spot}$ is the spot temperature, $T_{\rm star}$ is the effective temperature of the unspotted photosphere of the star, and $A_{\rm star}$ is the apparent area of the star (Maehara et al. 2012, 2017). The spot area in this equation is a lower limit because it is assumed that there is a reference hemisphere that has no (or negligible) starspot coverage (e.g., Fig. 3b in Notsu et al. 2013a). Notsu et al. (2019) used a relationship between star temperature and spot temperature (Maehara et al. 2017) for which $T_{\rm star}$ (in the range from 5600 to 6000 K) was taken from the *Kepler* catalog and $T_{\rm spot}$ was determined spectroscopically from intensities of the photospheric lines of sunspots.

Figure 21 (taken from Maehara et al. 2017) is a downward cumulative frequency distribution (solid line) showing the occurrence rate of active regions of a given spot area for solar-type stars (defined to be early G- and late F-type main-sequence stars

**Table 3** Parameters of Sun-like superflare stars

| KIC | $T_{\text{eff}}$ [a] (K) | $R^a$ $R_\odot$ | $P_{\text{rot}}$ [c] (day) | Amp [c] (ppm) | $m_{\text{Kepler}}$ [d] (mag) | $N_{\text{flare}}$ [e] | $E_{\text{max}}$ [e] (erg) | $A_{\text{spot}}$ [f] ($1/2 \times A_\odot$) | Flag[g] |
|---|---|---|---|---|---|---|---|---|---|
| 005648294 | $5653 \pm 198$ | $1.140_{(-0.082)}^{(+0.091)}$ | $22.336 \pm 0.085$ | 2543.81 | 14.757 | 1 | $2.4 \times 10^{34}$ | $4.3 \times 10^{-3}$ | 1 |
| 005695372 | $5791 \pm 203$ | $0.086_{(-0.060)}^{(+0.066)}$ | $21.195 \pm 0.130$ | 8881.39 | 14.054 | 3 | $7.7 \times 10^{33}$ | $8.9 \times 10^{-3}$ | – |
| 006347656[b] | $5623 \pm 197$ | $0.823_{(-0.057)}^{(+0.062)}$ | $28.441 \pm 0.598$ | 4110.04 | 14.86 | 4 | $2.2 \times 10^{34}$ | $3.6 \times 10^{-3}$ | 1 |
| 006932164 | $5731 \pm 201$ | $1.038_{(-0.093)}^{(+0.104)}$ | $22.631 \pm 0.077$ | 8538.25 | 15.883 | 1 | $4.0 \times 10^{34}$ | $1.3 \times 10^{-2}$ | – |
| 007435701 | $5812 \pm 203$ | $0.809_{(-0.057)}^{(+0.063)}$ | $20.648 \pm 0.053$ | 8039.18 | 15.09 | 1 | $9.1 \times 10^{33}$ | $6.7 \times 10^{-3}$ | – |
| 007772296 | $5616 \pm 197$ | $0.897_{(-0.063)}^{(+0.070)}$ | $23.257 \pm 0.083$ | 4136.53 | 14.88 | 3 | $1.1 \times 10^{34}$ | $4.3 \times 10^{-3}$ | 2 |
| 007886115 | $5729 \pm 201$ | $0.873_{(-0.060)}^{(+0.066)}$ | $20.351 \pm 0.110$ | 5280.17 | 14.96 | 1 | $9.0 \times 10^{33}$ | $5.2 \times 10^{-3}$ | – |
| 008090349 | $5760 \pm 202$ | $1.169_{(-0.083)}^{(+0.092)}$ | $24.898 \pm 3.822$ | 2482.89 | 14.675 | 1 | $1.8 \times 10^{34}$ | $4.4 \times 10^{-3}$ | – |
| 008416788 | $5889 \pm 206$ | $0.908_{(-0.061)}^{(+0.068)}$ | $22.519 \pm 0.042$ | 3300.19 | 13.582 | 5 | $1.7 \times 10^{34}$ | $3.4 \times 10^{-3}$ | 1 |
| 009520338 | $5947 \pm 208$ | $0.805_{(-0.058)}^{(+0.064)}$ | $23.857 \pm 0.093$ | 4138.47 | 15.27 | 1 | $9.9 \times 10^{33}$ | $3.4 \times 10^{-3}$ | – |
| 010011070[b] | $5669 \pm 198$ | $0.770_{(-0.053)}^{(+0.059)}$ | $23.970 \pm 0.623$ | 4423.92 | 14.949 | 2 | $1.6 \times 10^{34}$ | $3.4 \times 10^{-3}$ | 1 |
| 010275962 | $5782 \pm 202$ | $0.789_{(-0.058)}^{(+0.064)}$ | $26.117 \pm 1.426$ | 2323.69 | 15.296 | 2 | $1.9 \times 10^{34}$ | $1.9 \times 10^{-3}$ | – |
| 011075480 | $5953 \pm 208$ | $0.795_{(-0.057)}^{(+0.063)}$ | $21.651 \pm 0.278$ | 2001.46 | 15.108 | 1 | $9.3 \times 10^{33}$ | $1.6 \times 10^{-3}$ | – |
| 011141091 | $5756 \pm 201$ | $0.879_{(-0.062)}^{(+0.068)}$ | $23.446 \pm 0.373$ | 2345.32 | 14.982 | 1 | $1.5 \times 10^{34}$ | $2.3 \times 10^{-3}$ | – |
| 011413690 | $5886 \pm 206$ | $0.818_{(-0.062)}^{(+0.069)}$ | $24.238 \pm 0.246$ | 3232.87 | 15.725 | 1 | $3.0 \times 10^{34}$ | $2.7 \times 10^{-3}$ | 1 |
| 012053270 | $5971 \pm 209$ | $0.744_{(-0.053)}^{(+0.058)}$ | $21.796 \pm 1.587$ | 4338.8 | 15.017 | 1 | $3.1 \times 10^{34}$ | $3.0 \times 10^{-3}$ | – |

Table reproduced with permission from Okamoto et al. (2021), copyright by AAS

[a] Effective temperature and stellar radius values are taken from Gaia-DR2 in Berger et al. (2018). Stellar radii are shown in the unit of the solar radius ($R_\odot$)

[b] These 2 stars are also reported in Notsu et al. (2019)

[c] Rotation periods and average rotational variability are reported in McQuillan et al. (2014)

[d] Kepler-band magnitude (Thompson et al. 2016)

[e] $N_{\text{flare}}$ and $E_{\text{max}}$ are the number of superflares and maximum energy of superflares of each Sun-like star, respectively

[f] Area of the starspots on each Sun-like star in the unit of the area of the solar hemisphere ($1/2 \times A_\odot \sim 3 \times 10^{22}$ cm$^2$)

[g] Flag 1: It is possible the real rotation period value of the star is different from those reported in McQuillan et al. (2014) and used in this table, on the basis of our extra analyses. Then the rotation period value of these stars should be treated with caution, and it is possible these stars are not really Sun-like stars. Flag 2: This star is suspected to have a binary companion

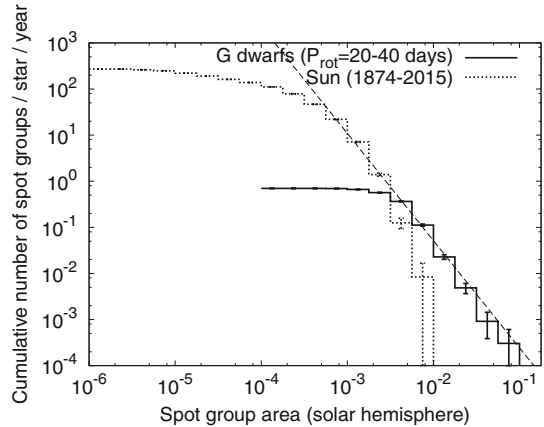

**Fig. 21** Cumulative frequency distribution of starspots on slowly rotating solar-type stars (solid line) and that of sunspot groups (dotted line). The thin dashed line represents the power-law fit (with index = $-2.3 \pm 0.1$) to the frequency distribution of starspots in the spot area range of $10^{-2.5}$ to $10^{-1.0}$. Image reproduced with permission from Maehara et al. (2017), copyright by the authors

that meet both of the following criteria: (1) 5600 K < $T_{\text{eff}}$ < 6300 K and log $g$ > 4.0 in Huber et al. (2014); and (2) 5100 K < $T_{\text{eff}}$ < 6000 K and log $g$ > 4.0 in Brown et al. (2011)). The corresponding solar distribution (dotted line) is shown for ∼ 140 years of solar data. The continuation of the extrapolated power-law fit to the distribution for solar-type stars (dashed line) through that of solar active regions suggests that both sunspots and larger starspots might be produced by the same physical process and that the Sun is capable of producing spot groups much larger than observed in modern times. Maehara et al. (2017) suggest that the steep drop-off of the solar distribution is due to "the lack of a 'super-active' phase on our Sun during the last 140 years". This seems problematic, however, because the four solar cycle interval of solar activity from ∼ 1945 to 1995, termed the modern grand maximum, is considered to be one of the strongest in the past 2000–4000 years (Solanki et al. 2004; Usoskin et al. 2006a; Muscheler et al. 2006; Usoskin 2017). Figure 21 indicates that a ∼ 30,000 μsh spot region, region ∼ 5 times larger than that of April 1947, would be expected to be observed once every ∼ 1000 years, with no evidence of a region approaching this size over the ∼ 400 years of telescopic observation of the Sun. That said, the time interval for which we have detailed knowledge of the Sun's variability is minuscule compared to the time-scale of stellar evolution. As discussed in Sect. 8.3.1 below, recent analysis of *Kepler* photometric and *Gaia* astrometric data by Reinhold et al. (2020) raises the possibility that our Sun is currently in a state of subdued activity in comparison with the bulk of Sun-like stars.

After Shibata et al. (2013), the energy (in erg) of a stellar flare ($E_{\text{flare}}$) in a sunspot group spot with magnetic flux density $B$, scale-length $L$, and area $A_{\text{spot}}$ has an upper limit determined by the total magnetic energy stored in a volume $A^{3/2}$ near the spot, i.e.,

$$E_{\text{flare}} \approx f E_{\text{mag}} \approx f \frac{B^2 L^3}{8\pi} \approx f \frac{B^2}{8\pi} A_{\text{spot}}^{3/2} \approx 7 \times 10^{32} \left(\frac{f}{0.1}\right) \left(\frac{B}{10^3 G}\right)^2 \left(\frac{A_{\text{spot}}}{3 \times 10^{19} \text{ cm}^2}\right)^{3/2}$$
$$\approx 7 \times 10^{32} (\text{erg}) \left(\frac{f}{0.1}\right) \left(\frac{B}{10^3 G}\right)^2 \left(\frac{A_{\text{spot}}/(2\pi R_\odot^2)}{0.001}\right)^{3/2}$$
(10)

where $f$ (0.1; Sect. 3.1.5) is the fraction of magnetic energy that is released during the flare and $R$ is the solar radius (to which the stellar radius is scaled). To scale the flare energy determined by Eq. (10) to SXR class, Shibata et al. (2013) assumed linear proportionality between these parameters from consideration of previous observational estimates of energies of typical solar flares (e.g., Benz 2008; also used by Gopalswamy et al. 2018). The linear scaling of Shibata et al. (2013) differs from that of Schrijver et al. (2012; Eq. (5) in this paper with an exponent of 0.72) as well as the similar scaling in Tschernitz et al. (2018; 0.79), both of which are based on the data in Kretzschmar (2011). Namekata et al. (2017) show that an exponent of 1.0 lies between those for an ordinary least squares (OLS) fit (0.84 ± 0.04) and a linear regression bisector fit (1.18 ± 0.04; Isobe et al. 1990) to a scatter plot of the logs of HMI white-light flare energy versus SXR peak flux for a sample of M- and X-class flares. It is not clear that the bisector fit is more appropriate in this case than the OLS fits in log–log space used by Schrijver et al., Tschernitz et al., and Namekata et al., which yielded slopes from 0.72 to 0.84. The Schrijver et al., Tschernitz et al., and Shibata et al. scalings are shown in Fig. 22 along with the Kretzschmar data points. The Tschernitz et al. and Schrijver et al. fits differ slightly because Schrijver did not correlate bolometric energy and GOES SXR class directly but first related TSI to SXR fluence and then converted SXR fluence to SXR class after Veronig et al. (2002a). The three scalings in Fig. 22 agree reasonably well for events in the X10-X1000 range of interest here, although Shibayama et al. (2013; their Appendix B) note an apparent disagreement between the observed frequency of nanoflares and the expected rate based on Kretzschmar's relation (Eq. (3) above) between bolometric energy and SXR energy.

Figure 23 from Okamoto et al. (2021) is a combined scatter plot of (1) the SXR classes of solar flares (right-hand y-axis) and (2) the bolometric energies of *Kepler* superflares on the 15 solar-type stars in Table 3 (left-hand axis) versus their spot areas. For both sets of flares, the energy to flare class scaling of Shibata et al. (2013) in Fig. 22 is used to infer the missing parameter—bolometric energy for solar flares and SXR class for stellar superflares. The SXR flare versus spot area comparison was patterned after Sammis et al. (2000).[8] For the stellar flare data points, spot areas were inferred from Eq. (9) with bolometric energy given by *Kepler* photometry. The solid and dashed slanting lines given by Eq. (10) represent the released flare energy (with $f = 0.1$) for a given de-projected spot area (inclination angle $i$ assumed to be 90°) on the x-axis and spot field strengths of 3 kG and 1 kG, respectively.

---

[8] The solar SXR peak fluxes are available at: https://www.ngdc.noaa.gov/stp/space-weather/solar-data/solar-features/solar-flares/x-rays/goes/xrs/; see Footnote 2 for sunspot area data.

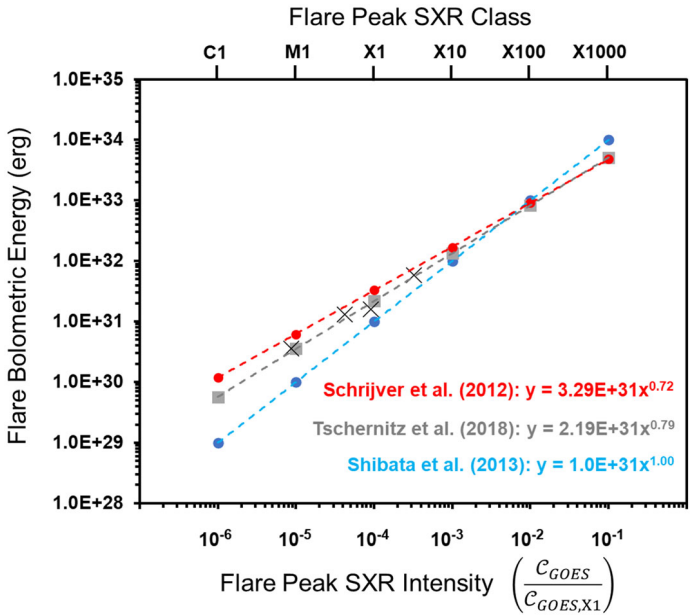

**Fig. 22** Three separate scalings of flare bolometric energy to SXR class: Schrijver et al. (2012), Tschernitz et al. (2018), and Shibata et al. (2013). The similar scalings of Schrijver et al. and Tschernitz et al. are both based on the data ("X" data points) from Kretzschmar (2011)

The apparent correlation in the peak SXR flare flux versus solar spot group area in Fig. 23 has been discounted by Hudson (2021) because of "a strong selection bias, resulting from the under-reporting of weaker GOES events due to higher background levels during active times (Wheatland 2010)". This selection effect will also apply to the flares observed by *Kepler*, for which the detection completeness of $10^{33}$ erg flares is 0.001 (Maehara et al. 2012). That said, the increase in the largest observed flare energies of stellar flares relative to those of solar flares in Fig. 23, in correspondence with the larger maximum spot areas inferred for stellar flares, is in accord with expectations—particularly because the starspot areas are likely underestimated due to the "null hemisphere" hypothesis. From this figure, Okamoto et al. (2021) conclude that "stellar superflares [represent] the magnetic energy release stored around the starspots, and the process is the same as that of solar flares".

### 3.2.5 Occurrence frequency of superflares on Sun-like stars

Figure 24 (adapted from Okamoto et al. 2021; see also Schrijver et al. 2012; Maehara et al. 2015; Lin et al. 2019; Notsu et al. 2019) shows the unified figure of occurrence frequency of flares as a function of flare energy, for solar flares and superflares on Sun-like stars (based on short time cadence *Kepler* data). There is an apparent smooth distribution from the smallest EUV solar flares all the way to the largest bolometric flares on Sun-like stars. It is remarkable that superflare frequency is roughly on the same line as that for solar flares, microflares, and nanoflares, $dN/dE \sim E^{-1.8}$, suggesting the same physical mechanism for both solar and stellar flares (see Sect. 3.1.3). In an early result for a sample of solar-type stars observed by the *Transiting Exoplanet Survey Satellite* (TESS; Ricker 2015), Tu et al. (2020) reported a slightly steeper ($-2.16 \pm 0.10$) slope, but within the uncertainties of the *Kepler* results. Recently, Aschwanden and Güdel (2021) obtained a slope of $-1.82 \pm 0.005$ for the fluence distribution of 162,262 flares on stars of types A-M and giants observed by *Kepler* at optical wavelengths (Yang and Liu 2019) verauss a slope of $-1.98 \pm 0.11$ for the (background- subtracted) peak fluxes of 338,661 GOES SXR flares (Aschwanden and Freeland 2012).

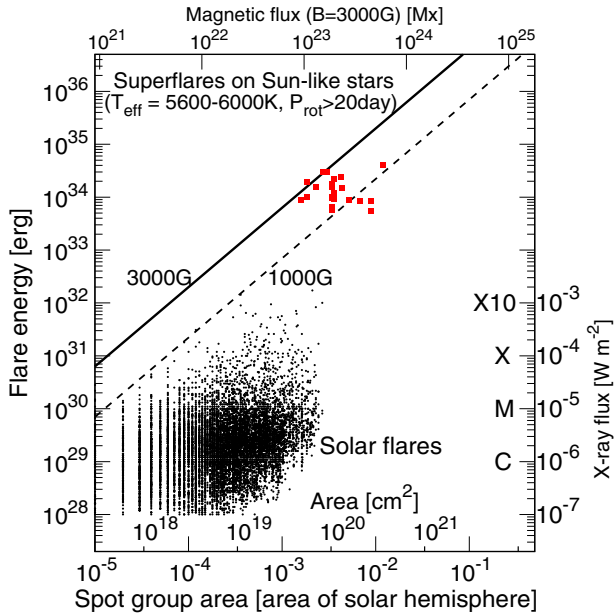

**Fig. 23** Scatter plot of flare energy versus sunspot area for solar flares (black dots) and superflares on solar-type stars (red squares) observed by *Kepler* (Table 3). The black solid and dashed lines correspond to the relationship in Eq. (10) between flare energy and spot area for $B = 3000$ and $1000$ G, respectively. For the *Kepler* superflares, the inclination angle ($i$) between the stellar rotational axis and the line-of-sight is assumed to be 90° (e.g., as is the case for the Sun viewed from Earth at the equinoxes). The measured SXR peak intensities for solar flares and calculated values are given on the right-hand y-axis. The values on the horizontal axis at the top show the total magnetic flux of a spot corresponding to the area on the horizontal axis at the bottom when $B = 3$ kG. Image reproduced with permission from Okamoto (2021), copyright by AAS

A power-law slope of $\sim -1.8$ pertains to a wide variety of solar emissions (Aschwanden et al. 2016) and is in theoretical agreement, within uncertainties, with the $-1.67$ value for hard X-ray ($> 10$ keV) peak fluxes obtained by Aschwanden (2012) from the SOC model for solar flares (Lu and Hamilton 1991). The similarity of slopes in these various studies may reflect the universality of the reconnection process in converting magnetic energy to radiative energy via particle acceleration in different types of settings (the Sun and other stars; Benz and Güdel 2010; Aschwanden and Güdel 2021) as well as in different types of flares (e.g., confined and eruptive; Harra et al. 2016; Tschernitz et al. 2018; cf., Cliver and D'Huys 2018). The Neupert effect implies a power-law slope of −2.0 for solar SXR flares (Aschwanden and Freeland 2012).

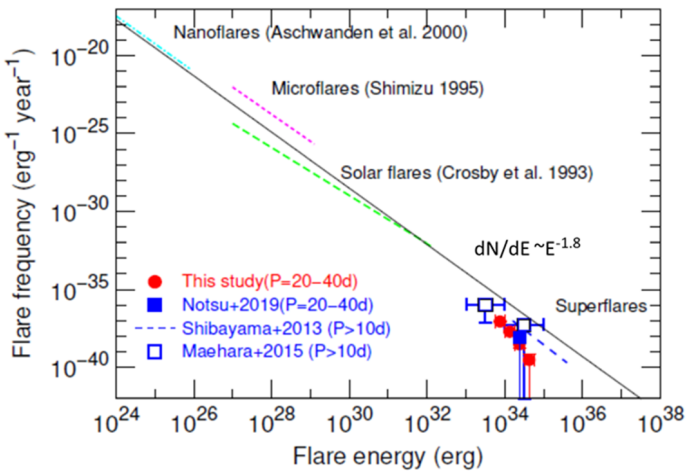

**Fig. 24** Comparison between the frequency distribution of superflares on Sun-like stars and solar flares. The red filled squares, blue filled square, blue dashed line, and blue open squares indicate the occurrence frequency distributions of superflares on Sun-like stars (slowly rotating solar-type stars with $T_{eff}$ = 5600–6000 K). The red filled circles correspond to the updated frequency values of superflares on Sun-like stars with $P_{rot}$ = 20–40 days, which are calculated with gyrochronology and sensitivity correction. Image adapted from Okamoto et al. (2021)

In their earlier analysis of a power-law distribution of Sun-like superflare stars (defined as G-type main sequence stars with $5800 \leq T_{eff} < 6300$ and rotation period > 10 days) observed during the first 500 days of the *Kepler* mission, Maehara et al. (2015) determined that the occurrence frequency of $10^{33}$ erg solar flares would be once every 500 years. Subsequently, Notsu et al. deduced that flares with energy of $> 10^{34}$ to $\lesssim 5 \times 10^{34}$ erg occur on old (~ 4.6 Gyr), slowly rotating ($P_{rot} \sim 25$ days) Sun-like stars approximately once every 2000–3000 years. For a larger sample size than that of Notsu et al. and using a gyrochronology correction, Okamoto et al. (2021) inferred that solar superflares with energies of $10^{34}$ erg (SXR class = X1000) could occur on slowly rotating Sun-like stars once every ~ 6000 years. This rate is preliminary, pending spectroscopic analysis of metallicity, surface gravity, binarity, and rotation period of the 15 (presumably) Sun-like stars in Table 3 on which it is based (see Appendix B in Okamoto et al.). In addition, the possible underestimation of the energy of *Kepler* superflares (Osten and Wolk 2015) noted in Sect. 3.2.2 will need to be addressed.

Five of the 15 Sun-like stars in Table 3 produced more than one such superflare during the ~ 4 years of *Kepler* observations that were considered. Statistically, this seems incongruent with ~ $10^{34}$ erg superflares happening only once in six millennia. This might suggest clustering of superflares on Sun-like stars (as witnessed for intense solar flares; e.g., Cliver et al. 2020b) but it also might mean that the sample of candidate Sun-like stars in Table 3 is contaminated by faster rotators (a possibility noted for five stars with flag = 1 in Table 3; see Appendix A in Okamoto et al.), underscoring the need for spectroscopic analysis. Until that is done, it is uncertain whether the stars in Table 3 are truly Sun-like. As Okamoto et al. (2021) explicitly note, "the number of Sun-like superflare stars ($T_{eff}$ = 5600–6000 K, $P_{rot} \sim 25$ days, and t $\sim$ 4.6 Gyr) that have been investigated spectroscopically and confirmed to be "single" Sun-like stars, is now zero …" Thus the question posed in the title of Notsu et al. (2019)—Do *Kepler* Superflare Stars Really Include Slowly Rotating [with rotation periods $\sim 25$ days] Sun-like Stars?—remains unanswered. Until it is answered affirmatively, the *Kepler* observations of superflares ($\geq 10^{33}$ erg) can only hint at the possibility that the Sun is capable of producing such flares.

While further spectroscopic work is required to substantiate the implications of superflares on Sun-like stars for extreme solar activity, observations of solar flares with bolometric energies within a half-decade of $10^{33}$ erg (Sect. 3.1.2 above) suggest that the Sun is capable of producing a threshold level superflare over time, while analyses of historical solar proton events based on cosmogenic radionuclides (Sect. 7 below) provide evidence that it has already done so.

## 4 Extreme solar radio bursts

Radio bursts from the Sun were discovered in late February 1942 when Hey (1946, with delayed publication due to wartime restrictions) correctly identified as solar radio emission what at first appeared to be enemy jamming of army radar receivers. More recently, radio frequency interference due to a great decimetric (range from 300 to 3000 MHz) burst on 6 December 2006 severely degraded the performance of Global Navigation Satellite Systems (including the U.S. Global Positioning System (GPS)), signaling a modern hazard of solar radio bursts (Gary 2008; Cerruti et al. 2008; Kintner et al. 2009; Carrano et al. 2009). In this section, after reviewing the climatology of solar radio bursts, we will focus on extreme radio bursts in the frequency range from 1.0 to 1.6 GHz range that includes the 1.58 GHz (L1 band) and 1.23 GHz (L2 band) operational frequencies of GPS.

### 4.1 Climatology of solar radio bursts

Following the pioneering work of Akabane (1956) on the peak flux density distribution of 3 GHz solar radio bursts observed at Ottawa, Toyokawa, and Tokyo, the most extensive studies of radio burst occurrence statistics have been conducted by Nita and colleagues (Nita et al. 2002, 2004a, b) and Giersch et al. (2017). Nita et al. (2002) analyzed a 40-year (1960–1999) compilation[9] by the National Geophysical Data Center (NGDC) of solar radio burst intensities to obtain

---

[9] https://www.ngdc.noaa.gov/stp/space-weather/solar-data/solar-features/solar-radio/radio-bursts/reports/fixed-frequency-listings/ (Now extended to 2010).

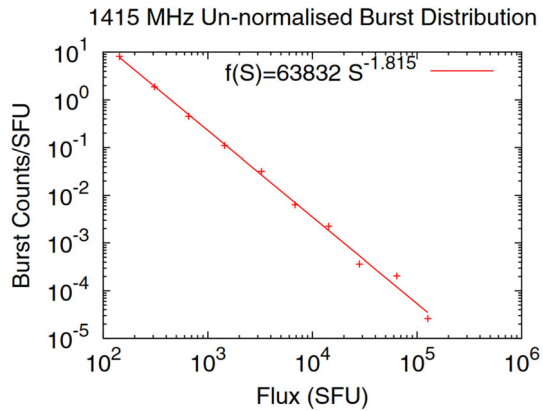

Fig. 25 Peak flux density distribution for solar $\sim 1.4$ GHz (L-band) bursts from 1996 to 2010. Image reproduced with permission from Giersch et al. (2017), copyright by AGU

occurrence frequency distributions for six discrete frequency groupings in the range from 0.1 to 37 GHz. Giersch et al. (2017) performed a similar analysis based on the updated NGDC compilation, but using only data from the U.S. Air Force (USAF) Radio Solar Telescope Network (RSTN) for the 1966–2010 interval for the eight fixed frequencies of the RSTN patrol.

Both Nita et al. (2002) and Giersch et al. (2017) stress that the data sets they used are incomplete, with nearly half of all events being missed. Nita et al. (2002) attribute the missing events to uneven geographical coverage while Giersch et al. suggest that breakdowns in "the process of report transmission [from the RSTN observatories] to [the National Geophysical Data Center (NGDC) or its National Center for Environmental Information (NCEI) successor] is likely to account for the 'missing' burst data." The power-law slopes of the peak flux density distributions obtained by Nita et al. (2002) for the six frequency ranges they considered range from $\sim -1.7$ to $-1.85$ versus $\sim -1.8$ to $-2$ for the eight RSTN frequencies examined by Giersch et al. For a shorter interval from 1994 to 2005, Song et al. (2012) obtain a range of slopes from $\sim -1.75$ to $-1.85$ for data from Nobeyama (http://solar.nro.nao.ac.jp/) for six frequencies from 1.0 to 35.0 GHz. Power-law slopes of $\sim -1.8$ are similar to those found for the hard X-ray burst count rate (Dennis 1985; Crosby et al. 1993). For bursts in the $\sim 1.2–1.6$ GHz frequency range employed by GPS, the slopes found by the various investigators are consistent with values of $-1.83 \pm 0.01$ for 1.0–1.7 GHz (Nita et al. 2002), $-1.815$ at 1.415 GHz (Giersch et al. 2017), and $-1.87 \pm 0.12$ at 1.0 GHz and $-1.83 \pm 0.07$ at 2.0 GHz (Song et al. 2012). The peak flux density distribution obtained by Giersch et al. (2017) for 1.415 GHz is given in Fig. 25. From this distribution, Giersch et al. estimated that a solar 1.4 GHz burst of peak flux density $3.2 \times 10^6$ solar flux units (sfu; 1 sfu = $10^{-22}$ W m$^{-2}$ Hz$^{-1}$) would occur about once per century while a burst of $6.1 \times 10^7$ sfu would occur once in a thousand years. The corresponding peak intensity values for 100-year and 1000-year bursts calculated from the 1.0–1.7 GHz distribution of Nita et al. (2002) are $1.2 \times 10^7$ sfu and $2 \times 10^8$ sfu, respectively (G. Nita, personal communication, 2018). Both sets of authors emphasized the uncertainty of the 1000-year events because of the large extrapolations involved (personal communication: Nita 2018; Giersch 2019).

### 4.2 Observations of extreme decimetric radio bursts

The great decimetric solar radio burst of 6 December 2006 that disrupted GPS operations was the largest burst yet observed from 1.0 to 1.6 GHz, with a peak flux density of $\sim 1 \times 10^6$ sfu recorded by the Frequency-Agile Solar Radiotelescope (FASR) Subsystem Testbed (FST; Liu et al. 2007) at Owens Valley (Gary 2008). A search by Klobuchar et al. (1999) for $\sim 1.4$ GHz bursts with peak flux densities $> 4 \times 10^4$ sfu observed by RSTN (and predecessor sites in Athens and Manila) that might affect the GPS identified 14 such events from 1967 to 1998 with peak flux densities ranging up to $8.8 \times 10^4$ sfu. The extreme burst observed at Owens Valley on 6 December 2006 prompted the following questions from Kintner et al. (2009): "Was the 6 December [solar radio burst; SRB] a simple outlier on a well-behaved statistical distribution such as, for example, a 100-year flood? … Or has the method of monitoring SRBs been inadequate so that the power of previous intense SRBs was underestimated?" In reference to the second question, Giersch et al. (2017) deduced a nominal saturation level of $10^5$ sfu at 1.415 GHz for the RSTN observatories (Gary 2008), with a station-to-station uncertainty on this value of over 50%.

In Figs. 26, 27 and 28, we show intensity versus time records of solar bursts at several frequencies that had intense ($> 80,000$ sfu) emission at 1.415 GHz. Figures 26 and 27 give time profiles for events on 28 May 1967 and 29 April 1973, respectively. Figure 28 contains the records of bursts on: (a) 6 December 2006, (b) 13 December 2006, (c) 14 December 2006, and (d) 21 April 2002. For the 6 December 2006 event (Fig. 28a), the Sagamore Hill 1.415 GHz radiometer saturated at a level of $\sim 1.4 \times 10^5$ sfu, well below the $\sim 1 \times 10^6$ sfu peak flux recorded at Owens Valley. For both the 13 December 2006 event and the 21 April 2002 event (Fig. 28b, d), the RSTN 1.4 GHz receivers at Palehua and Learmonth saturated at $\sim 10^5$ sfu.[10]

As shown in Table 4 below, only 14 1.0–1.6 GHz bursts with peak fluxes $\geq 80,000$ sfu were recorded from 1966 to 2015. The rarity of these large decimetric bursts and the fact that NOAA active region 10930 in December 2006 produced three such bursts (Fig. 28a–c) with peak fluxes $\gtrsim 10^5$ sfu (Gary 2008) suggests that a special circumstance is required for the occurrence of extreme $\sim 1.5$ GHz bursts. Giersch and Kennewell (2013) investigated the sunspot group and magnetic features of region 10930 to determine if any distinguishing features (e.g., spatial locations of minimum and maximum magnetic fields, magnetic field gradient between these locations, small-scale mixing of positive and negative polarities) were present. From a comparison of NOAA 10930 with a set of active regions from 2002 to 2008, they found no features that were unique to 10930, thus making such bursts "inherently unpredictable."

---

[10] https://www.ngdc.noaa.gov/stp/space-weather/solar-data/solar-features/solar-radio/rstn-1-second/; Giersch and Kennewell (2013).

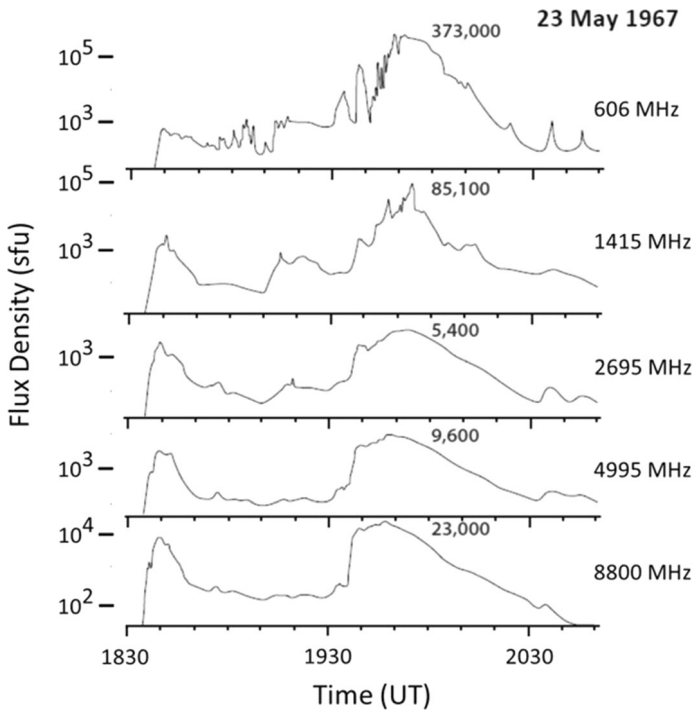

**Fig. 26** Sagamore Hill record at five frequencies (with peak fluxes given) for the great solar radio burst of 23 May 1967. Image adapted from Castelli et al. (1968)

While region 10930 may not have distinguished itself from the 2002–2008 comparison set of active regions, inspection of the great decimetric bursts themselves reveals a pattern that may provide insight into the conditions under which these extreme bursts arise. As noted by Cliver et al. (2011), all three of the events from region 10930 exhibited intense ($\gtrsim 10^5$ sfu) L-band (defined here to be $\sim$ 1.0–1.5 GHz) emission some tens of minutes after the impulsive phase of these flares (Fig. 28a–c, although the strongest such emission for the 13 December 2006 flare occurred during the impulsive phase (Fig. 28b). Similar delayed decimetric emission is also seen in Figs. 26, 27, and 28d. Delayed microwave bursts with spectral maxima at $\sim$ 3 GHz (Tanaka and Kakinuma 1962) and gradual or extended hard X-ray bursts with hardening spectra (Tsuneta et al. 1984; Dennis 1985; Tanaka 1987) that often accompany them, have both been interpreted (Cliver 1983; Cliver et al. 1986) in terms of electrons accelerated via reconnection and trapped on coronal loops in the standard CSHKP model for eruptive solar flares (Hudson and Cliver 2001; Shibata and Magara 2011). The late phase of such flares is characterized by the growth of post-flare loop systems (Bruzek 1964; Kahler 1977; Švestka et al. 1982) as loops are formed at successively greater heights in the wake of a CME (e.g., Forbes 2000; Lin and Forbes 2000; Hudson 2011; Benz 2017).[11] Because of their similar origins in delayed microwave bursts and gradual hard X-ray events, Cliver et al. (2011) suggested that the great decimetric bursts in December 2006 could also be interpreted in terms of magnetic reconnection and electron trapping in the CSHKP model. Tanaka and Kakinuma (1962) were the first to note the concurrence of delayed microwave and intense decimetric bursts.

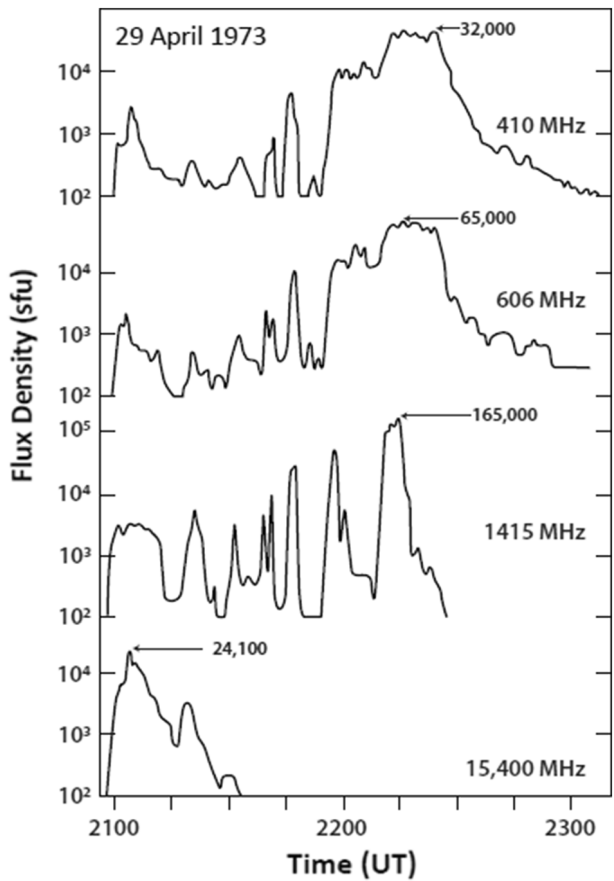

**Fig. 27** Sagamore Hill record at four selected frequencies (with peak fluxes given) for the great solar radio burst of 29 April 1973. Image adapted from Barron et al. (1980)

The list of great (peak intensity $\geq$ 80,000 sfu) 1.0–1.6 GHz bursts in Table 4 is based on searches of the NGDC website by Nita et al. (2002; for 1960–1999) and Giersch et al. (2017; 1966–2010) as well as our own search for large 1 GHz bursts on the Nobeyama website (1988–2015). Four of the eight additional cases (i.e.,

---

[11] As Švestka (2007) pointed out, the term post-flare loop is a misnomer; the reconnection that forms the flare loops is the primary flare energy release process. Švestka suggested "eruptive flare loop system" as a more accurate term to describe the phenomenon. Terms such as "post-eruption arcade" (e.g., Tripathi et al. 2004) also more correctly describe the loop systems.

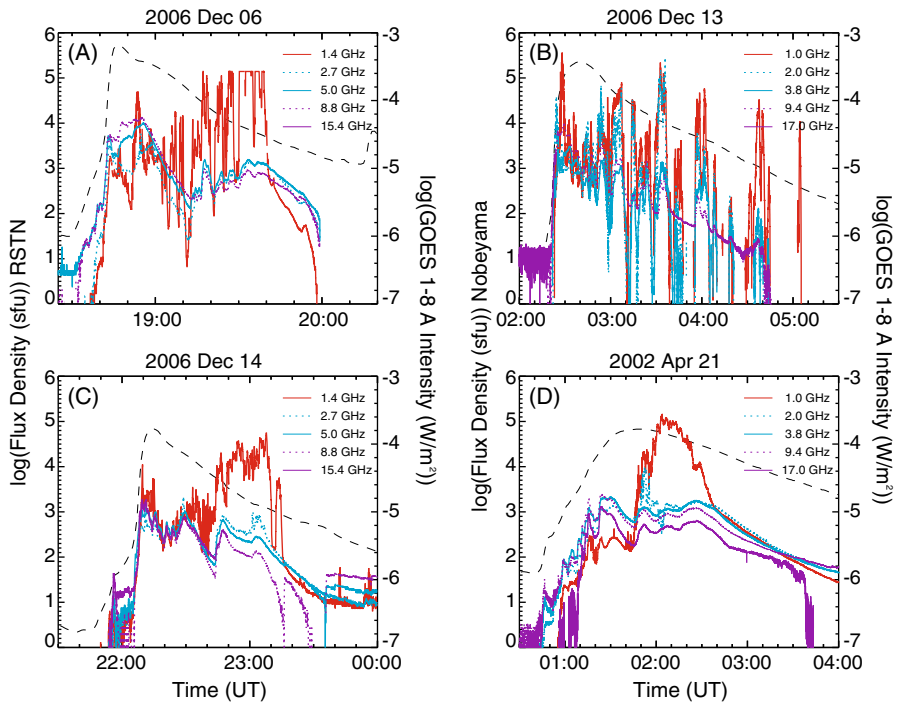

**Fig. 28** Time-intensity profiles of great ($\gtrsim 10^5$ sfu) decimetric 1.4 or 1.0 GHz bursts (red traces) along with bursts at other frequencies from $\sim 1$–17 GHz for: **a** 6 December 2006 (Sagamore Hill); **b** 13 December 2006 (Nobeyama); **c** 14 December 2006 (Learmonth); and **d** 21 April 2002 (Nobeyama). The dashed gray line in each of the panels is the time-intensity profile of the associated 1–8 Å SXR burst. Image reproduced with permission from Cliver et al. (2011), copright by AAS

those not shown in Figs. 26, 27, 28) listed in Table 4 (4 July 1974, 16 February 1979, 15 April 1990, and 5 March 2012) also appear to follow the characteristic (delayed) pattern seen in Figs. 26, 27 and 28, while in the other four cases (29 October 1968, 2 November 1992, 8 February 1993, 11 July 2005), intense decimetric peaks occurred relatively close (within 10 min) to the impulsive phase with no strong late phase emission.[12] Setting aside the 13 December 2006 radio burst in which strong ($10^5$ sfu) L-band emission occurred both during and after the impulsive phase, in 9 of the 13 other cases in Table 4, commensurately intense decimetric emission avoided the flare impulsive phase—widely considered to be the time of primary energy release in flares (e.g., Fletcher et al. 2011). Recently, Marqué et al. (2018) presented another example of such behavior in a radio burst on 4 November 2015 (Fig. 29) and listed another event not in Table 4 with a large (> 100,000 sfu) delayed decimetric burst on 24 September 2011 (Shakhovskaya et al. 2019). The spike burst at 2.65 GHz on 11 April 1978 reported by Slottje (1978) shares this timing characteristic. The frequently observed delay of the largest

---

[12] Time-intensity profiles can be seen at http://solar.nro.nao.ac.jp/norp/html/event/ for the 2 November 1992 and 8 February 1993 radio bursts.

**Table 4** 1.0–1.6 GHz radio bursts with peak fluxes ≥ 80,000 sfu (1966–2015). *Sources*: NGDC, 1960–2010 (https://www.ngdc.noaa.gov/stp/space-weather/solar-data/solar-features/solar-radio/radio-bursts/reports/fixed-frequency-listings/); Nobeyama, 1988–2015 (http://solar.nro.nao.ac.jp/norp/html/event/)

| Nos. | Date | Imp. phase[a] | Hα coords | SXR peak time (UT)[b] | SXR class[b] | 1.0–1.6 GHz Freq. (GHz)[c] | Peak time (UT) | Peak flux (sfu) | Delay (Min.)[d] | CME Spd. (km/s)[e] | Ref.[f] |
|---|---|---|---|---|---|---|---|---|---|---|---|
| 1 | 23 May 1967 | 18:40 | N30E25 | – | 2B | 1.4(SH) | 19:54 | $8.5 \times 10^4$ | 74 | N/A | (1,2) |
| 2 | 29 Oct 1968 | 15:22 | | – | 1N | 1.4(SH) | 15:24 | $8.1 \times 10^4$ | 2 | N/A | (2) |
| 3 | 29 Apr 1973 | 21:06 | N14W73 | – | 2B | 1.4(SH) | 22:14 | $1.6 \times 10^5$ | 68 | N/A | (2) |
| 4 | 4 Jul 1974 | 13:54g | S16W08 | – | 2B | 1.4(K) | 14:22 | $8.0 \times 10^5$ | 28 | N/A | (3) |
| 5 | 16 Feb 1979 | 01:51h | N16E59 | 02:00 | X2 | 1.4(M) | 02:32 | $8.8 \times 10^4$ | 41 | N/A | (4) |
| 6 | 15 Apr 1990 | 03:00 | N32E54 | 02:55 | X1 | 1.0(N) | 04:40 | $7.2 \times 10^5$ | 100 | N/A | (5) |
| 7 | 2 Nov 1992 | 02:54 | S23W90 | 03:08 | X9 | 1.0(N) | 03:02 | $\sim 1.0 \times 10^5$ | 8 | N/A | (5,6) |
| 8 | 8 Feb 1993 | 02:25i | S06E36 | 02:42 | C9 | 1.0(N) | 02:31 | $5.7 \times 10^5$ | 6 | N/A | (5,6) |
| 9 | 21 Apr 2002 | 01:23 | S14W84 | 01:51 | X1 | 1.0(N) | 02:04 | $1.5 \times 10^5$ | 41 | 2393 | (5,6) |
| 10 | 11 Jul 2005 | 16:39i | ? | 16:39 | ~C1 | 1.4(SV) | 16:36 | $1.2 \times 10^5$ | – 3 | 462? | (6) |
| 11 | 6 Dec 2006 | 18:44 | S06E63 | 18:47 | X6 | 1.4(OV) | 19:31 | $\sim 1.0 \times 10^6$ | 47 | N/A | (6,7) |
| 12 | 13 Dec 2006 | 02:29 | S06W24 | 02:40 | X3 | 1.0(N) | 02:28/03:35 | $4.4 \times 10^5$ | – 1/66 | 1774 | (5,6) |

**Table 4** continued

| Nos. | Date | Imp. phase[a] | Hα coords | SXR peak time (UT)[b] | SXR class[b] | 1.0–1.6 GHz Freq. (GHz)[c] | Peak time (UT) | Peak flux (sfu) | Delay (Min.)[d] | CME Spd. (km/s)[e] | Ref.[f] |
|---|---|---|---|---|---|---|---|---|---|---|---|
| 13 | 14 Dec 2006 | 22:09 | S06W46 | 22:15 | X1 | 1.6(OV) | 23:07 | $1.2 \times 10^5$ | 58 | 1042 | (6–8) |
| 14 | 5 Mar 2012 | ~03:50 | N17E52 | 04:09 | X1 | 1.0(N) | ~04:30 | $5.0 \times 10^5$ | ~40 | 1531 | (5) |

[a]Peak time of the first significant ($\gtrsim 10^3$ sfu) microwave ($\geq$ 3–30 GHz; cm-λ) burst

[b]Hα flare classifications from Solar-Geophysical Data (SGD) given before 1975; SXR classifications taken from https://www.ngdc.noaa.gov/stp/space-weather/solar-data/solar-features/solar-flares/x-rays/goes/xrs/

[c]Abbreviations for observatory names: SH = Sagamore Hill, K = Kiel, M = Manila, N = Nobeyama, SV = San Vito, OV = Owens Valley

[d]Delay from the peak of the first significant microwave peak to the peak of the delayed intense ($> 8 \times 10^4$ sfu) L-band emission

[e]Linear speed from https://cdaw.gsfc.nasa.gov/CME_list/

[f]1—Knipp et al. 2016, 2—Barron et al. 1980, 3—SGD (No. 360), Pt. 1, p. 30 (2.8 GHz burst profile), 4—NGDC website (see Footnote 10), 5—Nobeyama website, 6—Cliver et al. 2011, 7—Gary 2008, 8—Carrano et al. 2009

[g]other reported values are closer to $8.0 \times 10^4$ sfu

[h]Peak microwave flux from ~600–850 sfu

[i]Peak microwave flux < 100 sfu (see Cliver et al. 2011)

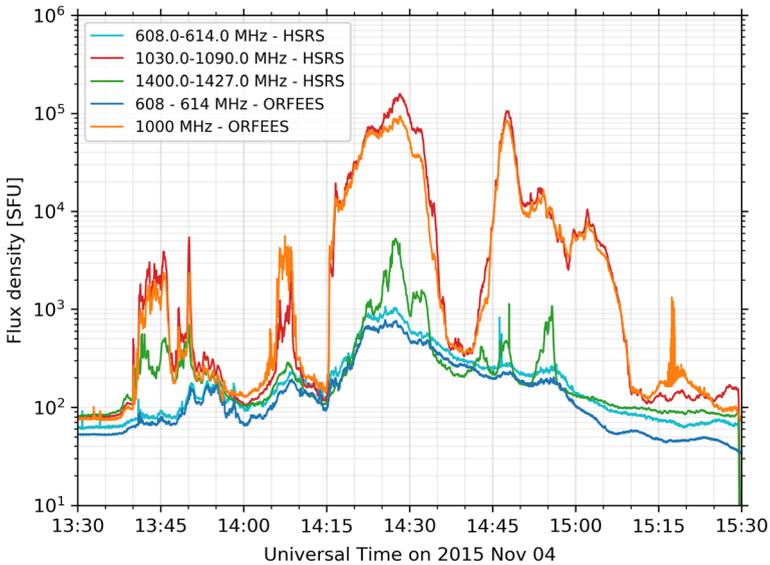

**Fig. 29** Decimetric emission at several discrete frequencies for a radio burst on 4 November 2015 observed by the Humain Solar Radio Spectrometer (HSRS) in Humain, Belgium and the Observations Radio pour FEDOME et l'Étude des Éruptions Solaires (ORFEES) spectrograph in Nançay, France). Image reproduced with permission from Marqué et al. (2018), copyright by the authors

decimetric bursts relative to the flare impulsive phase suggests that a special circumstance is required for their generation.

### 4.3 Are out-sized decimetric spike bursts due to electron cyclotron maser emission?

The evolution of the decimetric burst emission in Fig. 28(c; at 1.4 GHz), (d; 1.0 GHz) is instructive. In both cases the decimetric emission more or less tracks that at centimeter wavelengths (cm-$\lambda$; 3–30 GHz) through the impulsive phase. After $\sim$ 22:40UT in 26(c) and 01:40 UT in 26(d), however, the $\sim$ 1 GHz emission dwarfs that at $\sim$ 4–5 GHz by 1–2 orders of magnitude, before dropping abruptly to a level that closely tracks the higher frequency emissions. This behavior suggests an additional component of the decimetric emission, specifically a coherent emission contribution (Melrose 2017) because of the high peak fluxes (brightness temperatures), either plasma emission or electron cyclotron maser (ECM; see Treumann 2006; Fleishman 2006, for reviews) emission, riding atop an incoherent gyrosynchrotron (GS) component (Nindos 2020) that is dominant at centimeter wavelengths. This behavior is pronounced in the $\sim$ 1000 MHz emission in the event on 4 November 2015 in Fig. 29 which intermittently rapidly increases and decays by up to three orders of magnitude, returning to levels characteristic of encompassing frequencies at $\sim$ 600 MHz and $\sim$ 1400 MHz.

The cumulative density function for 1 GHz in Fig. 30 from Song et al. (2012), which is unaffected by receiver saturation, is consistent with a transition from gyrosynchrotron emission to a dominant coherent process for events with peak fluxes $> 10^4$ sfu which lie above the extension of the power-law distribution to higher intensities, in contrast to the behavior at higher frequencies up to 35 GHz where the data points fall below the power-law fit for peak fluxes $< 10^4$ sfu. Thus extreme decimetric bursts suggest a different emission mechanism than that responsible for merely large events. The high $X_{\min}$ value determined by the maximum likelihood estimator method (Clauset et al. 2009) for the 1 GHz bursts in Fig. 30 results in the power-law fit being based on only 53 events versus 139 events for 2 GHz and a larger uncertainty in the slope of the fit at 1 GHz. That said, the largest 1 GHz peak flux density in the figure is more than an order of magnitude larger than the corresponding value at 2 GHz. The GS spectrum for large cm-$\lambda$ bursts characteristically has its maximum at $\sim$ 5–10 GHz (e.g., Castelli et al. 1967; Guidice and Castelli 1975; Stähli et al. 1989), tapering down to a minimum in the decimetric range. The markedly larger maximum flux value at 1 GHz relative to that at 2 GHz in Fig. 30 supports the picture of an additional non-GS (coherent) emission mechanism for the largest bursts at 1 GHz.

Following Gary (2008), Cliver et al. (2011) interpreted the delayed decimetric peak in Fig. 28a in terms of ECM emission. Figure 31 contains a $\sim$ 40 s time sample (with gaps) of the Owens Valley FST record from 1.0–1.5 GHz for the 6 December 2006 burst showing a proliferation of intense narrow-band (3–4 MHz) spikes with durations less than 20 ms that are a signature of coherent radio emission. Wang et al. (2008) reported spike emission at 2.6–3.8 GHz for the 13 December 2006 radio burst and also attributed it to electron cyclotron maser emission. ECM emission driven by a loss cone instability was proposed by Wu and Lee (1979) to account for terrestrial auroral kilometric radiation (AKR) and subsequently developed by Holman et al. (1980) and Melrose and Dulk (1982) to explain solar millisecond spike bursts (Dröge 1977; Slottje 1978) such as seen in Fig. 31.

ECM emission is based on direct amplification of free-space electromagnetic waves in a plasma with a non-thermal electron population.[13] The amplification results from a linear plasma instability that can occur when the local electron cyclotron frequency ($f_{ce}$) exceeds the local plasma frequency ($f_{pe}$), i.e., $f_{ce} \gtrsim f_{pe}$, where $f_{ce}$ (MHz) $= eB/2\pi m_e$ and $f_{pe}$ (MHz) $= (e/2\pi)(n_e/m_e\varepsilon_0)^{1/2}$. However, the effective condition, viz., $f_{ce} >> f_{pe}$, for fundamental ECM emission to escape the Sun's atmosphere is more stringent because of resonant gyromagnetic absorption at the second harmonic layer in the weaker fields of the overlying thermal plasma (Treumann 2006; Holman et al. 1980; Melrose 2017). Calculations of such gyromagnetic absorption (Melrose and Dulk 1982; McKean et al. 1989) indicate that it can effectively suppress ECM emission. Various suggestions have been proposed to overcome this difficulty, with none yet generally accepted (Melrose 1999, 2009; Treumann 2006; Ning et al. 2021a, b).

---

[13] Treumann (2006) points out that the term electron-cyclotron maser is unfortunate/misleading because neither quantum effects, nor energy levels, nor elementary population inversions are involved in ECM radiation. The usage stems from the resemblance between molecular masers and plasma emissions of high brightness temperature.

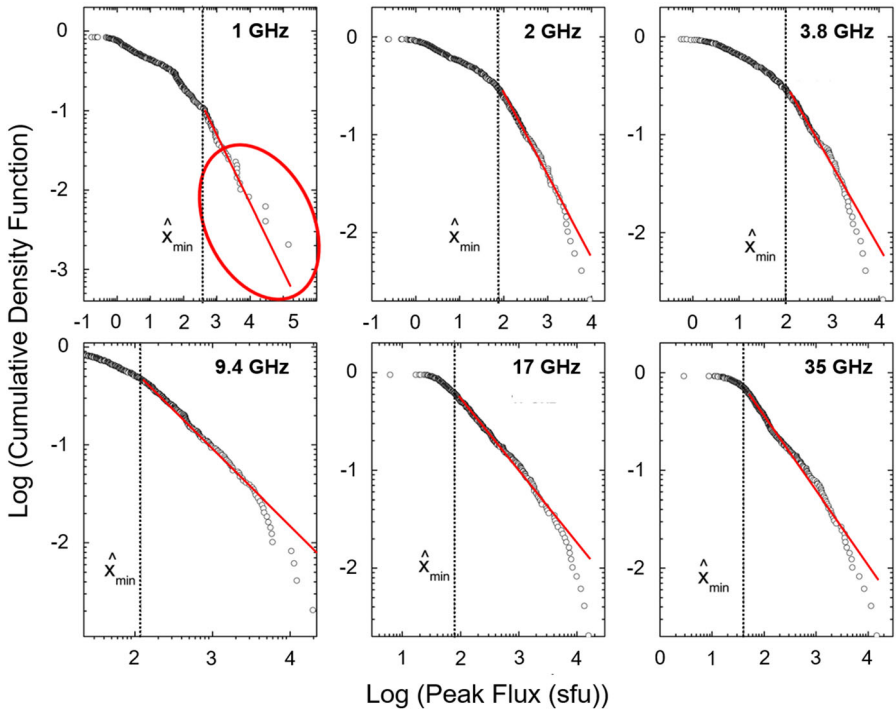

**Fig. 30** Cumulative density functions for solar radio bursts recorded at Nobeyama Observatory from 1994 to 2005 at six separate frequencies from 1 to 35 GHz. The red oval in the 1 GHz panel highlights the unusual behavior relative to the other frequencies on the tail of the distribution. $X_{min}$ is the smallest value used for the power-law fit. Image adapted from Song et al. (2012), copyright by AAS

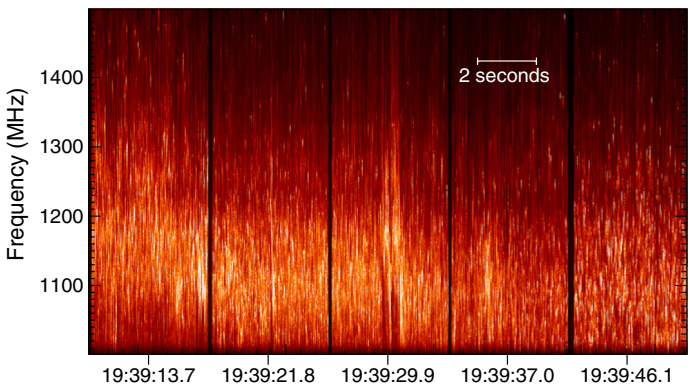

**Fig. 31** Detail of Owens Valley FST observation of the extreme decimetric burst on 6 December 2006. The figure shows four-second scans of burst intensity in right circular polarization, with 4 s gaps between them (not shown). Time labels on the bottom axis are for the center of each 4 s scan. The emission consists of a proliferation (many hundreds per second) of narrow-band (3–4 MHz) spike bursts with durations less than 20 ms. Image reproduced with permission from Cliver et al. (2011), copyright by AAS

Beginning with the assumption that ECM emission arises from an unstable "horseshoe" distribution (Ergun et al. 2000), rather than a loss-cone-driven instability, Melrose and Wheatland (2016) argued that the $f_{ce} \gg f_{pe}$ condition can be satisfied if deep density cavities (Calvert 1981; Hilgers 1992; Alm et al. 2015) driven by parallel electric fields (Temerin et al. 1982; Ergun et al. 2001) which make ECM emission possible in terrestrial aurora, can exist in the coronal arcades of eruptive flares. Cliver et al. (2011) made a similar suggestion. Quoting from Melrose and Wheatland (2016), "The formation of a density cavity … suggests a new possibility for allowing a fraction of the ECME to escape through the second-harmonic layer. This possibility arises if the density cavity extends to above the second-harmonic absorption layer. This is the case when the flux tube in which the acceleration (and associated density depletion) occurs extends to a height where *B* has decreased by a factor of two from its value at the emission point of the ECME. The radiation then passes through the second-harmonic layer in the low-density cavity. The gyromagnetic absorption coefficient is proportional to the density of thermal electrons and hence would be anomalously weak in an anomalously low-density region. If the density in the cavity is orders of magnitude lower inside the flux tube than outside it, as is the case of AKR, then gyromagnetic absorption at the second harmonic would be unimportant. The fraction of the ECME that escapes would then be the fraction that is ducted along the low-density flux tube to above the second-harmonic layer." The vertical extension of a density cavity suggested by Melrose and Wheatland as a pre-requisite for the escape of ECM emission is in keeping with the evolution of loop arcades in the standard CSHKP model of eruptive flares. Such loop systems can reach heights $\sim 10^5$ km (Kahler 1977; Švestka et al. 1982). The outward motion of the post-eruption arcade associated with the great decimetric burst on 21 April 2002 (Fig. 28d), as derived from TRACE 195 Å images and X-ray source centroids from the *Ramaty High-Energy Solar Spectroscopic Imager* (RHESSI), is shown in Fig. 32 (adapted from Gallagher et al. 2002).

The pattern of higher-energy emissions from the most recently formed loops (e.g., Švestka et al. 1987; Anzer and Pneuman 1982) is a standard feature of the CSHKP model (Fig. 10). Our speculative working hypothesis to account for delayed out-sized decimetric burst such as shown in Figs. 26, 27, 28 and 29, based on analogy with AKR in Earth's magnetosphere (where such characteristics of the emission region as a strong electric field, density cavity, and horseshoe electron distribution, can be observed in situ), is given in schematic form in Fig. 33, taken from Cliver et al. (2011).

The decimetric spike emission in Fig. 31 could also be produced by plasma emission (Zheleznyakov and Zaitsev 1975; Kuijpers et al. 1981; Bárta et al. 2011a, b; Karlický and Bárta 2011; Feng et al. 2018; Karlický et al. 2021), the accepted mechanism for solar metric type II and type III bursts as well as for fine structure in Type IV bursts (e.g., Kaneda et al. 2017). Marqué et al. (2018) write, "[The ECM emission] mechanism is usually expected to not operate in the solar corona, because of the high electron plasma frequency. The exceptional occurrence of the strong radio burst near 1000 MHz [in reference to an event on 4 November 2015 they analyzed (Fig. 29)] could of course be explained by an exceptional situation of plasma parameters in the corona, so that electron cyclotron maser emission might arise. The argument was put forward by Régnier (2015) and Cliver et al. (2011). However, the identification of well-known fine structures, like fiber bursts and zebra patterns, that are observed in type IV bursts, but usually at lower frequencies where it is still more unlikely that the cyclotron frequency exceeds the plasma frequency, makes such an exceptional situation improbable. Instead, it rather argues for coherent plasma emission."

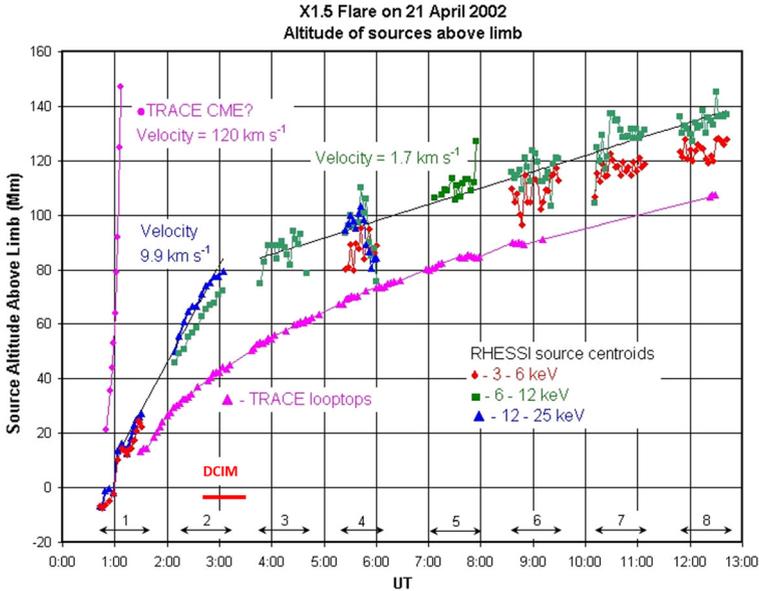

**Fig. 32** Radial growth of the post-eruption loop system for the flare associated with the great decimetric (DCIM) burst on 21 April 2002 based on TRACE 195 Å and RHESSI images. X-ray source centroids at 3–6, 6–12, and 12–25 keV are indicated by diamonds, squares, and triangles, respectively. The red horizontal bar indicates the interval of peak decimetric emission ($\sim$ 01:45–02:30 UT). Image adapted from Gallagher et al. (2002)

For either interpretation (plasma emission or ECM emission), an exceptional situation of plasma parameters seems a necessity to explain the rarity of outsized decimetric spike bursts in the context of the standard CSHKP model for eruptive flares. The vast majority of such flares are not accompanied by such emission. Such a special condition, e.g., the formation of a density cavity permitting the $f_{ce} \gg f_{pe}$ inequality required for ECM emission generation and escape, can also account for the rapid switch on/off behavior of the intense delayed component of certain $\sim$ 1 GHz bursts. The observed/inferred late superimposed spike emission in great decimetric bursts (Figs. 26, 27, 28, 29) implies that in certain cases extreme events are not achieved merely by scaling up an event—adding more of the same (in this case incoherent gyrosynchrotron emission)—but by crossing a physical threshold after which out-sized growth of the event is based on a different emission mechanism from that associated with smaller events.

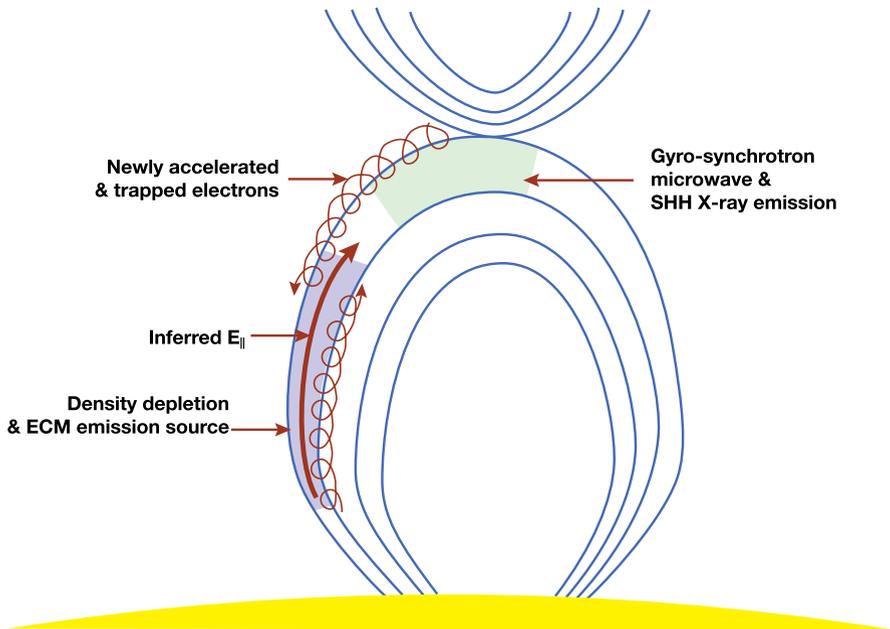

**Fig. 33** Schematic showing how ECM decimetric emission might arise in post-eruption loops as a result of strong late-phase reconnection and electron acceleration in a field-aligned potential drop (and density depletion) in conjunction with delayed coronal hard X-ray and microwave bursts. Image reproduced with permission from Cliver et al. (2011), copyright by AAS

Such events are referred to as Dragon Kings (Sornette and Quillon 2012), a double metaphor indicating their extreme size and peculiar nature. Dragon Kings have the following characteristics: (1) they do no not belong to the same population as other such events as manifested by the size distributions for 1 GHz bursts in Fig. 30; and (2) they appear as a result of amplifying mechanisms (in this case coherent radio spike emission; Fig. 31) that are not fully active for the rest of the population. The intense delayed $\sim$ 1 GHz solar radio bursts satisfy both criteria.

## 5 Coronal mass ejections (CMEs)

As shown in Fig. 12 in Sect. 3.1.5, the kinetic energy of CMEs dominates the energy budget of large eruptive flares, accounting for $\sim$ 75% (or more, Sect. 3.1.5) of the total energy released (Emslie et al. 2012; cf. Aschwanden et al. 2017). Not coincidentally, CMEs are the drivers for the two major space weather phenomena— solar particle events and geomagnetic storms (Kahler 1992; Gosling 1993; Green et al. 2018; Schrijver and Siscoe 2012). In fact, Gopalswamy et al. (2018) have shown that CME speed can be used to organize the full range of heliospheric and terrestrial effects of solar eruptions (Fig. 34). One of the geo-effective CMEs that occurred during the sequence of strong solar activity in October–November 2003 is shown in Fig. 35.

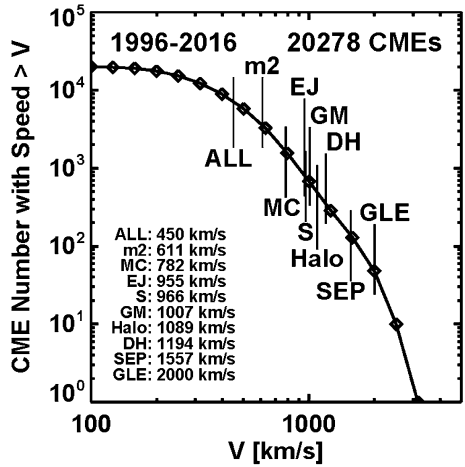

**Fig. 34** Cumulative frequency distribution function for plane-of-sky speeds of SOHO LASCO CMEs showing the organization of interplanetary and terrestrial phenomena by this parameter. The average speeds of CME populations responsible for various coronal and interplanetary phenomena are indicated by vertical lines. Abbreviations refer to CMEs that are associated with: metric type II bursts (m2), magnetic clouds (MC; Klein and Burlaga 1982), interplanetary CMEs (ICMEs; Russell 2001; cf. Burlaga 2001) without flux rope structure (EJ, for ejecta), shocks detected in situ in the solar wind (S), non-recurrent geomagnetic storms (GM), decametric-hectometric type II bursts (DH), NOAA class S1 SEP events (with peak > 10 MeV fluxes > 10 proton flux units), and ground level enhancement (GLE; > 500 MeV) proton events detected by ground-based neutron monitors. Image reproduced with permission from Gopalswamy (2018), copyright by Elsevier

Webb and Howard (2012) reviewed observations of CMEs and Temmer (2021) recently discussed them from a space weather perspective. See Vršnak (2021) for a recent review of the origins and interplanetary propagation of CMEs.

### 5.1 CME climatology

Gopalswamy (2018) constructed a CME speed distribution for $\sim$ 27,000 coronal transients observed by SOHO LASCO (Brueckner et al. 1995) from January 1996 to March 2016. As was the case for sunspot group areas (Fig. 6) and flares (Fig. 8), Gopalswamy found that a modified exponential function provided a better fit to the data than a power law over the full range of observations. Riley (2012) had previously noted a well-defined "knee" in the CME speed distribution.

Exponential and power-law fits to CME speed and kinetic energy distributions are shown in Fig. 36 (Gopalswamy 2018). From the speed distribution (left panel), Gopalswamy obtained 100-year (1000-year) CME speeds of 3800 km s$^{-1}$ (4700 km s$^{-1}$) based on the exponential fit. For the power-law fit the corresponding values are 4500 km s$^{-1}$ and 6600 km s$^{-1}$. The 100-year exponential-based speed is only $\sim$ 10% higher than that of the fastest CME observed by LASCO, 3387 km s$^{-1}$ (mass = $9.5 \times 10^{15}$ g) for an event on 10 November 2004 that was associated with a W49 flare. The CME kinetic energy values (right panel) for 100-year (1000-year) events are $4.4 \times 10^{33}$ erg ($10^{34}$ erg) based on the modified exponential function. The most energetic LASCO CME occurred on 9 September 2005 ($4.2 \times 10^{33}$ erg; mass = $1.6 \times 10^{17}$ g; speed = 2257 km s$^{-1}$; W67).

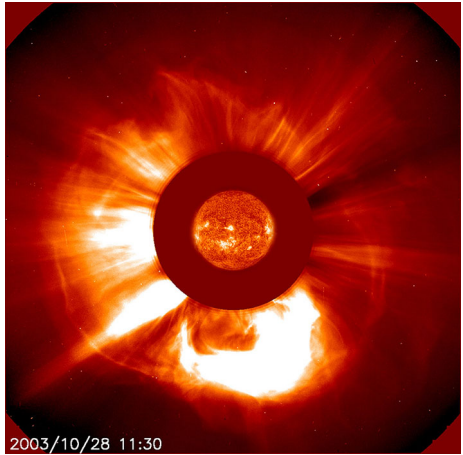

**Fig. 35** CME observed on 28 October 2003 during the Halloween storms (Gopalswamy et al. 2005a, b; Webb and Allen 2004). The measured linear speed was 2,459 km s$^{-1}$ versus an average CME speed of $\sim$ 450 km s$^{-1}$. The CME was associated with both an extreme geomagnetic storm and a severe solar proton event (G5 and S4, respectively on the NOAA Space Weather Scales (http://www.swpc.noaa.gov/noaa-scales-explanation). (LASCO CME image and Extreme-ultraviolet Imaging Telescope (EIT; Delaboudinière et al. 1995) image of solar disk. Image reproduced with permission from https://sci.esa.int/web/soho/-/47806-lasco-c2-image-of-a-cme, copyright by ESA & NASA

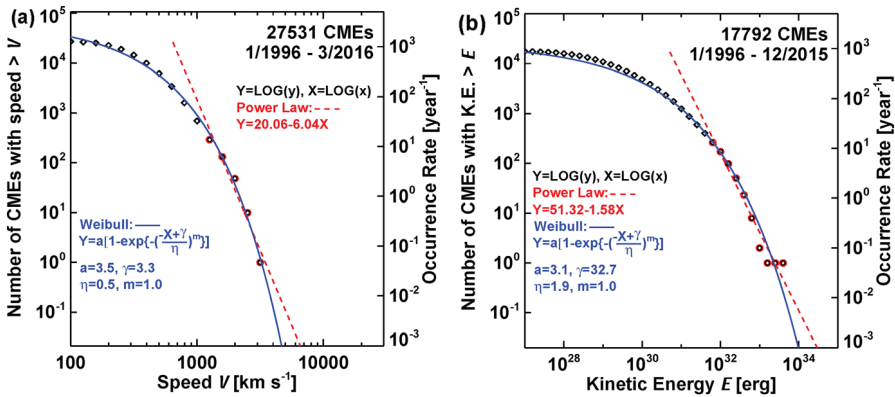

**Fig. 36** **a** Downward cumulative distribution (left hand axis) of the number of CMEs from January 1996 to March 2016 with speeds greater than a given value V (black diamond and red circle data points). This annualized distribution (right hand axis) is fitted with a modified exponential function (solid blue line) and power-law (dashed red line; for the tail of the distribution (red circle data points) to give the corresponding annual occurrence frequency distribution (OFD). The fit equations and parameters are given in the figure panel. **b** Same as **a** for CME kinetic energies ($E$) from 1996 to 2015. Image adapted from Gopalswamy (2018)

Gopalswamy (2011, 2018) obtained an estimate of the fastest possible CME of 6700 km s$^{-1}$—comparable to the 1000-year estimate from Fig. 36 using a power-law fit—based on a spot group area (6000 μsh) similar to that of April 1947 with a uniform field strength of $6.1 \times 10^3$ G (Livingston et al. 2006; cf. Sect. 3.1.7(a)) to yield a total magnetic energy of $\sim 4 \times 10^{36}$ erg. Using the rules of thumb from Emslie et al. (2012; Sect. 3.1.5), this would imply a flare with bolometric energy $\sim 10^{35}$ erg and a SXR class of $\sim$ X70,000 via Eq. (5). If we assume a 6:1 ratio of bolometric to CME energy, then the corresponding flare would still have a SXR class of $\sim$ X30,000 ($\sim 5.7 \times 10^{34}$ erg). For the largest possible SXR flare (X180) flare from Sect. 3.1.7 based on the April 1947 spot group, the corresponding CME would have a total energy (kinetic plus potential) of $8.4 \times 10^{33}$ erg for a 6:1 CME to flare apportionment.

The search for stellar CMEs is a rapidly developing field (e.g., Odert et al. 2017; Moschou et al. 2017, 2019; Argiroffi et al. 2019; Koller et al. 2021; Namekata et al. 2021). Studies by Harra et al. (2016) and Veronig et al. (2021) suggest that the transient dimming signatures on the Sun left by CMEs (Hudson et al. 1996; Sterling and Hudson 1997; Dissauer et al. 2019; Jin et al. 2020) may provide the best evidence of stellar eruptions. Notably, Veronig et al. (2021) used EUV and X-ray observations from the *Extreme Ultraviolet Explorer* (Bowyer and Malina 1991), *Chandra* (Weisskopf et al. 2000) and *XMM-Newton* (Jansen et al. 2001) to identify 21 CME candidates via dimmings on 13 different Sun-like and late-type flaring stars. Approximately half of the dimmings were found on three stars—the young and rapidly rotating K0V star AB Dor (five events), the young M0Ve star AU Mic (three) and Proxima Centauri (two)—with the others on G- to M-type pre-main-sequence and main-sequence stars. The 21 dimmings exceed the total number of stellar CME detections previously reported.

### 5.2 Fast transit interplanetary coronal mass ejections (ICMEs)

Before CMEs were discovered (in 1971), their existence was inferred, if not fully appreciated, by the timing delay between solar flares and magnetic storms at Earth (first hinted at by the 1859 event; Stewart, 1861) and spectroscopically via Doppler shifts or direct off-limb (Fényi 1892) observations of eruptive prominences (Švestka and Cliver 1992; Cliver 1995, and references therein). In terms of extreme events, the short timing delays ($\sim$ 15–30 h) between sudden commencements of great historical geomagnetic storms (caused when CME-driven shocks strike the magnetosphere) and their associated flares are important because they provide a significant extension of the data base for major CMEs. The shortest ICME transit time yet recorded, 14.6 h (Cliver et al. 1990b), was that for an eruptive flare on 4 August 1972. The flare was preceded by four hours by two closely-spaced sudden commencements at Earth signifying ICMEs (that can be linked to large flares from the same region on 2 August; Pomerantz and Duggal 1974) that presumably "pre-conditioned" the interplanetary medium (e.g., Lugaz et al. 2017), resulting in the record low transit time. From the empirical shock arrival time model of Gopalswamy et al. (2005b, c), a 4700 km s$^{-1}$ CME (100-year event; exponential

Table 5　Historical fast transit events

| Nos. | Flare date | UT | Location | Area | SC Date | SC UT | $\Delta T$ | $V_{\text{inf}}$ | Ref.[h] |
|---|---|---|---|---|---|---|---|---|---|
| 01 | 1 Sep 1859 | 1118 | N20W12 | 2300 | 2 Sep | 0448 | ≤ 17.1 | 2356 | N, HH |
| 02 | 15 Jul 1892 | 1700 | S31W22 | 835 | 16 Jul | 1230[e] | 19.5 | 2144 | H, N |
| 03 | 10 Sep 1908 | 0536 | S06W18 | 491 | 11 Sep | 0947 | 28.2 | 1605 | H |
| 04 | 24 Sep 1909 | 1006 | S05W08 | 631 | 25 Sep | 1143 | 25.6 | 1728 | H, N |
| 05 | 10 Nov 1916 | 1542 | N24E18[c] | 192 | 11 Nov | 1912 | 27.5 | 1636 | N |
| 06 | 14 Feb 1917 | 1606 | S23E44[c] | 82 | 15 Feb | 1200 | 19.9 | 2108 | N |
| 07 | 25 Jan 1926 | 2000 | N21W17 | 3385 | 26 Jan | 1648[f] | 20.8 | 2033 | N |
| 08 | 31 Jul 1937 | 1642 | N24E67[d] | 1104 | 1 Aug | 2136 | 28.9 | 1575 | N |
| 09 | 16 Jan 1938 | 0040 | N17E31 | 3116 | 16 Jan | 2235 | 21.8 | 1958 | CS, N, Ca |
| 10 | 15 Apr 1938 | 0830 | N27W12 | 1045 | 16 Apr | 0542 | 21.2 | 2002 | Cb |
| 11 | 28 Feb 1941 | 0930[a] | N12W14 | 683 | 1 Mar | 0354 | 18.4 | 2253 | CS, Ca, N1 |
| 12 | 17 Sep 1941 | 0836 | N11W09 | 1896 | 18 Sep | 0448 | 19.8 | 2117 | N, CS, Ca |
| 13 | 28 Feb 1942 | 1242 | N07E03 | 1865 | 1 Mar | 0812 | 19.5 | 2144 | N, Ca |
| 14 | 6 Feb 1946 | 1628 | N27W19 | 4799 | 7 Feb | 1018 | 17.8 | 2320 | Ca, Cb |
| 15 | 25 Jul 1946 | 1504 | N21E16 | 4279 | 26 Jul | 1842 | 27.6 | 1631 | Cb |
| 16 | 20 Jan 1957 | 1100 | S30W18 | 557 | 21 Jan | 1254 | 25.9 | 1712 | Cb |
| 17 | 9 Feb 1958 | 2108 | S12W14 | 756 | 11 Feb | 0124 | 28.3 | 1600 | Cb |
| 18 | 10 May 1959 | 2102 | N18E47 | 662 | 11 May | 2324 | 26.4 | 1688 | Cb |
| 19 | 14 Jul 1959 | 0325 | N17E04 | 1314 | 15 Jul | 0800 | 28.6 | 1587 | Cb |
| 20 | 16 Jul 1959 | 2114 | N16W31 | 1775 | 17 Jul | 1642 | 19.5 | 2144 | Cb |
| 21 | 12 Nov 1960 | 1315 | N28W01 | 1519 | 13 Nov | 1023 | 21.2 | 2002 | CS, Ca, E |
| 22 | 4 Aug 1972 | 0620 | N14E08 | 1107 | 4 Aug | 2054 | 14.6 | 2847 | Ca, Cb |
| 23 | 14 Jul 2000 | 1024[b] | N22W07 | 460 | 15 Jul | 1417 | 27.9 | 1670[g] | G |
| 24 | 26 Oct 2003 | 1741[b] | N04W43 | 1350 | 28 Oct | 0130 | 31.8 | 1537[g] | G1 |
| 25 | 28 Oct 2003 | 1106[b] | S20E02 | 2120 | 29 Oct | 0600 | 18.9 | 2459[g] | G1 |
| 26 | 29 Oct 2003 | 2041[b] | S19W09 | 2610 | 30 Oct | 1620 | 19.7 | 2029[g] | G1 |

Table reproduced with permission from Gopalswamy et al. (2005b), copyright by AGU

[a]Based on a crochet in the Abinger magnetic traces (Newton 1941)

[b]Time of CME onset at 1 $R$s (solar limb)

[c]The area of the associated active region is rather small, so the level of confidence on the flare association is low

[d]The flare longitude makes the association questionable, although one cannot rule out intense storms from off-center CMEs (e.g., Gopalswamy 2002)

[e]Hale (1931) gives a second more violent storm at 1730 UT, which would have resulted in a longer transit time (24.5 h) and hence a smaller inferred CME speed (1788 km/s)

[f]Newton (1943) gives the transit time as 24 h, even though the difference between the listed flare and geomagnetic storm onsets is only 20.8 h

[g]Events 23–26 are from the SOHO era, where 23 is the Bastille Day event and 24–26 are from Gopalswamy et al. (2005b)

[h]Ca is Cliver et al. (1990a); Cb is Cliver et al. (1990b); CS is Cliver and Svalgaard (2004); E is Ellison et al. (1961); G is Gopalswamy et al. (2002); G1 is Gopalswamy et al. (2005b); H is Hale (1931); HH is Hayakawa et al. (2022); N is Newton (1943); N1 is Newton (1941); See Footnote 2 for data source for group spot areas

fit in Fig. 36a) would have a transit time of 11.8 h versus an asymptotic time in the model of 11.6 h.

Gopalswamy et al. (2005b) compiled a list of fast transit ICMEs (defined as events with Sun-Earth transit times $\lesssim$ 30 h) that is reproduced here as Table 5. The only such event that we are aware of that has occurred since is the backside event on 22–23 July 2012, where the travel time from the Sun to $\sim$ 1 AU was 18.6 h (Russell et al. 2013; Baker et al. 2013; Liu et al. 2014a, b; Gopalswamy et al. 2016). The short transit time in this event has been attributed to a preceding CME that reduced the drag force (Vršnak et al. 2013) due to the interaction of the ICME and the ambient solar wind by reducing the density and increasing the flow speed in the ambient medium (Liu et al. 2014b; Temmer and Nitta 2015). Including the September 1859 eruption (transit time $\leq$ 17.1 h; W12), 19 of the 31 fast transit events in Table 5 originated within 30° of solar central meridian.

Active region magnetic fields are the key determinant of peak CME speed in the corona, prior to deceleration. Vršnak (2021) writes, "Statistically, fast and impulsively-accelerated CMEs originate from strong-field regions, and start to accelerate at low heights (Vršnak 2001; Vršnak et al. 2007; Bein et al. 2011) [See also Dere et al. 1997]. This is consistent with the hypothesis that stronger CME accelerations are driven by stronger magnetic fields, as the Lorentz force is the main driver of the eruption. … the kinetic energy of the eruption comes from the free energy stored in the magnetic field, and … it can be concluded that $\rho v^2/2 < B^2/2\mu_0$ [where $\rho$ = the CME plasma mass-density and $\mu_0$ = the magnetic permeability of free space], i.e., that the CME kinetic energy density cannot exceed the total magnetic energy density, implying $v_{CME} < v_A$, where $v_A$ represents the Alfvén speed within the CME body (for details see Vršnak 2008, and Sect. 2.2.3 in Green et al. 2018). Thus, in stronger fields an eruptive structure can basically achieve a higher speed."

## 6 Geomagnetic storms and aurorae

While geomagnetic storms are not solar events per se, extreme storms are Earth's natural detection system for powerful ICMEs with strong embedded southward-pointing magnetic fields. Systematic geomagnetic observations originated nearly 200 years ago (Cawood 1979; Chapman and Bartels 1940) and detailed auroral observations are available for a few centuries before that, providing a long-term indirect record of extreme solar activity.

In this section, we will use the minimum hourly Dst index (Sugiura 1964; Sugiura and Kamei 1991) during a storm as the measure of storm intensity. Because of the threat of geomagnetically induced currents (GICs) to the power grid (Pirjola 2000; Molinski 2002; Schrijver et al. 2014; 2015), $dB/dt$ (where $B$ is the ground magnetic field) is increasingly used as a measure of storm strength (Kataoka and Ngwira 2016; Pulkkinen et al. 2017). Thomson et al. (2011) used extreme value statistics (Coles 2001; Beirlant et al. 2004) to calculate that a magnetic storm with a d$H$/d$t$ variation (where $H$ is the horizontal component of $B$) of 1000–4000 (1000–6000) nT/minute at 55–60 north geomagnetic latitude could be expected once every 100 (200) years. Storm strength and GIC amplitude, which is strongly dependent on local ground conductivity (e.g., Love et al. 2019b; Lucas et al. 2020), are not closely related (Pulkkinen et al. 2012), however. For example, Huttunen et al. (2008) write that for GIC measurements in Finland, "The largest GIC of the solar cycle 23 (57.0 A on 29 October 2003) [recorded on the Finnish natural gas pipeline network] took place when Dst was barely at the intense storm level while the largest Dst storm of the solar cycle 23 (on 20 November 2003 with Dst minimum of − 422 nT) was associated with much lower-amplitude GIC (23.8 A)." In Sect. 6.3, we show that the extreme Dst storms considered here are characteristically accompanied by auroral effects—indicative of rapid geomagnetic field variations—extending to the low magnetic latitudes of the four stations (Hermanus, Kakioka, Honolulu, San Juan) used to determine Dst.

### 6.1 Climatology of extreme geomagnetic storms

Riley and Love (2017) classified storms with minimum Dst values in the range from − 600 nT < Dst < − 250 nT as severe (see also Tsurutani et al. 1992; Gonzalez et al. 1994; Lakhina et al. 2012) and those with minimum Dst < − 600 nT as extreme. Several determinations of waiting times have been made for storms with minimum Dst values of ∼ − 600 nT, corresponding to the lowest Dst value of − 589 nT observed in modern times (on 14 March 1989 storm; Allen et al. 1989), and − 850 nT, the minimum hourly Dst value inferred by Siscoe et al. (2006) for the 2 September 1859 storm. Results of these studies are given in Table 6. The calculated waiting times in the table for a storm with minimum Dst ∼ − 600 nT range from 25 to 60 years with a median value of 55 years depending on the data interval considered, the assumed form of the Dst peak intensity distribution, and the analysis techniques employed. The corresponding values for a storm with minimum Dst of − 850 nT are a range from 49 to 333 years with a median of 93 years. Estimates for the minimum Dst value of 100-year storms range from − 542 nT (Love 2020) based on a lognormal distribution to − 1100 nT (Kataoka and Ngwira 2016; power-law distribution). Gopalswamy's (2018) estimates for a 100-year storm are − 603 nT (modified exponential distribution) and − 774 nT (power-law distribution). For a 1000-year event he obtained − 845 nT (exponential function) and − 1470 nT (power law). As we will see below, there is evidence that the 1000-year estimate of − 845 nT has been exceeded three times within the last ∼ 160 years, so in this case the power law may be more appropriate.

Gonzalez et al. (2011) and Baker et al. (2013) have presented evidence based on the "fast transit" ICMEs of 4 August 1972 and 22–23 July 2012 (see Sect. 5.2 and Table 5) that suggest the possibility of far larger storms than one with a minimum Dst of ∼ − 850 nT. For the 4 August 1972 event, Gonzalez et al. (2011) estimated that if the ICME magnetic field had been southward, the associated geomagnetic storm would have had a minimum Dst of ∼ − 1400 nT because of its measured/ inferred high CME speed (Zastenkar et al. 1978; Cliver et al. 1990a). For the July

**Table 6** Waiting time studies for extreme geomagnetic storms with Dst ≲ − 600 nT

| References | Data interval | Dst threshold (nT) | No. of events per 100 years | Waiting time (years) | Method/approach |
|---|---|---|---|---|---|
| Tsubouchi and Omura (2007) | 1957–2001 | − 589 | 1.67 | 60 | Extreme value theory |
| Love (2012) | 1859–2011 | − 589 | 1.961 | 56 | Poisson model; frequentist and Bayesian inference |
|  |  | − 1760 | 0.654 | 159 |  |
| Love et al. (2015) | 1957–2012 | − 589 | 4.03 | 25 | Lognormal distribution; MLE |
|  |  | − 600 | 3.79 | 26 |  |
|  |  | − 850 | 1.13 | 88 |  |
|  |  | − 589 | 1.86 | 54 | Lognormal distribution; weighted least squares |
|  |  | − 600 | 1.72 | 58 |  |
|  |  | − 850 | 3.6 | 278 |  |
| Riley and Love (2017) | 1957–2016 | − 850 | 2.03 | 49 | PLD; K-S; bootstrapping, likelihood ratio test |
|  | 1964–2016 | − 850 | 1.03 | 97 |  |
|  |  | − 850 | 0.03 | 333 | Lognormal distribution |
| Love (2020) | 1957–2016 | − 542 | 1 | 100 | Upper-limit lognormal |
|  | 1957–2016 | − 591 | 1 | 100 | Extreme value distribution |
| Love (2021) | 1902–2016 | − 663 | 1 | 100 | Rank statistics; Weibull model |
| Gopalswamy et al. (2018) | 1957–2016 | − 603 | 1 | 100 | Weibull distribution |
|  |  | − 845 | 1 | 100 | PLD |

MLE, maximum likelihood estimator; PLD, power law distribution; K–S, Kolmogorov–Smirnov test

2012 backside ICME, a model for the 22–23 July 2012 storm based on solar wind observations that assumed optimal solar wind magnetosphere coupling conditions (based on seasonal and time-of-day orientation of Earth's magnetic dipole; Russell and McPherron 1973; Cliver et al. 2000, 2002; Temerin and Li 2002) yielded a minimum Dst value of − 1182 nT (Baker et al. 2013). As we will see below, the low-latitude aurorae of September 1859 and February 1872 suggest minimum hourly Dst values from ∼ 1200–1250 nT. Vasyliūnas (2011) has argued that the maximum strength of a geomagnetic storm will be limited to ∼ − 2500 nT by the inability of Earth's dipole field to balance the mechanical stresses on magnetospheric plasma beyond a certain point. Recently, from a consideration of an "ICME in a sheath" as a storm driver, Liu et al. (2020) obtained a comparable limiting Dst value of ∼ − 2000 nT for an extreme geomagnetic storm.

### 6.2 Identified geomagnetic storms (1500-present) with Dst < − 500 nT

Until recently, there were only two well-documented cases of geomagnetic storms comparable to or more intense than the 14 March 1989 event (Allen et al. 1989; Yokoyama et al. 1998; Kappenman 2006; Pulkkinen et al. 2012; Boteler 2019): 2 September 1859 (Loomis, 1859, 1860, 1861; Stewart, 1861; Kimball 1960; Tsurutani et al. 2003; Akasofu and Kamide 2005; Siscoe et al. 2006; Green and Boardsen 2006; Green et al. 2006; Gonzalez et al. 2011; Cliver and Dietrich 2013; Tsurutani et al. 2018) and 14 May 1921 (Silverman and Cliver 2001; Kappenman 2006; Cliver and Dietrich 2013). Detailed auroral descriptions were available for three other low-latitude aurora events: 29 August 1859 (Loomis, 1859, 1860, 1861; Stewart, 1861; Kimball 1960; Green and Boardsen 2006; Green et al. 2006), 4 February 1872 (Silverman 2008), and 25 September 1909 (Silverman 1995). Lately, a newly identified trove of historical auroral observations (e.g., local treatises, chronicles, diaries, and newspapers), with an emphasis on those from East Asia, have been used to identify and document intense storms on 8 March 1582 (Hattori et al 2019), 15 February 1730 (Hayakawa et al. 2018b), 17 September 1770 (Hayakawa et al. 2017), 31 October 1903 (Hayakawa et al. 2020b), 1 March 1941 (Hayakawa et al. 2021), and 28 March 1946 (Hayakawa et al. 2020a). In addition, such newly-found auroral observations have been brought to bear on the events of August 1859 (Hayakawa et al. 2018c, 2019b), September 1859 (Hayakawa et al. 2016; Hayakawa et al. 2018c, 2019b, 2020c, 2022), February 1872 (Hayakawa et al. 2018a), September 1909 (Hayakawa et al. 2019a), and May 1921 (Hayakawa 2020). In a further recent development, minimum Dst values have been constructed from historical magnetic records for the September 1909 (Love et al. 2019a), May 1921 (Love et al. 2019c), and October 1903 (Hayakawa et al. 2020b) storms. A contemporary drawing of the 1770 aurora (Fig. 37) and a map of locations in East Asia from which that aurora was reported (Fig. 38) are representative of the recent auroral research.

It would be useful to be able to estimate the strength of the three newly reported pre-nineteenth century geomagnetic storms (1582, 1730, 1770) as well as for those in August 1859 and February 1872, for comparison with more recent events. This can be done via a relationship between the lowest magnetic latitude of overhead aurorae (as opposed to the lowest latitude from which the aurora was observed) and minimum Dst values of the associated magnetic storms. The first such comparison of these parameters we are aware of was made by Akasofu and Chapman (1963). With the exception of March 1989, the determination of either of these parameters for the extreme events considered here can be problematic. In particular, the assignment of a minimum equatorward boundary to a storm is fraught with uncertainty. First, there are subjective decisions that must be made for early auroral records (e.g., Is enough detail provided to make a given observation credible? Is the phenomenon really an aurora rather than a forest fire, a comet, zodiacal light, or an atmospheric optical effect (e.g., Kawamura et al. 2016; Usoskin et al. 2017)? Second, the availability of an auroral observation is dependent on the distribution of recording observers convolved with local weather conditions and, more recently, light pollution. Third, auroral observations will often be based on untutored

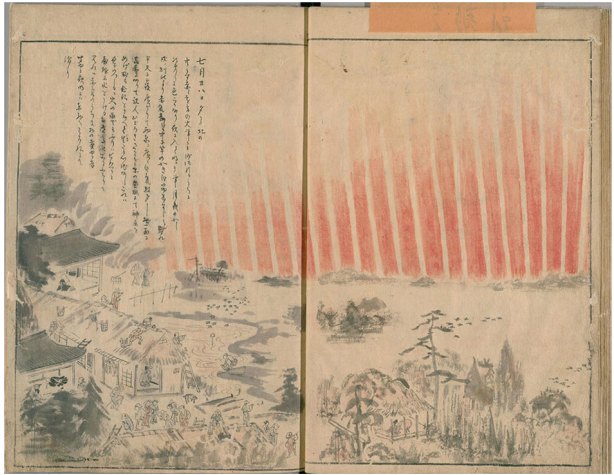

**Fig. 37** Contemporary drawing from Japan of the aurora of 17 September 1770. Image reproduced with permission from Hayakawa et al. (2017), copyright by AAS

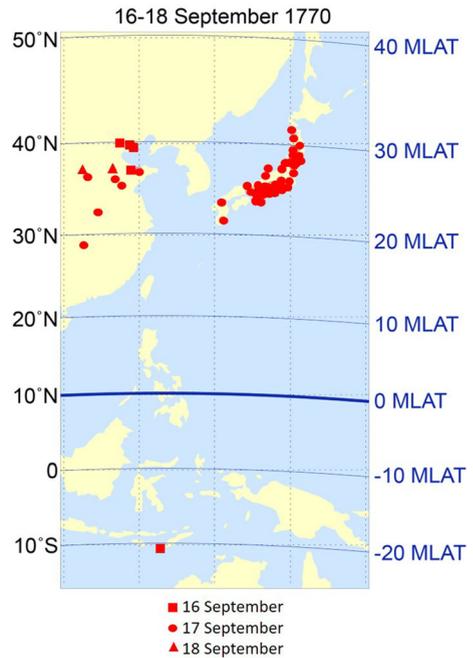

**Fig. 38** Locations in East Asia from which the aurora of 16–18 September 1770 was reported. Image reproduced with permission from Hayakawa et al. (2017), copyright by AAS

observation of a highly dynamic phenomenon that requires attention to detail, in particular, the elevation of the aurora above the horizon in a given direction. Finally, a physical height of the aurora above the Earth's surface must be assumed (for a poorly observed phenomenon—low-latitude aurora—distinguished by its rarity) and the assignment of magnetic coordinates is based on a model for the evolution of Earth's field spanning centuries. In their determinations of the lowest magnetic latitude of overhead aurora for historical events, Hayakawa, Ebihara, and colleagues considered maximum auroral heights of 400 km (Roach et al. 1960; Ebihara et al. 2017) [and also 800 km for the 1909 storm (Loomis, 1861; see Shea and Smart 2006, p. 374 ff.)] and used the *gufm1* magnetic field model (Jackson et al 2000) back to 1590 and the *Cals3k.4b* model (Korte and Constable 2011) before then.

The main difficulty in the determination of the peak intensity of pre-twentieth century magnetic storms is the relative scarcity of early magnetic observations, and the fact that, even when such observations were made, they frequently were driven off-scale. For this, and other reasons (e.g., Akasofu and Kamide 2005), estimates for the minimum hourly Dst value of the September 1859 storm have ranged from $-625$ nT (Siscoe et al. 2006) to $-1760$ nT (Tsurutani et al. 2003).

Following Akasofu and Chapman (1963), comparisons of the lowest latitude of overhead aurorae with their peak magnetic index values have been made by various authors (e.g., Feldstein and Starkov 1967; Lui et al. 1975; Gussenhoven et al. 1981; Schulz 1997). Here, we employ the relationship obtained between these parameters in Yokoyama et al. (1998), in part because their analysis encompassed a greater range of storm intensity than was the case for earlier studies. Even then, the curve Yokoyama et al. obtained (empirically determined from their Fig. 7 by Hayakawa 2020), indicated by open black circles in Fig. 39 and given by

$$\text{Dst(min)} = -2200\cos(\text{ILAT})^6 + 12, \qquad (11)$$

where ILAT (in degrees) is the invariant magnetic latitude (O'Brien et al. 1962) of the most equatorward extent of overhead aurora,[14] was limited to storms with minimum Dst $> -350$ nT. An extrapolation of this curve was used to determine minimum Dst for the five low latitude aurorae for which magnetic measurements either were not available (1582, 1730, 1770) or were inadequate (August 1859, 1872). In addition, recently uncovered auroral observations from South America (Hayakawa et al. 2020c) suggest a significant change — from $\sim -850$ to $-1050$ nT (Cliver and Dietrich 2013) to $-1200$ nT—in the minimum Dst value for the 2 September 1859 storm. In each of these six cases minimum Dst was $< -500$ nT. Auroral and magnetic parameters and contextual data for the 12 documented

---

[14] Yokoyama et al. (1998) used corrected geomagnetic latitude (CGL; Gustafsson et al. 1992) instead of ILAT. CGL has constant values of magnetic latitude along geomagnetic field lines while ILAT is nearly, but not strictly, invariant along field lines (Richmond 1995). Both CGL and ILAT can deviate by up to several degrees from Magnetic Apex latitude at the lowest latitudes ($\sim 25°$) of observed overhead aurorae (VanZandt et al. 1972; Laundal and Richmond 2017).

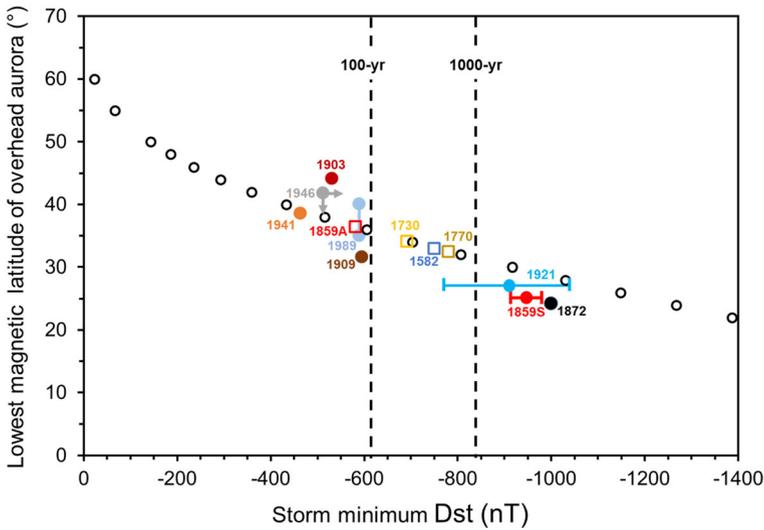

**Fig. 39** Lowest magnetic latitude (ILAT) of overhead aurora for extreme storms versus storm minimum Dst value—an extension of a plot from Yokoyama et al. (1998). The Yokoyama et al. plot was based on 423 storms from 1983 to 1991 (all with Dst values $\gtrsim -350$ nT) for which the magnetic latitude of lowest aurora was determined from *DMSP/F2* precipitating electron data (Gussenhoven et al. 1981, 1983). Filled circle points indicate great storms with minimum Dst values based on direct magnetic measurements. Open squares on the smooth curve of black points given by Eq. (11) are based on the most-equatorward latitude at which aurora was observed overhead (assuming a peak altitude of 400 km). The vertical dashed lines indicate estimates of Dst for 100- and 1000-year storms (Gopalswamy et al. 2018)

geomagnetic disturbances exceeding (or possibly exceeding for the March 1941 storm) this threshold for storm strength are given in Table 7.

As a test of the reliability of Eq. (11), we calculated a minimum Dst value (given in italics in Table 7) for the post-1872 storms for which a Dst determination could be made (although for non-standard station sets for all but the 1989 event). For these six cases, the Dst values based on Eq. (11) (these are not plotted in Fig. 39) agreed reasonably well with those based on magnetic records, given the practical problems in determining the ILAT of overhead aurora delineated above. That said, the uncertainties are large enough that we regard the storm intensities obtained by this method as suggestive rather than definitive.

In regard to the events of most interest for this review, the great storms of September 1859, May 1921, and February 1872 storms: Hayakawa et al. (2022) calculated a minimum Dst value of $-949(\pm 30)$ nT from the Colaba observations for 1859; Love et al. (2019c) used four alternative stations to obtain a minimum Dst of $-907 (\pm 132)$ for 1921; no Dst index has yet been estimated for 1872—Eq. (11) gives a minimum Dst of $-1250$ nT, but taking the Colaba magnetogram into account (Hayakawa et al. 2018a, 2020), suggests a conservative value of $\sim -1000$ nT, at the high end of the uncertainty range for the 1859 and 1921 storms.

Table 7 Auroral parameters for magnetic storms with minimum Dst values $\lesssim -500$ nT

| Storm peak date | Storm peak U.T. (mag. station)[f] | Lowest lat. auroral peak time (U.T.) | Lowest lat. auroral peak time (L.T.) | Place name of lowest lat. auroral obs | Geographic coords. of most equatorward auroral obs. Lat. | Geographic coords. of most equatorward auroral obs. Lon. | Lowest mag. lat. of auroral observation | Overhead auroral ILAT (400 km) | Place name/elev. angle of obs. used for ILAT determination | Min. Dst value (nT)[g] | Refs. (notes) |
|---|---|---|---|---|---|---|---|---|---|---|---|
| 8 March 1582 | – | ~ 13–15 | 22–24 | Búngo | N33° 14' | E131° 36' | N28.8° | 33.0° | Kofukuji / 90° | − 750 | 1 |
| 15 Feb 1730 | – | ~ 13–17 | 22–26 | Kyoto | N34° 59' | E135° 47' | N25.8 | 34.2° | Kaminoyama / 90° | − 690 | 2 |
| 17 Sep 1770 | – | ~ 12.5–16.5 | 20–24 | Dòngtínghú | N28° 51' | E112° 37' | N18.8° | 32.5° | Toyoda / 35° | − 780 | 3 |
| 28 Aug 1859 | 21–10 (RGO) | ~ 08–09 (29) | 03–04 (29) | Panama | N08° 59' | W79° 31' | 20.2° | 36.5° | Havana / 90° | − 580 | 4 |
| 2 Sep 1859 | 6 (Colaba) | | | Vessel Dart | S19° | W149° | S17.3° | 25.1° | Santiago / ≥ 90°, ≈60° | − 949 (± 30) (− 1200) | 4(a) |
| 2 Sep 1859 | 6 (Colaba) | 08.5– | 22.0? (1) | Honolulu | N21° 18' | W157° 51' | N20.5° | | | | |
| 4 Feb 1872 | ~ 19 (Colaba) | ~ 19 | 23.5 | Shàoxīng | N30° 00' | E120° 35' | N18.7° | 24.2° | Jacobabad/90° | ~ − 1000 (− 1250) | 5(b) |
| 31 Oct 1903 | ~ 15 (Dst*) | ~ 11–15 | 20–24 | Sydney | S33° 52' | E151° 12' | S42.2° | 44.1° | Sydney / 90° | − 531 (− 290) | 6 |
| 25 Sep 1909 | 12.5, 18.5 (Dst*) | 16.7–17 | 01.7–02 (26) | Matsuyama | N33° 51' | E132° 47' | N23.2° | 31.6° | Matsuyama / 30° | − 595 (− 830) | 7(c) |
| 15 May 1921 | 5.5 (Dst*) | 5.75–6.5 | 18.25–19 | Apia | S13° 50' | W171° 45' | S16.2° | 27.1° | Apia / 22° | − 907 (± 132) (− 1080) | 8 |
| 1 March 1941 | 16 (Dst*) | 16:05–18:30 | 01:05–03:30 (2) | Oshidomari | N45° 14' | E141° 13' | 35.0° | 38.5° | Wakkanai / 90° | − 464 (− 490) | 9 |

Table 7 continued

| Storm peak date | Storm peak U.T. (mag. station)[f] | Lowest lat. auroral peak time (U.T.) | Lowest lat. auroral peak time (L.T.) | Place name of lowest lat. auroral obs | Geographic coords. of most equatorward auroral obs. Lat. | Geographic coords. of most equatorward auroral obs. Lon. | Lowest mag. lat. of auroral observation | Overhead auroral ILAT (400 km) | Place name/elev. angle of obs. used for ILAT determination | Min. Dst value (nT)[g] | Refs. (notes) |
|---|---|---|---|---|---|---|---|---|---|---|---|
| 28 Mar 1946 | ~ 14 (Dst*) | 11.6–14.3 | 18.6–21.3 | Watheroo | S30° 19' | E115° 52' | S41.8° | ≤ 41.8° | Watheroo / corona | ≤ − 512 (≤ − 370) | 10(d) |
| 14 Mar 1989 | 01 (Dst) | ~ 01 | ? | DMSP | – | – | – | 35–40.1° | DMSP | − 589 (− 430 to − 650) | 11(e) |

References: (1) Hattori et al. (2019), Hayakawa (2020); (2) Hayakawa et al. (2018), Hayakawa (2020); (3) Willis et al. (1996), Ebihara et al. (2017), Hayakawa et al. (2017), Hayakawa (2020); (4) Carrington (1859), Hodgson (1859), Loomis (1859, 1860, 1861), Secchi (1859), Stewart (1861), Bartels (1937), Kimball (1960), Tsurutani et al. (2003), Cliver and Svalgaard (2004), Siscoe et al. (2006), Green and Boardsen (2006), Green et al. (2006), Tyasto et al. (2009), Cliver and Dietrich (2013), Hayakawa et al. (2016, 2018c, 2019b, 2020c, 2022); (5) Chapman (1957), Silverman (2008), Hayakawa et al. (2018a), Valach et al. (2019); (6) Hayakawa et al. (2020b); (7) Silverman (1995), Hayakawa et al. (2019a), Love et al. (2019a); (8) Angenheister and Westland (1921), Silverman and Cliver (2001), Kappenman (2006), Cliver and Dietrich (2013), Love et al. (2019c); Hapgood (2019), Hayakawa (2020); (9) Karinen and Mursula (2005, 2006), Hayakawa et al. (2021); (10) Karinen and Mursula (2005, 2006), Hayakawa et al. (2020a), (11) Allen et al. (1989), Rich and Denig (1992), Hayakawa (2020)

(a) While the date of the observation from Honolulu is most likely 1 September, it is not given explicitly in the report.

(b) Aurorae were reported from Bombay (Chapman 1957; cf. Preece, in Fron. 1872) and Aden (Chapman 1957) as well as from Khartoum and Gondokoro (Fron 1872), with the report from Bombay deemed credible; those from Aden and Khartoum plausible, and that from Gondokoro uncertain (Hayakawa et al. 2018a)

(c) Silverman (2008) suggests that unsubstantiated reports of aurora at Singapore (− 10.0°) magnetic latitude, Batavia (− 17.5°), Coco Islands (− 23.2°), Rodriquez (− 27.1°), and Durban (− 31.1°) were due to conflation with telegraph outages.

(d) See Fig. 7 in Kimball (1960) in regard to the distinction between overhead aurora and corona.

(e) Low latitude auroral boundary based on *Defense Meteorology Satellite Program* (DMSP) observations of precipitating electrons (Rich and Denig 1992; Yokoyama et al. 1998)

[f] (Dst*) indicates a non-standard Dst magnetic station set

[g] Values in italics/parentheses based on Eq. (11) using ILAT for a maximum auroral height of 400 km

As noted in Sect. 3.1.7 for the great spot groups of cycle 18, even very large groups, with peak areas during their disk passage ≳ 4500 μsh, are not a sufficient condition for an extreme magnetic storm. Nor are they necessary. The magnetic storm in May 1921 with a minimum value of ∼ − 900 nT (Love et al. 2019c) originated in Greenwich spot group 9334 with peak area of 1709 μsh during its disk passage. The area of the spot group responsible for the 1872 storm is estimated to have been about half this size (Hayakawa et al. 2022, in preparation). Magnetic storm strength is more strongly dependent on flare location (with optimum geoeffectiveness for flares located near central meridian), orientation of the magnetic field in the causative CME (leading edge southward), and the seasonal variation of the Sun-Earth geometry (equinoctial) (Cliver and Hayakawa 2020).

### 6.3 Timing relationship between great storms and aurora: global substorms

Magnetograms for the three largest storms in Table 7—September 1859, 1872, 1921—are given in Figs. 40, 41 and 42, respectively. In each of the three figures, the interval of maximum auroral activity (based on subjective reported aspects such as brightness/extent/dynamics/observer reaction) is indicated. In each case, there is evidence that auroral activity occurs near the time of a minimum in the low-latitude geomagnetic horizontal (H) component. For the September 1859 event, the sharp downward spike in the Colaba magnetogram in Fig. 40 occurred in concert with auroral activity over a wide range of latitudes in North America (Green and Boardsen 2006) and South America (Hayakawa et al. 2020c). The red bar denoting the interval of intense auroral activity in the American sector bounds the timings of the deep sharp excursion in Colaba and strong magnetic variations observed in Rome (Secchi 1859; Blake et al. 2020) and Ekaterinburg (Tyasto et al. 2009).

The February 1872 event (Silverman 2008; Hayakawa et al. 2018a) also presents evidence, although not as clear cut, of the simultaneity of storm peak and higher latitude auroral activity in a great storm. In that case the onset of auroral activity in northern India (Chapman 1957) was accompanied by the insertion of a deflector magnet at the Colaba Observatory (magnetic latitude = 10.0° N in 1872) near Bombay to keep the magnetic traces on scale in the H magnetogram, resulting in the loss of approximately 15 min of data (Fig. 41). The equatorward extent of overhead

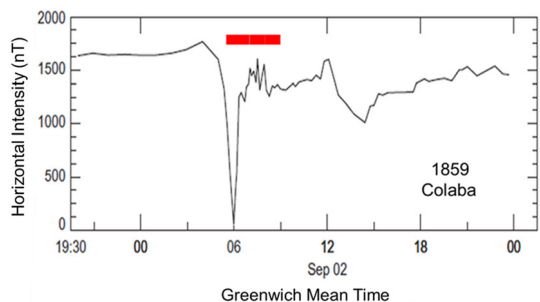

**Fig. 40** Magnetogram from Colaba for the extreme magnetic storm of 2 September 1859. The red bar corresponds to the time of intense auroral activity (∼ 05–09 UT) in the American sector. Image adapted from Tsurutani et al. (2003)

aurora in this storm (N24.2°) suggests that this is the strongest storm in the last 450 years.

The abrupt onset of the 1872 aurora near the time of insertion of the deflector magnet is made clear by the correspondent to the Times of India from Jacobabad (magnetic latitude = 19.9° N in 1872) who wrote, "As I was returning home around half-past 11 p.m., a sudden change from darkness to light was noticed as bright as the full moon. I was amazed; the conversion of Saint Paul came vividly before me—in fact, I was terrified by the sudden change; my dog became motionless and seemed to tremble. I thought it must have been a fire, no, the whole place was magically illuminated. … Its shape was an arch, though not quite so perfect, shooting from the east horizon to the zenith and very nearly at right angles to the magnetic meridian." (Excerpted from Chapman 1957). The timing coincidence of the auroral onset at Jacobabad and the insertion of the deflector magnet at Colaba strongly suggests that the H-trace went off scale to lower H-values, but we do not know what minimum value was achieved during the data gap. The listed H-range, which appears to be measured from the peak amplitude of the sudden commencement, is given as 1023 nT in the inset box.

A similar correspondence of aurora and storm peak was found for the May 1921 event. The red oval in the magnetogram from Apia, Samoa (magnetic latitude = 16.2° S in 1921) for the May 1921 event in Fig. 42 encompasses a positive magnetic bay of $\sim$ 400 nT (Cliver and Dietrich 2013). The increase in H rules out a ring current effect. This is the only example we have for an extreme storm where a feature on a low latitude magnetogram corresponds to precisely timed and described auroral activity observed from the same site. Excerpting from Cliver and Dietrich (2013): "At Apia, on 15 May 1921, Angenheister and Westland (1921) reported an auroral arc that spanned $\sim$ 25° in the southern skies from 5:45 to 6:30 UT [6:15–7:00 p.m. local time]. The arc, 'of a glowing red color', was centered approximately on the magnetic meridian and had a peak altitude of 22° [placing the lowest magnetic latitude of overhead aurora at 27.1° (Table 7)]. They noted that, 'The point of the greatest intensity appeared to move from east to west at about 6 h 20 m. Greenwich time …' and that no signs of the light were seen after 6:30 UT."

This is the lowest magnetic latitude ($\sim$ 15°) from which an aurora has yet been credibly observed. Although this event has its own puzzle—the aurora was observed only to the south from Auckland (Silverman and Cliver 2001)[15]—the authoritative report by trained observers and the coincidence in time between the magnetic bay in Fig. 42 and the description of the aurora mark this as an important observation for our understanding of low-latitude aurorae.

The coincident timing of auroral/ionospheric current on low-latitude magnetograms in the three largest storms for which magnetic records exist focuses attention on the role of field-aligned currents in such storms (e.g., Cid et al. 2015). The idea that such currents might contribute to Dst in extreme storms has its origin in the observation of the deep negative excursion in the Colaba H-trace for the 1859

---

[15] Cliver and Dietrich (2013) suggest the gap in the aurora between Apia and Auckland was caused by separate bands of aurora observed from these two sites, with reference to DMSP observations for the March 1989 aurora showing a similar latitudinal rift (Allen et al. 1989).

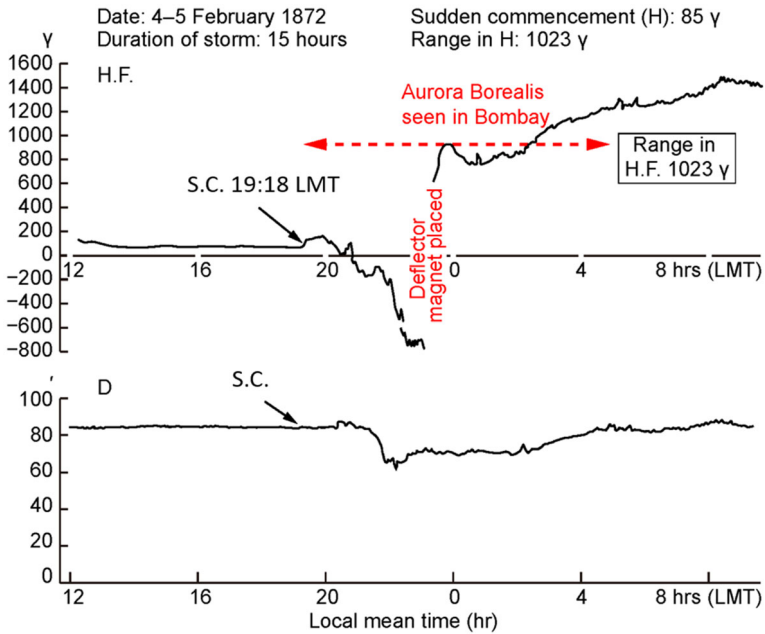

**Fig. 41** Magnetogram from Colaba for the extreme geomagnetic storm of 4 February 1872. The deflector magnet was installed near the time of the sudden appearance of the aurora at Jacobabad in order to keep the "light stylus" on the photographic recording paper. The long duration of the aurora as seen from Bombay is questionable and may refer to the interval of telegraph disturbance. Local Mean Time (LMT) = UT + 4:51. Abbreviations: H.F., horizontal force; D, declination; S.C., sudden commencement of storm. Image adapted from Fleming (1954)

event (Siscoe et al. 2006; Green and Boardsen 2006; Cliver and Dietrich 2013). For the same event, the magnetometer station in Rome recorded a reported rapid H-change of $\sim$ 3000 nT and a swing in declination of 4° 13' in less than one hour that occurred close in time to the negative spike in the Colaba magnetogram (Secchi, 1859; Cliver and Dietrich 2013; Blake et al. 2020). Siscoe et al. (2006) framed the problem, "The issue regarding the Bombay magnetogram for the 1859 storm is whether its unprecedented negative excursion resulted from ionospheric currents or magnetospheric currents. If it resulted from ionospheric currents, then the size of the excursion is not so exceptional, but the fact that ionospheric currents could profoundly affect a magnetogram at such low latitude remains an exceptional aspect of the storm. Such an interpretation would seem to imply that overhead auroras might have reached the latitude of Bombay, yet against this inference, Green and Boardsen (2006) report that auroral records for the storm indicate that overhead auroras came no closer to Bombay than 10° latitude. If instead of ionospheric currents the deep negative excursion in the Bombay magnetogram resulted from magnetospheric currents, then we learn that in the case of superstorms the hourly averaged Dst index could be under-representing the actual extent of H-depression that we normally associate with the ring current." Based on the 1859, 1872, and 1921 events, we suggest that low-latitude auroral effects are a common

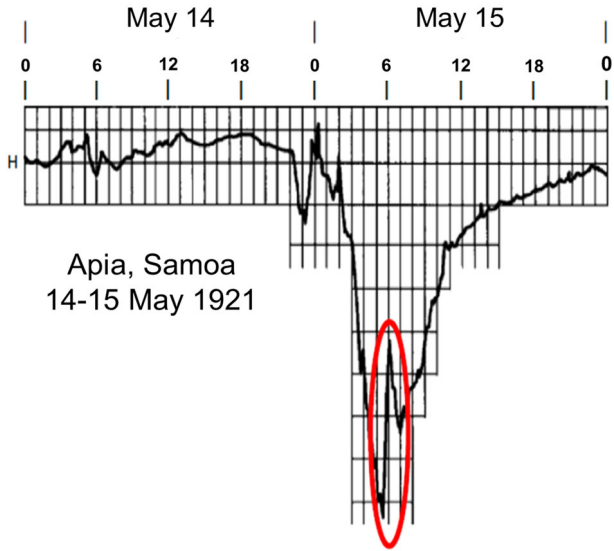

**Fig. 42** Magnetogram from Apia, Samoa for the extreme geomagnetic storm of 15 May 1921. The red oval indicates a positive bay coincident with the observation of aurora from the magnetometer station. Image adapted from Angenheister and Westland (1921) and Cliver and Dietrich (2013)

aspect of exceptional geomagnetic storms. From these three events it also appears that overhead auroras or their associated field-aligned currents can affect magnetometers located $\sim 10°$ equatorward in latitude. It is this "global substorm" aspect of extreme geomagnetic storms, bringing large d$B$/d$t$ variations to a large fraction of the Earth's population, that makes such storms the pre-dominant space weather threat.

## 7 Extreme solar energetic proton (SEP) events

### 7.1 Acceleration of protons at the Sun

Solar energetic proton events were first detected in 1942 (Lange and Forbush 1942a, b; Berry and Hess 1942; Forbush 1946), though not recognized as such at the time, by ionization chambers that were used to monitor galactic cosmic ray intensity. Thus, in some of the earlier literature (e.g., Meyer et al. 1956), SEP events are referred to as solar cosmic ray events. The first solar proton events originated in the same active region (Fig. 43) that produced the first solar radio signals detected at Earth (Hey 1946).

Since their discovery, the understanding of how SEPs are accelerated has gone from tacit acceptance that they are accelerated in flares—the default position (e.g., Forbush 1946; Meyer et al. 1956) prior to the discovery of CMEs—to a growing recognition/belief, with key early contributions by Wild et al. (1963), Lin (1970), and Kahler et al. (1978), that the protons observed in space, even at high-energies,

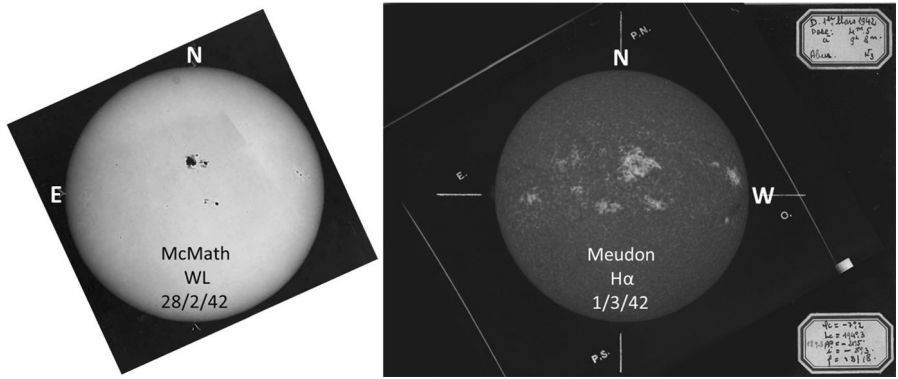

**Fig. 43** The large active region near disk center on 28 February 1942 (RGO 14015; Jones 1955) was notable for the first observations of both solar radio (Hey 1946) and particle emission (Forbush 1946). It also had the strongest magnetic field (6100 G) recorded at Mt. Wilson from 1917 to 2004 (Livingston et al. 2006). Left: White-light image from McMath-Hulbert Solar Observatory on 28 February; Right: Ca II K1v image from Meudon on 1 March

are primarily accelerated by CME-driven shock waves (Reames 2009, 2013, 2015; Mewaldt et al. 2012; Desai and Giacalone 2016; Cliver 2016; Bruno et al. 2018). A schematic showing the flare and shock acceleration sites in an eruptive flare is given in Fig. 44. The flare-resident acceleration process, likely stochastic in nature (e.g., Petrosian 2012; Benz 2017), is driven by reconnection at the X-type neutral point which evolves into a neutral current sheet (Fig. 10). It is responsible for upward-moving electrons that reveal themselves via fast-drift type III radio emission (Fig. 11) and downward "precipitating" electrons that give rise to hard X-ray and microwave emission and heat the chromosphere to fill the reconnected loops with hot plasma that radiates in soft X-rays—the Neupert effect (Sect. 3.1.4).

Arguments favoring CME-driven shocks (Cliver 2020) over a flare process include: (1) the strong association of large high-energy SEP events with fast CMEs and low-frequency (decametric-hectometric type II) shocks; (2) the association of certain large SEP events with weak flares; (3) the rapid arrival of high-energy protons from poorly-connected eruptive flares; and less directly, (4) the three-to-one (or more) ratio of CME energy to bolometric energy in eruptive flares, and a well-developed theoretical framework for diffusive shock acceleration. In addition, Fig. 44 shows the intrinsic advantage of CME-driven shock waves—they accelerate on open field lines whereas protons accelerated via reconnection in the flare are at least initially trapped in either the post eruption arcade or in the body of the CME. Flare-associated electrons are observed via type III bursts during the impulsive phase of eruptive flares so the picture (shock acceleration on open field lines; flare acceleration on closed) is not as neat as shown in Fig. 44, but the association of type III bursts with eruptive flares is not particularly strong. Cane and Reames (1988) find that of 685 eruptive flares (accompanied by a type II and/or a type IV radio burst) from 1961 to 1983, 37% lacked associated type III emission and for another 27% the accompanying type III emission was weak.

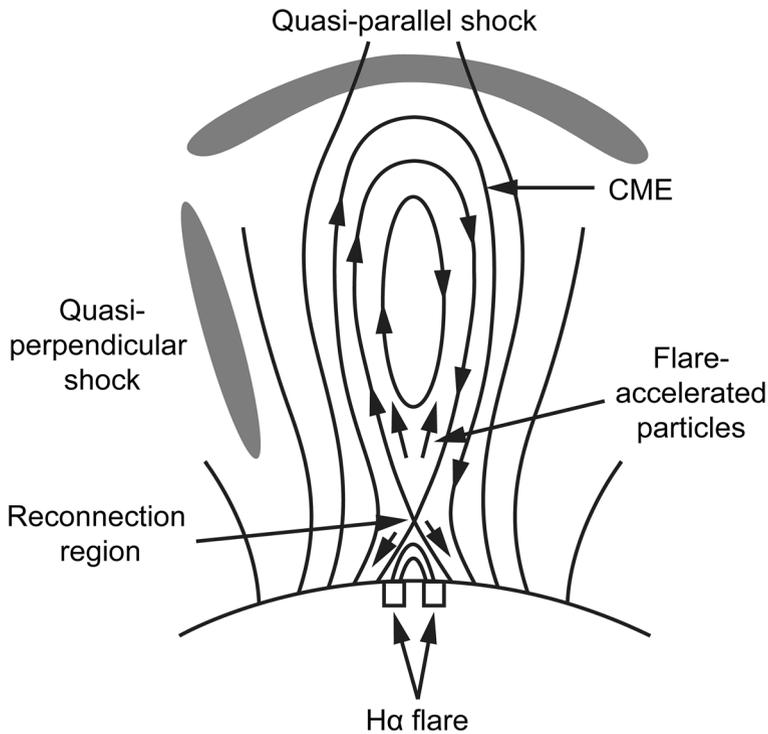

**Fig. 44** Standard CSHKP-type model for eruptive flares showing the flare-resident particle acceleration site at an X-type reconnection point above the flare loops and below the disconnected CME (over time, the X-point will develop into a neutral current sheet between the flare loops and the CME) and CME-driven shocks (quasi-parallel and quasi-perpendicular). Image adapted from Cliver et al. (2004)

The strongest evidence that a flare-resident acceleration process can contribute to the proton events observed in space is provided by remote gamma-ray observations of eruptive flares. Forrest and Chupp (1983) provided the first such evidence—prompt 4–8 MeV gamma ray line emission observed by the Gamma-ray Spectrometer (Forrest et al. 1980) on the Solar Maximum Mission (SMM; Bohlin et al. 1980) during the impulsive phases of flares on 7 and 21 June 1980, requiring the rapid (within seconds) acceleration of protons to tens of MeV. Subsequently, Forrest et al. (1986) and Chupp et al. (1987) reported pion-decay emission during the impulsive phase of a flare on 3 June 1982, indicating proton acceleration to $> 300$ MeV. More recently, from *Fermi* Large Area Telescope (LAT; Atwood et al. 2009) $\gamma$-ray observations of the impulsive 12 June 2010 flare, Ackermann et al. (2012) determined that the bulk of the $> 100$ MeV protons were accelerated with a delay of $\sim 10$ s from the $> 300$ keV electrons. So there is little doubt that protons can be accelerated to high energies by a flare-resident acceleration process. The question is: Can enough escape to make a significant contribution to the high-energy SEP events observed in space?

The 3 June 1982 slare observed by SMM also had a delayed onset phase of pion-decay-dominated emission (Forrest et al. 1986; Chupp et al. 1987). In the early 1990s the *Compton Gamma-ray Observatory* (Gehrels et al. 1993) observed flares for which such post-impulsive phase emission could be sustained for hours (e.g., Kanbach et al. 1993; Akimov et al. 1996). There is growing evidence that such emission is thought to result from precipitating protons accelerated by a coronal shock (Frost and Dennis 1971; Ramaty et al. 1987; with both of these papers reaching back to Wild et al. 1963; Vestrand and Forrest 1993; Cliver et al. 1993; Pesce-Rollins et al. 2018; Ackermann et al. 2017; Klein et al. (2018; with reservations); Hudson 2018 (CME-driven shock acceleration on non-flaring closed loops); Winter et al. 2018; Omodei et al. 2018; Jin et al. 2018; Kahler et al. 2018; Gopalswamy et al. 2018, 2021)—with delayed proton acceleration/trapping on large-scale flare loops (Grechnev et al. 2018; de Nolfo et al. 2019) as the principal competing alternative.

Comparisons of impulsive and extended phase γ-ray emissions for flares observed by Fermi LAT have been used to address the question of which phase of proton acceleration (impulsive (flare) or delayed (CME-driven shock)) is the dominant source of the large high-energy SEP events at 1 AU. From an analysis of 30 *Fermi* LAT events observed from 2011 to 2015, Share et al. (2018) calculated that the number of > 500 MeV protons required to produce a late-phase pion decay signal was 10 times greater than the number accelerated during the impulsive phase. For a sample of eight associated SEP events, they found that the average number of > 500 protons required to produce the late phase gamma ray emission was only about 15% of the number of protons observed in space, although systematic uncertainties could increase this percentage by up to a factor of five. Subsequently, from a comparison of the numbers of > 500 MeV protons inferred from Fermi LAT observations ($N_{\gamma\text{-ray}}$) with those observed in space ($N_{\text{SEP}}$) by the *Payload for Matter–Antimatter Exploration and Light Nuclei Astrophysics* (PAMELA, Adriani et al. 2014) for 14 events, de Nolfo et al. (2019) found a poor correlation (r = 0.15), with the ratio $N_{\gamma\text{-ray}}/N_{\text{SEP}}$ varying from $\sim 2 \times 10^{-3}$ to $\sim 10^3$. More recently, Gopalswamy et al. (2021) argued, on the basis of corrected values for both $N_{\gamma\text{-ray}}$ (based on the visibility of limb events) and $N_{\text{SEP}}$ (based on the effect of flare latitude on SEP propagation) in the de Nolfo et al. sample, that the two parameters are highly correlated (r = 0.77). The "flare vs. shock" debate for the largest high-energy SEP events (e.g., Vashenyuk et al. 2011; McCracken et al. 2012; Grechnev et al. 2015; Cliver 2016; Bazilevskaya 2017; Klein and Dalla 2017; Kocharov et al. 2018; Struminsky 2018; Kouloumvakos et al. 2020; Kocharov et al. 2020, 2021; Hutchinson et al. 2022) is far from over and will continue to inform understanding of SEP acceleration at the Sun.

As shown in Fig. 44, a CME—the defining characteristic of an eruptive flare—can drive two kinds of shocks. The radially outward motion of the CME drives a bow shock while the lateral expansion of the CME works as a three-dimensional piston to drive a shock across the face of the Sun (Vršnak and Cliver 2008). If the outward motion dominates the lateral, a driven shock manifested by a slow-drift metric type II burst will arise near the "nose" of the CME. CMEs with significant impulsive over-expansion will create an EUV wave in the corona and, if strong

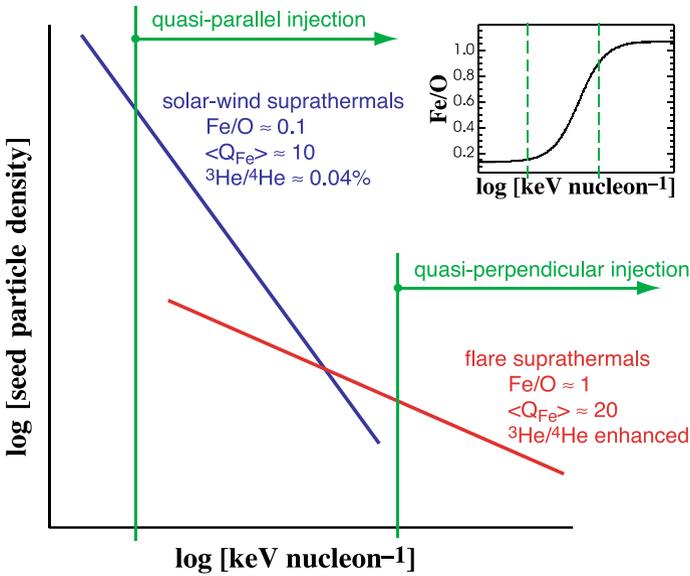

**Fig. 45** Schematic representation of the seed populations for shock accelerated SEPs. Because of their higher energies, flare suprathermals, with enhanced Fe/O ratios and Fe charge states, are more accessible to quasi-perpendicular shocks. The inset shows how the Fe/O ratio in the seed population varies with energy, with the vertical dashed lines corresponding to the solid green lines in the larger figure. Image reproduced with permission from Tylka et al. (2005), copyright by AAS

enough, a Moreton wave in the chromosphere (Vršnak, 2016). Because the direction of motion of the shock is along the magnetic field, a bow shock is referred to as a quasi-parallel shock, while the shock driven by the lateral expansion of the CME is termed a quasi-perpendicular shock. The angle between the upstream magnetic field and the shock normal vector is labelled $\theta_{Bn}$. Quasi-perpendicular acceleration, for shocks with large $\theta_{Bn}$, is a variant of the diffusive shock acceleration at parallel shocks (Jokipii 1982, 1987).

From in situ observations of low-energy interplanetary electron and proton events, Tsurutani and Lin (1985) found that essentially every quasi-perpendicular ($\theta_{Bn} \gtrsim 70°$) shock produced a shock spike in proton and electron fluxes, while all quasi-parallel ($\theta_{Bn} \lesssim 50°$) shocks in their sample with shock speeds $\gtrsim 250$ km s$^{-1}$ produced a proton energetic storm particle event (a slow rise beginning several hours before the shock). They noted that a significant ambient population of low energy electrons and ions (termed shock "seed" particles) were present in the interplanetary medium prior to every shock considered.

Tylka et al. (2005) stressed the importance of shock geometry and seed particles close to the Sun for understanding the spectral and compositional variability of solar energetic particle events observed at 1 AU. Figure 45, taken from their paper, shows the interplay of these two parameters. Tylka et al. attributed hard-spectra SEP events for which the Fe/O ratio increased with ion energy to quasi-perpendicular shocks, which have a high energy injection requirement (Jokipii 1987), favoring suprathermals characterized by high Fe/O ratios and Fe charge states associated with small

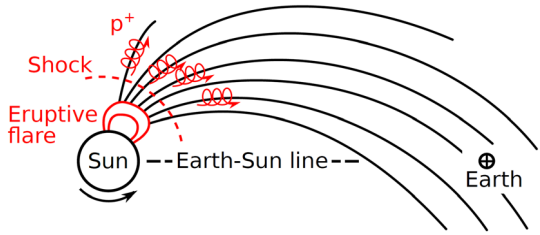

**Fig. 46** Schematic showing CME-driven bow-shock (dashed-line) acceleration of protons in the corona and interplanetary space on open Parker spiral field lines connecting to Earth. Image reproduced with permission from Kusano et al. (2020), copyright by IOP

impulsive solar flares (Reames 1999, 2013). Quasi-parallel shocks can operate on relatively low energy seed particles such as solar wind suprathermals and produce softer spectra proton events characterized by low Fe/O ratios and Fe charge states. Most large events will have contributions from both types of shocks, leading to the broad range of variability in composition/charge-states/spectra/time profiles observed in large SEP events. For the better-defined cases of the two types of large gradual SEP events, Tylka and Lee (2006) were able to use their shock formulation to reproduce the organization of SEP elemental abundances by charge/mass ratio discovered by Breneman and Stone (1985) and provide the first theoretical explanation for this effect. Recently, Cliver et al. (2020a) presented evidence that shock geometry could account for the variation of shock spectra with parent flare solar longitude in high-energy proton events. Neither of these two results are readily explained in terms of a dominant flare-resident acceleration process for large gradual SEP events.

Not every eruptive flare—with a CME fast enough to drive a coronal/ interplanetary shock—leads to a large SEP event, but only those connected to Earth by the interplanetary magnetic field. Nominally, because of the Parker spiral in the heliospheric field, the most advantageous location for an eruption to produce a SEP event at Earth is $\sim$ W55, i.e., 55° from solar central meridian, the angular distance an active region rotates toward the west limb in the $\sim$ 4 days it takes for the solar wind, with average speed $\sim$ 400 km s$^{-1}$, to propagate to Earth (Fig. 46). That said, SEP events, even those at high-energy, can be seen over a wide range ($\sim$ 240°; E90-W150) of solar longitudes because of the broad extent of CME-driven shocks (Cliver et al. 1995, 2005; Lario et al. 2014; Gómez-Herrero et al. 2015; cf. Dresing et al. 2012). Such shocks are initially driven by the lateral expansion of the CME; when the over-expansion stops, the shocks become freely-propagating waves (Uchida 1968; Warmuth 2015).

Even for nominally well-connected eruptive flares, the relation between SEP event strength and parent flare size (e.g., Grechnev et al. 2015) or CME speed (Kahler 2001) is rather loose, in part because large soft-spectrum proton events can arise in rather weak flare events, viz., filament eruptions outside of active regions (e.g., Gopalswamy et al. 2015; Cliver et al. 2019). Such events tend to be associated with weaker flares and more-slowly accelerating CMEs from the $\sim$ W35-75 zone of good connection for which the quasi-parallel shock acceleration should be dominant. Hard-spectrum SEP events favor west-limb CMEs for which lateral expansion drives a quasi-perpendicular shock toward the field line connected to Earth while the outward CME motion is transverse to the Sun-Earth line, reducing the time which the quasi-parallel bow shock is connected to the field line to Earth (Cliver et al. 2020a). Other factors that can affect correlations between SEP event size and eruptive flare/CME parameters include: episodes of eruptive flares with fast CME-driven shocks that increase the pre-event SEP background of seed particles for later eruptions in a series (Kahler 2001; Cliver 2006b), coronal/interplanetary magnetic field topology (Richardson et al. 1991; Kong et al. 2017), and converging CME-driven shocks (Pomerantz and Duggal 1974; Kallenrode and Cliver 2001a, b; Gopalswamy et al. 2004; Lario and Karelitz 2014).

The principal measures of SEP event size are the peak intensity or flux ($f$) and the integral fluence ($F$; event-integrated flux). The fluence of protons with energy above 30 MeV, designated $F(>30$ MeV) or simply $F_{30}$, is a standard measure (e.g., Shea and Smart 1990), primarily because of its applicability to radiation dose calculations. In Table 8 we present lists of the ten strongest SEP events during modern times in terms of: (a) $F_{30}$; (b) $F_{200}$ (a representative fluence value for large events that can be robustly estimated from the cosmogenic-isotope proxy data nearly independently of the exact intensity spectrum; Kovaltsov et al. 2014); and (c) a ground level enhancement (GLE) quasi-fluence, $F_{GLE}$, viz., the neutron monitor (NM; Simpson et al. 1953) count rate increase (% above the galactic cosmic ray background) integrated over the entire duration of a GLE, averaged for 30 polar sea-level neutron monitors, given in units of %*h (Asvestari et al. 2017a).[16] For the $F_{30}$ column in Table 8, the list consists of both compound events (including two episodes of large events within a 3-month interval) and individual SEP events. Episodes of strong eruptive flares from a single active region that involve converging shocks, as occurred, e.g., in August 1972 and October 1989, are particularly effective for producing large fluence events at low energies.

As was the case for geomagnetic storms, great spot groups are neither necessary nor sufficient for an out-sized SEP event. The strongest high-energy SEP event of the modern era, the GLE on 23 February 1956, originated in Greenwich sunspot group 17351 that had a maximum area of 1734 μsh during its disk passage while the five great active regions of cycle 18 (1944–1954), each with sunspot areas $\gtrsim$ 4500 μsh, gave rise to a single GLE—a notable event associated with the large spot group in July 1946, but one that Duggal (1979) estimated was only about 10% as intense as the 1956 GLE.

### 7.2 Occurrence frequency distribution for directly observed > 30 MeV SEP events

Using GOES proton data from 1987 to 2016, Gopalswamy (2018) determined 100- and 1000-year events for the fluence of > 30 MeV events. Note that these

---

[16] A ground level enhancement (GLE) Is formally defined as follows (Poluianov et al. 2017): "A GLE event is registered when there are near-time coincident and statistically significant enhancements of the count rates of at least two differently located neutron monitors, including at least one neutron monitor near sea level and a corresponding enhancement in the proton flux measured by a space-borne instrument. Relatively weak SEP events registered only by high-altitude polar neutron monitors, but with no response from cosmic-ray stations at sea level, can be classified as sub-GLEs."

**Table 8** Largest SEP events during the space age, rank ordered for $F_{30}$, $F_{200}$, and $F_{GLE}$

| Rank | Date | $F_{30}$ ($10^9$ cm$^{-2}$) (a) | Rank | Date | $F_{200}$ ($10^7$ cm$^{-2}$) (b) | Rank | Date | $F_{GLE}$ (%*h) (c) |
|---|---|---|---|---|---|---|---|---|
| (1) | 2, 4, & 7 Aug 72 | 8.4 | (1) | 23 Feb 56 | 14.0 | (1) | 23 Feb 56 | 5202 ± 104 |
| (2) | 12, 15, & 20 Nov 60 | 6–9.8 | (2) | 12 Nov 60 | 6.4 | (2) | 29 Sep 89 | 1189 ± 60 |
| (3) | Aug–Oct 89(d) | 7.2 | (3) | 19 Oct 89 | 5.5 | (3) | 12 Nov 60 | 677 ± 25 |
| (4) | Sep–Nov 01(e) | 5.5 | (4) | 14 Jul 00 | 3.4 | (4) | 24 Oct 89 | 576 ± 27 |
| (5) | 14 Jul 00 | 4.3 | (5) | 29 Sep 89 | 3.1 | (5) | 15 Nov 60 | 552 ± 106 |
| (6) | 10, 14, & 16 Jul 59 | 4.0 | (6) | 15 Nov 60 | 3.0 | (6) | 19 Oct 89 | 411 ± 15 |
| (7) | Oct–Nov 03(f) | 3.8 | (7) | 24 Oct 89 | 2.2 | (7) | 20 Jan 05 | 385 ± 55 |
| (8) | 08 Nov 00 | 3.1 | (8) | 20 Jan 05 | 2.2 | (8) | 15 Apr 01 | 170 ± 15 |
| (9) | 23 July 12(g) | 2.1 | (9) | 17 Jul 59 | 1.6 | (9) | 28 Oct 03 | 110 ± 7 |
| (10) | 23 Feb 56 | 1.4 | (10) | 22 Oct 89 | 1.6 | (10) | 28 Jan 67 | 110 ± 3 |

References: (a) ($F_{30}$) (1) Jiggens et al. (2014); (2) Webber et al. (2007; low), Shea and Smart (1990; high); (3) Smart et al. (2006b); cf. Jiggens et al. (2014) for October 1989; (4) Smart et al. (2006b) and NGDC (https://www.ngdc.noaa.gov/stp/satellite/goes/dataaccess.html); (5) Smart et al. (2006b); (6) Usoskin et al. (2020b); (7,8) NGDC; (9) Gopalswamy et al. (2016); (10) Usoskin et al. (2020b)
(b) ($F_{200}$) Usoskin et al. (2020b) for 1956; Kovaltsov et al. (2014)
(c) ($F_{GLE}$) Asvestari et al. (2017a); integral of the excess above the galactic cosmic ray background over the entire duration of the event
(d) SEP events on 12 and 16 Aug, 29 Sep, and 19, 22, and 24 Oct 89
(e) events on 24 Sep and 4 and 22 Nov 01
(f) events on 28 and 29 Oct and 2 Nov 03
(g) Backside solar event observed by STEREO-A (Gopalswamy et al. 2016)

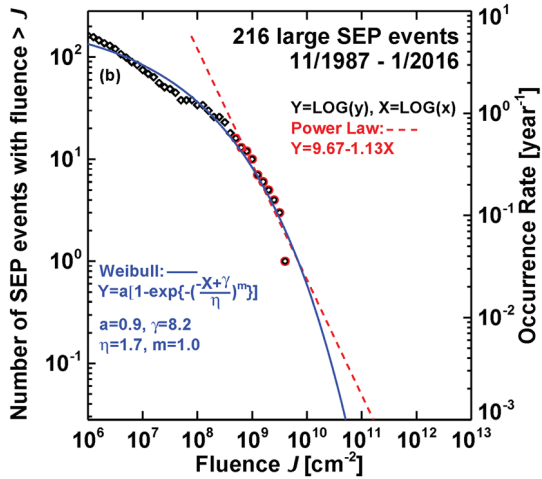

**Fig. 47** Downward cumulative distribution (left hand axis) of the number of solar proton events from November 1987 to January 2016 with > 30 MeV fluences greater than a given value J (black diamond and red circle data points). This annualized distribution (right hand axis) is fitted with a modified exponential function (solid blue line) and power-law (dashed red line; for the tail of the distribution (red circle data points)) to give the corresponding annual occurrence frequency distribution (OFD). The fit equations and parameters are given in the figure. Image adapted from Gopalswamy (2018)

distributions (based on https://umbra.nascom.nasa.gov/SEP/) are for individual SEP events and do not include composite events as in Table 8. The downward cumulative OFD for the > 30 MeV fluence is shown in Fig. 47. The 100-year (1000-year) event in this parameter is 1.6 (5.1) $\times 10^{10}$ cm$^{-2}$, for a modified exponential function fit. The exponential fit in Fig. 47 does not pass through the last data point. Ellison and Ramaty (1985) and Band et al. (1993) functions forced through this data point yield 100- and 1000-year estimates factors of 3–5 lower (Gopalswamy 2018). Both the largest single (14 July 2000; $4.3 \times 10^9$ cm$^{-2}$) and compound (4–7 August 1972; $8.4 \times 10^9$ cm$^{-2}$) SEP events in Table 8 fall short of the 100-year event. The power-law fit in Fig. 47 yields estimates of 100-year and 1000-year $F_{30}$ events of 2.1 and $16 \times 10^{10}$ cm$^{-2}$, respectively. From a correlation between > 30 MeV SEP fluence and flare SXR fluence, Cliver and Dietrich (2013) obtained a worst-case (assuming a composite SEP event plus shock) fluence for the Carrington event of $\sim 10^{10}$ cm$^{-2}$ (with the $\pm 1\sigma$ uncertainty spanning a range from $\sim 10^9$–$10^{11}$ cm$^{-2}$).

### 7.3 SEP event proxies

#### 7.3.1 Nitrate concentration in polar ice (a discredited proxy)

Dreschhoff and Zeller (1990, 1998) proposed that nitrate concentrations in polar ice cores could be a proxy for strong SEP events (with $F_{30} > 10^9$ cm$^{-2}$). Atmospheric ionization caused by solar energetic particles can affect atmospheric chemistry and particularly lead to the formation of nitrates. These nitrates constitute only a minor fraction of the nitrate produced in Earth's atmosphere. Other major, non-cosmogenic sources are thunderstorm activity in the tropical regions, biomass burning, and anthropogenic emissions (Legrand and Mayewski 1997). Therefore, a clean signal can be expected only in the polar region, especially in Antarctica, which is relatively isolated from the world by the polar vortex during winter seasons. From comparison of transient nitrate concentrations in ice cores (one from Greenland (1561–1991) and two shorter cores from Antarctica (1905–1991)) with > 30 MeV solar proton fluences for known SEP events from 1942 to 1989, McCracken et al. (2001) obtained an empirical conversion factor between these two parameters. This conversion factor was used to identify a list of 70 strong > 30 MeV SEP events in the Greenland ice core from 1561 to 1950. The strongest nitrate peak during this interval appeared to be associated with the Carrington event in 1859 (Sect. 1.1) with an inferred $F_{30}$ value of $\sim 2 \times 10^{10}$ cm$^{-2}$.

Subsequent independent studies did not confirm the relation between the nitrate spikes and SEP events. Wolff et al. (2012) analyzed nitrate records from 14 well-dated ice cores from Greenland (6) and Antarctica (8) that did not include the GISP2 H core analyzed by McCracken et al. and concluded that "the nitrate event identified as 1859 in the GISP2 H core … is most likely the same event that more recent Greenland cores identify at 1863. The parallel event in other [Greenland] cores, as well as all other significant nitrate spikes in those cores, has an unequivocal fingerprint of a biomass burning plume [viz., co-located spikes in ammonium, formate, black carbon and vanillic acid]." The conclusion of Wolff et al. (2012) was supported by Duderstadt et al. (2016) and Wolff et al. (2016), but disputed by Smart et al. (2014, 2016), and Melott et al. (2016). The ultimate test for the nitrate method was performed when the strong SEP events of 774 AD and 993 AD were discovered (see next section). If high nitrate composition was indeed a SEP event proxy with the empirical conversion factor proposed by McCracken et al. (2001), very strong spikes would have been observed for these events. Sukhodolov et al. (2017) analyzed four ice cores with quasi-annual resolution, two each from Greenland and Antarctica, for years around 774 AD and found no noticeable nitrate spike in any of the series. In addition, Mekhaldi et al. (2017) analyzed several high-resolution ice cores from both Greenland and Antarctica for the periods around 774 AD, 993 AD, and 1956 AD, corresponding to the strongest cosmogenic and modern SEP events, and found no evidence for notable nitrate spikes during any of these periods. Thus, the nitrate proxy for strong SEP events must be dismissed.

### 7.3.2 Cosmogenic radionuclides in tree rings and ice cores

The only means yet known to find clear signatures of major SEP events in the past is based on cosmogenic radionuclides—specifically, radiocarbon ($^{14}$C; half-life = $5.73 \times 10^3$ year), beryllium-10 ($^{10}$Be; $1.36 \times 10^6$ year), and chlorine-36 ($^{36}$Cl; $3.01 \times 10^5$ year)—produced by solar energetic particles in the Earth's atmosphere. Of these radionuclides, $^{14}$C and $^{10}$Be are the isotopes primarily used to reconstruct solar variability on a multi-millennial time scale via variability of galactic cosmic rays (GCRs; Beer et al. 2012; Usoskin 2017), either from individual records or in a composite approach (Steinhilber et al. 2012; Wu et al. 2018).

Radiocarbon is produced in the atmosphere mostly by the reaction $^{14}$N (n,p)$^{14}$C, often called neutron capture. Atmospheric neutrons are generated in nucleonic cascades induced by primary cosmic ray, or high-energy solar, particles. The above reaction is the main sink of thermalized neutrons in the atmosphere; almost all thermal neutrons lead to production of radiocarbon. Upon production, radiocarbon is oxidized to form carbon dioxide $^{14}$CO$_2$ which takes part in the global carbon cycle, including mixing in the atmosphere, ocean circulation, and exchange between different carbon reservoirs (e.g., Roth and Joos 2013). Radiocarbon is typically measured as $\Delta^{14}$C, i.e., the normalized and corrected (for isotope fractionation) ratio of $^{14}$C to $^{12}$C, in living or dead trees, which allows for absolute dating via dendrochronology. The data are available with 1-, 5- or 10-year time resolutions as a multi-sample composite IntCal dataset (Reimer et al. 2009, 2013, 2020) covering the Holocene (last $\sim$ 12,000 years) and extending, with lower resolution—based also on lake/marine sediments and speleothems (stalactite and stalagmites)—to the last fifty millennia. Modern state-of-the-art models (Masarik and Beer 2009; Kovaltsov et al. 2012; Poluianov et al. 2016) can compute the radiocarbon production by cosmic rays and reconstruct solar variability in the past (e.g., Usoskin et al. 2016). However, radiocarbon data cannot be used for reconstructions of solar variability after 1955–1956 because of the bomb-effect (production of $^{14}$C in the atmosphere by nuclear bomb tests; e.g., Hua and Barbetti 2014) and is problematic/difficult to use for this purpose since the late nineteenth century because of the Suess effect (increasing use of fossil fuels dilutes $^{14}$C in the atmosphere; Suess 1955). Finally, $^{14}$C cannot be used as a proxy for solar activity for years earlier than about 12 millennia ago, i.e., before the Holocene period of fairly stable climate, because the carbon cycle is not well known during the ice age and subsequent deglaciation.

Cosmogenic $^{10}$Be is produced as a result of spallation of oxygen and nitrogen nuclei by energetic particles. Production of beryllium-10 by cosmic rays has been modelled by several groups (e.g., Masarik and Beer 1999; Webber et al. 2007; Kovaltsov and Usoskin 2010; Poluianov et al. 2016). After production, beryllium is thought to attach to aerosols that are subject to both wet and dry gravitational sedimentation. The $^{10}$Be isotope is usually measured in ice cores from Greenland or Antarctic ice sheets that can go back a million years or more. Dating of ice cores is not absolute and may be uncertain by up to a decade or more during the early Holocene (Adolphi and Muscheler 2016). Because beryllium is not globally mixed in the atmosphere, a detailed transport model of the 3D atmosphere is needed (e.g., Heikkilä et al. 2013), but even then an uncertainty of the order of 20–30% remains because of regional atmospheric dynamics, geomorphology of the site, and details of the local deposition (Sukhodolov et al. 2017).

A third cosmogenic isotope, $^{36}$Cl, produced as a result of spallation of atmospheric $^{40}$Ar by energetic particles, is involved in the chlorine cycle, and can be measured in polar ice-cores along with $^{10}$Be. Because of its smaller production rate (roughly one-tenth that of beryllium), more difficult measurement and complicated atmospheric transport, $^{36}$Cl is not typically used for reconstructions of long-term solar activity. However, because of its greater sensitivity to lower energy ($\sim 30$ MeV) primary particles than either $^{14}$C or $^{10}$Be, $^{36}$Cl provides a useful diagnostic for the spectra of historical SEP events (Mekhaldi et al. 2021). The ratio of $^{36}$Cl to $^{10}$Be measured in the same ice core for the same time period can give a rough estimate of the hardness of the spectrum of primary energetic particles in the energy range of 30–200 MeV. For SEP events, this ratio varies from $\sim 1.2$ for very hard events to $\sim 6$ for soft-spectrum events (Webber et al. 2007; Mekhaldi et al. 2015; Mekhaldi and Muscheler 2020).

Although the background level of cosmogenic isotopes is defined by GCRs, extreme SEP events may also leave their signatures in the isotope records. This idea has long been discussed and implemented (e.g., Lingenfelter and Ramaty 1970; Castagnoli and Lal 1980; Masarik and Reedy 1995) but the earlier estimates of the SEP signatures in cosmogenic data were uncertain and differed by up to two orders of magnitude. Modern systematic modelling was first performed by Usoskin et al. (2006b) and Kovaltsov et al. (2012) who demonstrated that major SEP events can be identified in the cosmogenic radionuclide records. A search of all available datasets by Usoskin and Kovaltsov (2012) identified a list of about 20 candidates for large SEP events during the Holocene. They suggested that major events took place ca. 780 AD (Miyake et al. 2012) and 1460 AD. The former is now generally accepted as a solar event (cf. Frolov et al. 2018), while the latter was likely caused by a volcanic eruption in 1458 (Sigl et al. 2014, 2015), with aerosol loading leading to rapid removal of beryllium in the polar region, emptying the stratospheric reservoir (Baroni et al. 2011, 2019). Such volcano-induced beryllium spikes can be identified in ice cores through enhanced sulphate concentration. Volcanic eruptions are not expected to affect global $^{14}$C, and indeed no significant enhancement of $^{14}$C was observed ca. 1460 AD. As discussed below, other cosmogenic-based events (or candidate events) have been identified following the Miyake et al. (2012) discovery event.

### 7.4 SEP events deduced from cosmogenic radionuclide records

#### 7.4.1 The greatest known SEP event: 774 AD

The largest SEP event identified in the cosmogenic radionuclide record thus far was discovered by Miyake et al. (2012) in high-resolution data from two Japanese cedar trees, for which a fast increase of $\Delta^{14}$C was found from 774 to 775 AD followed by a gradual decline (Figs. 3, 48). The fast-rise/slow-decay time profile is similar to that for large SEP events as shown at $> 30$ MeV energies for proton events beginning on 4 and 7 August 1972 in Fig. 50 below, but is $\sim 1000$ times longer because of the carbon cycle that moves $^{14}$C from creation in the atmosphere to ingestion by a tree. The reality of the event was confirmed shortly thereafter by $^{14}$C data from many different trees around the world, including the southern hemisphere, as well as measurements of $^{10}$Be and $^{36}$Cl in both Greenland and Antarctic ice cores (Usoskin et al. 2013; Jull et al. 2014; Güttler et al. 2015; Miyake et al. 2015; Rakowski et al. 2015; Mekhaldi et al. 2015; Sukhodolov et al. 2017; Uusitalo et al. 2018; Büntgen et al. 2018). Current estimates of the globally-averaged net atmospheric production of $^{14}$C for the 774–775 AD event (see Table 9) lie in the range of $(1.3–2.2) \times 10^8$ cm$^{-2}$ (Usoskin et al. 2013; Pavlov et al. 2013a,b; Güttler et al. 2015; Mekhaldi et al. 2015), more than three times the annual production by GCRs for moderate solar activity (modulation potential of 468 MV; Wu et al. 2018). This range is due to different carbon cycle models used to convert the measured $\Delta^{14}C$ into a production rate. The most advanced model (Büntgen et al. 2018) yields production of $1.9 \pm 0.1 \times 10^8$ cm$^{-2}$.

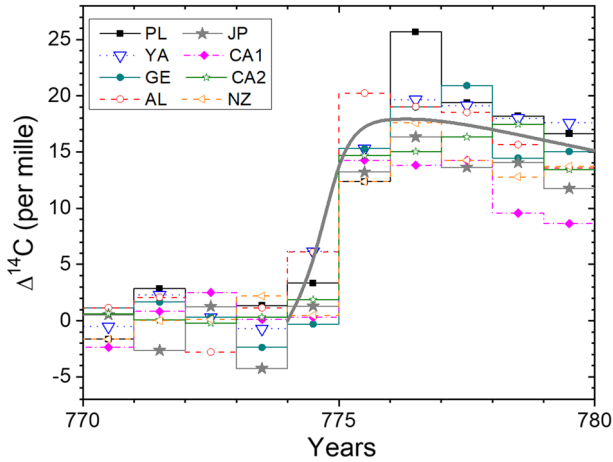

**Fig. 48** Baseline adjusted radiocarbon response $\Delta^{14}C$ in different trees around the year 774–775 AD as denoted in the legend: PL—Poland (Rakowski et al. 2015), YA—Yamal peninsula, Russia (Jull et al. 2014), GE—Germany (Usoskin et al. 2013), AL—Altai, Russia (Büntgen et al. 2018), JP – Japan (Miyake et al. 2012), CAL1 and CAL2—two sites in California, USA (Jull et al. 2014; Park et al. 2017), NZ—New Zealand (Güttler et al. 2015). The gray curve represents the model response to an instant production of $^{14}C$ by a SEP event with a hard spectrum (Usoskin et al. 2013) taking place in mid-774 AD. Adapted from Uusitalo et al. (2018)

**Table 9** Different estimates of the global radiocarbon production for the 774 AD event ($^{14}Q_{774}$), as published in different sources, using different carbon-cycle models

| $^{14}Q_{774}$ ($\times 10^8$ cm$^{-2}$) | Carbon-cycle model | Refs. |
| --- | --- | --- |
| (1.1–1.5) | 6-Box | Usoskin et al. (2013) |
| 1.7 | 6-Box | Pavlov et al. (2013a, b) |
| 2.2 | 11-Box | Güttler et al. (2015) |
| 2.16 | Box-diffusion | Mekhaldi et al. (2015) |
| 2.18 | 11-Box | Uusitalo et al. (2018) |
| 1.88 ± 0.1 | 22-Box | Büntgen et al. (2018) |

Several hypotheses were proposed to explain the enhancement from 774–775 AD. Miyake et al. (2012) considered both gamma-ray emission from a nearby supernova and a hard-spectrum SEP event from a superflare and noted difficulties for both: (1) "a supernova that occurred relatively recently and relatively near Earth should still be tremendously bright (in radio, X-rays and [gamma-ray line emission from decay of] $^{44}Ti$), and such an object is not observed", and (2) "it is believed that a super flare has never occurred on our Sun, due to the absence of an historical record (such as a record of aurora and mass extinction caused by the expected destruction of the ozone layer) and theoretical expectations" (based on the view at that time that hot Jupiters were required for large flares on solar-type stars, e.g., Schaefer et al. 2000, Ip et al. 2004; Lanza 2008). Also in 2012, Eichler and Mordecai (2012) suggested that a large, long-period comet (comparable to Hale–Bopp) impacting the Sun could cause a solar flare sufficiently large to account for the $^{14}$C increase. This hypothesis has received little attention, likely because of its speculative nature. In 2013, Pavlov et al. (2013a, b) and Hambaryan and Neuhäuser (2013) proposed that gamma-ray bursts (GRBs), representing the most energetic impulsive energy releases observed in the universe (up to $10^{54}$ erg, assuming isotropy; Gehrels and Mészáros 2012; Berger 2014) as sources of the 774–775 AD event. Both Pavlov et al. (2013a) and Hambaryan and Neuhäuser argued that the GRBs would need to: (1) originate nearby in the Milky Way (based on the energy requirements) and (2) be of the short-duration type ($< 2$ s) associated with the merger of two compact objects (two neutron stars or a neutron star and a black hole) to form a black hole (because long-duration ($> 2$ s) GRBs, due to explosions of nearby hyper-massive stars would imply an observable remnant when none was observed—similar to the objection raised by Miyake et al. 2012, to a regular supernova). Hambaryan and Neuhäuser (2013) concluded that the 774–775 AD event provided the first evidence for a short GRB in our galaxy while Pavlov et al. (2013b) pointed to galactic SN remnant W49B as a possible association for a long ($> 2$ s) GRB source.

The SEP event origin hypothesis gained traction in 2013, as the initial energy requirement inferred by Miyake et al. (2012) for the 774–775 AD $^{14}$C event was re-evaluated. Miyake et al. (2012) estimated the energy in solar energetic protons needed to produce the $^{14}$C event in 774–775 AD to be $\sim 10^{35}$ erg. Melott and Thomas (2012) argued for a reduced SEP event energy (by two orders of magnitude from $\sim 2 \times 10^{35}$ erg to $\sim 2 \times 10^{33}$ erg) but they did so by linking the 774 AD event to a CME with an opening angle of only 24°. A more realistic CME with angular width of 90° (Gopalswamy et al. 2001, 2012; Moschou et al. 2019) for the large 774 AD event would only reduce the emitted proton energy by a factor of $\sim 7$ to $3 \times 10^{34}$ erg. Moreover, from Fig. 12 above (Emslie et al. 2012), this would imply a flare bolometric energy of $\sim 3 \times 10^{35}$ erg. Subsequently, Usoskin et al. (2013), in a paper subtitled "The Sun is to Blame", modelled the 774–775 $^{14}$C event using a family of three carbon cycle models. Correction of a geometrical error (confusing the Earth's cross section and surface area) reduced the Miyake et al. estimate of the net $^{14}$C production by a factor of $\sim 4$–5, making it compatible with a solar source (a proton spectrum of the 1956 GLE scaled up by a factor of 45), in accord with a 25–50 multiple suggested by the analysis of Usoskin and Kovaltsov (2012). Usoskin et al. (2013) regarded such a source (which could consist of a single solar event or an episode of eruptive flares) as being "strong, but not inexplicably strong". It would be far less energetic (see Sect. 7.8) than the initially suggested $\geq 10^{35}$ erg causative solar event (Miyake et al. 2013; Shibata et al. 2013).

Cliver et al. (2014) and Neuhäuser and Hambaryan (2014) argued against the SEP hypothesis. Cliver et al. noted that the 1956 GLE spectrum on which the 45-fold multiple for the 774 AD event was based, would imply a $F_{30}$ value of $\sim 8 \times 10^{10}$ cm$^{-2}$, $\sim 10$ times larger than that of the strongest 3-month interval of SEP activity in the modern era (Table 8). Such an $F_{30}$ value fell outside the corresponding cumulative frequency distribution of Kovaltsov and Usoskin (2014). In addition, they noted that reconstructed solar activity ca. 774–775 AD was relatively weak in comparison with the well-observed interval of strong activity from $\sim$ 1945–1995 (Bazilevskaya et al. 2014). Among other arguments, Neuhäuser and Hambaryan (2014) reported that the 1956 event was not detected in 1-year resolution records of $^{10}$Be concentration (cf. McCracken and Beer 2015). As shown by Usoskin et al. (2020b), the 1956 event was a factor of 15 lower than a single-proxy observational threshold and would not be detectable in a 1-year record. In regard to the high multiple of 45 suggested by Usoskin et al. (2013) for the spectrum of the 1956 GLE to account for the 774 SEP event, this factor was later found to be an underestimate (see Sect. 7.4.2).

Additional hypotheses for the 774–775 AD event were put forward by Liu et al. (2014a) and Neuhäuser and Neuhäuser (2015). Liu et al. (2014a) suggested that the 774–775 AD event was due to a comet impact at Earth. This conjecture was dismissed by Usoskin and Kovaltsov (2015; see also Overholt and Melott 2013, and Neuhäuser and Hambaryan 2014) because a comet of a realistic mass/size to account for the observed $^{14}$C increase would have produced a disastrous, unobserved, geological/biological impact on Earth. As an alternative explanation for 774–775 AD event, Neuhäuser and Neuhäuser (2015) proposed abrupt GCR demodulation. The appeal of such a scenario is that neither a rare cosmic energy release nor a very large (and/or unusual) solar flare is required—although the Sun is still to blame. The optimum scenario for such demodulation would be a deep rapid-onset solar minimum following a very strong 11-year cycle, with the strong cycle fixing the GCR baseline and the deep minimum setting the ceiling. It is doubtful if such a modulation could account for either the fast rise (Fig. 48) or the intensity (Kovaltsov et al. 2012; Poluianov et al. 2019) of the 774–775 AD event. Even a complete switch-off of solar modulation (viz., applying the local interstellar spectrum of GCR to Earth) would lead to only a $\sim 60\%$ increase in annual $^{14}$C production versus the inferred $\sim 400\%$ enhancement for 774–775 AD (Mekhaldi et al. 2015).

The SEP event scenario for the 774–775 AD event is preferred over the main competing galactic GRB hypothesis for three reasons: (1) the latitudinal gradient and inferred global symmetry of the cosmogenic signal (Sukhodolov et al. 2017; Uusitalo et al. 2018; Büntgen et al. 2018) (if a GRB source were located near to Earth's equatorial plane, the expected latitudinal gradient would be opposite to the observed polar enhancement of the cosmogenic signal; Uusitalo et al. 2018),[17] (2) the significant response in $^{10}$Be (e.g., Mekhaldi et al. 2015; Sukhodolov et al. 2017) which is not expected for a GRB (Pavlov et al. 2013a), and (3) the identification of four additional cosmogenic nuclide events in 993–994 AD (Miyake et al. 2013, 2014; Fogtmann-Schulz et al. 2017; O'Hare et al. 2019), $\sim 660$ BC (Park et al. 2017; O'Hare et al. 2019), 7176 BC (Paleari et al. 2022; Brehm et al. 2022), and 5259 BC (Brehm et al. 2022). In addition to these five confirmed SEP events, three other candidates are known presently that need to be independently confirmed; 1052 AD and 1279 AD (Brehm et al. 2021) as well as 5410 BC (Miyake et al. 2021). We note that another candidate event of 3372 BC (Wang et al. 2017) has been recently dismissed by an independent analysis (Jull et al. 2021). These events and candidates imply a higher occurrence rate than expected for a galactic GRB (e.g., Hambaryan and Neuhäuser 2013; Thomas et al. 2013). The above lines of evidence have established the SEP paradigm for the 774 AD SEP event (Miyake et al. 2020a).

---

[17] An off-equator GRB source would result in a non-symmetric $^{14}$C signal because of limited cross-talk between the hemispheres.

Various analyses imply that the 774 AD proton event occurred around the boreal summer of 774 AD (Sukhodolov et al. 2017 (early autumn); Büntgen et al. 2018 (summer); Uusitalo et al. 2018 (late spring)) as a single SEP event, or as a short (shorter than 2–3 months) sequence of events (Güttler et al. 2015)—similar to episodes of high-energy SEP activity in 1960 and 1989 (Table 8). The 774–775 AD $^{14}$C event was so strong and clear that it is used as a marker (tie point) for archaeological dating (e.g., Wacker et al. 2014; Dee and Pope 2016; Büntgen et al. 2017; Oppenheimer et al. 2017).

### 7.4.2 Energy spectra of cosmogenic-based SEP events

The energy spectrum of the 774 AD event must have been very hard to produce such a strong signal in cosmogenic isotopes (Usoskin and Kovaltsov et al. 2012; Thomas et al. 2013). Thus, as noted above, Usoskin and Kovaltsov (2012) and Usoskin et al. (2013) modelled the 774–775 AD event in the $^{10}$Be and $^{14}$C record in terms of a 25 to 50-fold multiple of the 1956 GLE, the strongest high-energy event recorded by neutron monitors. Subsequently, Mekhaldi et al. (2015) used three different isotopes ($^{14}$C, $^{10}$Be, and $^{36}$Cl) with different response functions, to recreate the proton spectrum of the 774 AD event (Fig. 49) which has a spectral index ($S_I = \log (F_{30}/F_{200})$) of $\sim 0.8$. $S_I$ values obtained from the O'Hare et al. (2019) spectra for the 993 AD and $\sim 660$ BC SEP events are comparable. The spectra of these first three historical events are comparable to those of the hardest spectra GLEs: GLE No. 5 on 23 February 1956 ($S_I = 1.0$) and GLE No. 69 on 20 January ($\sim 1.2$) (Asvestari et al. 2017a, b; Cliver et al. 2020a, b; Usoskin et al. 2020a, b; International GLE Database (https://gle.oulu.fi)). The recently reported 7176 BC event has also been confirmed to be a hard spectrum event (Brehm et al. 2021, 2022; Paleari et al. 2022). These hard spectra imply that while the high-energy SEP fluence of the 774 AD event was extreme, the fluence in the 10–30 MeV range was not enhanced commensurately. The smaller increase at $F_{30}$ may be related to the so-called "streaming limit" (Reames and Ng 2010; Lario et al. 2008) in which proton-generated waves upstream of a CME-driven shock suppress the escape of the low-energy part of the SEP spectrum (Asvestari et al. 2017a) or they may result from a west limb origin (as was the case for the 1956 GLE) that favors a dominant near-Sun quasi-perpendicular shock geometry leading to a hard proton spectrum (Cliver et al. 2020a). In contrast, GLEs arising near disk center such as GLE No. 24 (4 August 1972) tend to have soft spectra (Smart et al. 2006a; Cliver et al. 2020a). The extreme softness of the August 1972 spectrum likely also reflects preceding CME activity (Kallenrode and Cliver 2001a,b). Figure 50 (adapted from Cliver and Dietrich 2013) shows the 30 MeV proton flux versus time profile for the sequence of SEP events in August 1972. The black vertical lines mark the time of sudden commencements at Earth, signalling the arrival of shock waves, and the arrow indicates the Hα flare maximum at 06:35 UT. It can be seen that the shock waves roughly bound the strong pulse of $> 30$ MeV protons. This component (red-hatching in Fig. 50) from the 4 August 1972 event accounts for $\sim 70\%$ of $F_{30}$ in this composite event. Pomerantz and Duggal (1974) suggested that the pulse is due to proton acceleration/trapping by converging shock waves. The GLE onset was approximately six hours after the Hα flare maximum.

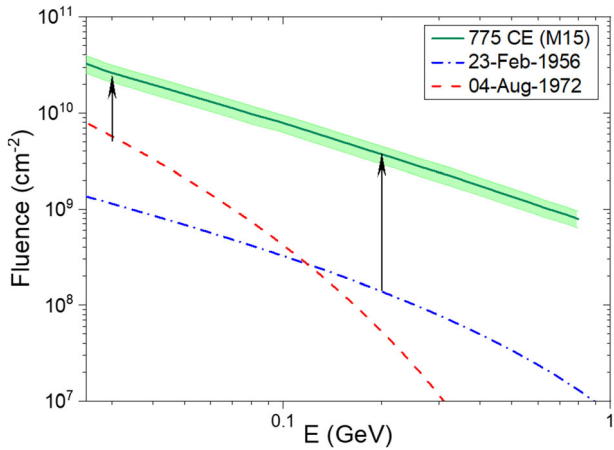

**Fig. 49** Event-integrated fluences of solar energetic protons for the event of 774 AD (Mekhaldi et al. 2015, green line with the shaded 68% confidence interval), as well as for the major GLE events with the known (Koldobskiy et al. 2021) hardest (23 February 1956; blue dashed curve) and softest (4 August 1972; red dotted curve) energy spectra. The 774 AD spectrum above $\sim 200$ MeV is based on an extrapolation. The two black arrows denote the enhancement of SEP fluence at 30 MeV and 200 MeV relative to the soft spectrum event of 4 August 1972 and the hard spectrum event of 23 February 1956, respectively. Image modified after Mekhaldi et al. (2015)

Because of its inferred fast transit CME (Table 5) and origin in an active region near solar central meridian, the 1859 event is a good candidate to produce a large soft-spectrum GLE (Smart et al. 2006a; Cliver and Dietrich 2013). To date, analyses of ice cores for $^{36}$Cl, formed in Earth's atmosphere by impinging $> 30$ MeV solar protons, have not provided evidence for a significant low-energy proton event associated with the Carrington flare (F. Mekhaldi, personal communication, 2022).

An overview of different estimates of the annual global $^{14}$C production ($Q_{774}$) by the 774 AD event (the number of $^{14}$C atoms produced on average per cm$^2$ of the Earth's surface) is given in Table 9. It ranges from 1.1 to 2.2 ($\times 10^8$ cm$^{-2}$) with more accurate recent models yielding a production rate of $\sim 2 \times 10^8$ cm$^{-2}$. The modelled global $^{14}$C production for GLE No. 5 ($Q_{1956}$) ranges between $2.5 \times 10^6$ atoms cm$^{-2}$ (Usoskin et al. 2006b; Pavlov et al. 2014) and $3.04 \times 10^6$ atoms cm$^{-2}$

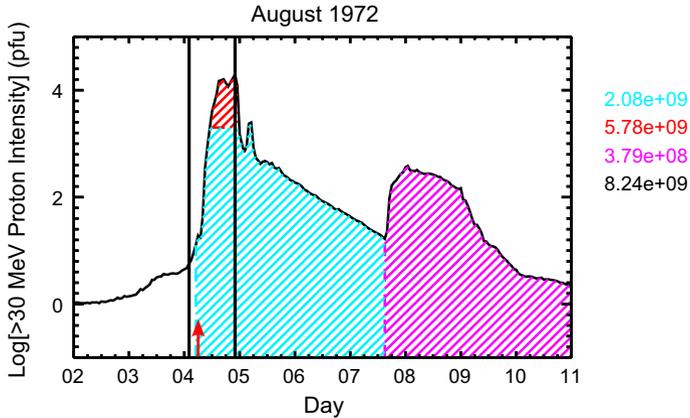

**Fig. 50** >30 MeV SEP flux time profiles for 2–11 August 1972. The high fluxes associated with the 4 August 1972 event were bounded by strong geomagnetic storm sudden commencements (vertical black lines) at 01:19 UT (and 02:20 UT) and 20:53 UT. The red arrow indicates the Hα flare maximum. Red-hatching is a guesstimate of the shock contribution to the SEP event; blue- and magenta-hatching indicate the fluences attributed to the 4 and 7 August eruptive flares, respectively. The numbers on the right give the fluences for the various components (and total fluence in black) in units of cm$^{-2}$. Image reproduced with permission from from Cliver and Dietrich ([2013](#)), copyright by the authors. (Original from Solar-Geophysical Data, No. 338, Part I, pp. 120–121, 1972)

(Usoskin et al. [2020b](#)). Thus the ratio $Q_{774}/Q_{1956}$ ranges from 36 (110/3.04) to 88 (220/2.5). By accounting for the different geomagnetic field strengths (dipole moment of $\sim 8 \times 10^{22}$ and $\sim 10^{23}$ A m$^2$ for 1956 and $\sim$ 1774 AD, respectively; Usoskin et al. [2016](#)), the global $^{14}$C production of the 1956 event would have been $\sim$ 10–20% lower if it occurred in 774 AD, i.e., in a range from $\sim$ 2.1 to $\sim$ 2.6 corresponding to $Q_{774}/Q_{1956}$ ratios of 42 to 105 (or $\sim$ 40–100). The $Q_{774}/Q_{1956}$ ratio is used to scale the proton spectrum of GLE No. 5 to that of 774 AD, with corresponding ratios for the other cosmogenic-nuclide based SEP events. An assessment of the ratio of the strength of the known historical events in comparison to the strongest directly observed GLE No. 5 is shown in Fig. [51](#). The red dashed line depicts the current detection threshold by the cosmogenic-isotope method (Usoskin et al. [2020b](#)); no events below this line can be reliably identified from proxy data.

Until recently (e.g., Miyake et al. [2020a](#)), the scaling factor of the 774 AD event was often given as 25–50, following the less certain early range considered by Usoskin and Kovaltsov ([2012](#)), despite several subsequent studies (Table [9](#)) indicating higher $Q_{775}/Q_{1956}$ ratios. The Usoskin and Kovaltsov estimate of $Q_{774}$, as well as those of Usoskin et al. ([2013](#)) and Pavlov et al. ([2013a, b](#)), were based on preliminary $^{14}$C data, did not consider the secular variation of the geomagnetic field, and used a 6-box carbon-cycle model (e.g., Oeschger et al. [1974](#)) leading to lower estimates. The 6-box model is normally used for slow GCR and geomagnetic changes but is not appropriate for fast changes in $^{14}$C production. Later, more sophisticated 11-box models (Güttler et al. [2015](#); Mekhaldi et al. [2015](#); Uusitalo et al. [2018](#)) gave a higher $Q_{774}$ value of $\sim 2.2 \times 10^8$ atoms cm$^{-2}$. However, these models did not distinguish between the terrestrial hemispheres. The 22-box model

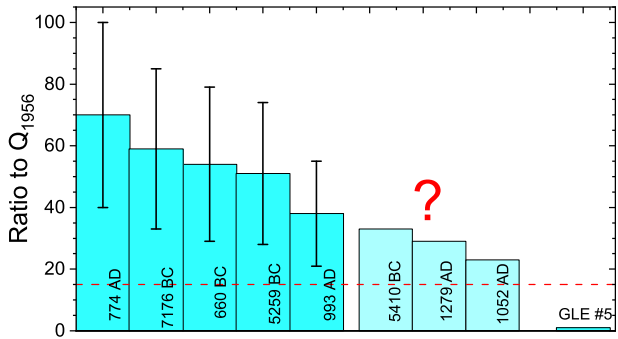

**Fig. 51** Ratio of the modelled annual radiocarbon production $Q$ of known (blue) and candidate (light blue) historical SEP events (according to Büntgen et al. 2017; O'Hare et al. 2019; Sakurai et al. 2020; Brehm et al. 2021; Miyake et al. 2021; Paleari et al. 2022; Brehm et al. 2022) relative to that of GLE No. 5 (Usoskin et al. 2020b). The horizontal red dashed line depicts the current sensitivity of the single-proxy cosmogenic-isotope method in SEP event detection (Usoskin et al. 2020b). Uncertainties include the variety of the published estimates. The question mark indicates that the 5410 BC, 1279 AD, and 1052 AD candidate events are not yet confirmed

applied by Büntgen et al. (2018) is, in fact, the same 11-box model but considers both hemispheres separately, thus 22 boxes. It yields $Q_{774}$ of $1.88 \times 10^8$ atoms cm$^{-2}$, the current standard. As noted above, the $Q_{1956}$ value and SEP event spectrum have also evolved, from a $Q$ of $2.5 \times 10^6$ atoms cm$^{-2}$ (Usoskin et al. 2006b) increasing to $3.04 \times 10^6$ atoms cm$^{-2}$ (Usoskin et al. 2020b), with a decrease in the $^{14}$C yield function (Poluianov et al. 2016) more than offset by a modification of the proton spectrum (including an increase in $F_{30}$ from $10^9$ cm$^2$ to $\sim 1.4 \times 10^9$ cm$^{-2}$) to yield a $Q_{774}/Q_{1956}$ ratio of 62. Adjusting for the stronger geomagnetic field in $\sim 774$ AD yields a ratio of $\sim 70$ (188/2.58). The next step in the refinement of $Q_{774}$ will be to perform full dynamical atmospheric modelling of the carbon cycle for an instant $^{14}$C production event. Work is in progress on such a model.

### 7.5 Constraining the $F_{30}$ OFD with lunar data

Cosmogenic isotopes are produced by energetic particles in situ in lunar soil and rocks. Because the Moon is not protected by a magnetic field or an atmosphere, low-energy particles with energy of tens of MeV can produce cosmogenic isotopes in shallow layers of the lunar surface. Because the flux of such low-energy particles is dominated by SEP events, the lunar surface acts as a primitive spectrometer, so that the upper shallow layers are mostly affected by abundant but less energetic SEPs, while deeper layers are irradiated by more energetic GCRs. An analysis of the cosmogenic isotope depth profile in lunar samples may reveal the time-integrated flux of SEPs over the timescale of millennia and even millions of years (e.g., Jull et al. 1998; Reedy 1998; Nishiizumi et al. 2009). Lunar samples were collected primarily (382 kg) during the manned Apollo missions in 1969–1972, with a small amount (321 g) of material brought to Earth by Soviet *Luna* unmanned landers.[18]

---

[18] The Chinese Chang'e-5 lunar probe returned $\sim 2$ kg of regolith to Earth in December 2020.

The samples were analyzed on Earth, including measurements of such isotopes as $^{14}$C (half-life $5.73 \times 10^3$ year), $^{41}$Ca ($1.03 \times 10^5$ year), $^{81}$Kr ($2.29 \times 10^5$ year), $^{36}$Cl ($3.01 \times 10^5$ year), $^{26}$Al ($7.17 \times 10^5$ year), $^{10}$Be ($1.36 \times 10^6$ year), and $^{53}$Mn ($3.74 \times 10^6$ year).

Although lunar soil and rocks make it possible, in contrast to terrestrial archives, to estimate the spectrum of SEPs on a long-time scale, individual SEP events cannot be identified in this way. The isotopes are produced in situ, and thus are not transported, deposited, and archived in a stratified (i.e., datable) archive. The measured abundance of an isotope is a balance between production by energetic particles and decay, over a timescale of a few isotope half-lives. Thus, only the mean flux of SEPs can be directly estimated from lunar data. Different groups, using different isotopes, provided several estimates (see Nishiizumi et al. 2009 and Kovaltsov and Usoskin 2014) of the long-term average of the > 30 MeV flux as $(0.8–1.7) \times 10^9$ (cm$^2$ year)$^{-1}$, with a mean (over different reconstructions) value of $(1.18 \pm 0.14) \times 10^9$ (cm$^2$ year)$^{-1}$. This is consistent with the mean > 30 MeV SEP flux measured during the instrumental era, viz., $(1.1 \pm 0.4) \times 10^9$ (cm$^2$ year)$^{-1}$, with mean fluxes for individual solar cycles varying from $3 \times 10^8$ to $2.2 \times 10^9$ (cm$^2$ year)$^{-1}$ (Reedy 2012).

### 7.6 A composite > 30 MeV cumulative occurrence frequency distribution

In this section, we add the cosmogenic-based SEP events and lunar data to the $F_{30}$ OFD in Fig. 47. Since individual SEP events cannot be resolved in the proxy data, which has annual resolution, annual fluence is often considered (Usoskin and Kovaltsov 2012) in such distributions. This is valid particularly for strong events, which dominate annual fluences over a large number of weaker events (e.g., Bazilevskaya et al. 2014). Years with the strongest directly measured $F_{30}$ fluence were 1972 ($\sim 9 \times 10^9$ cm$^{-2}$) and 1989 ($\sim 8 \times 10^9$ cm$^{-2}$). For comparison the largest close clusters of events (within one month) during these years were as follows: August 1972 (3 events; $8.4 \times 10^9$ cm$^2$); and September–October 1989 (4 events; $5.6 \times 10^9$ cm$^{-2}$). The annual $F_{30}$ values of SEPs for the years from 1954–2020 (cycles 19 through 24) are shown in Fig. 52 as open triangles. One can see that the distribution is quite flat (approximately a power law with the spectral index of about $-0.4$) for weak and moderate fluences ($F_{30} < 5 \times 10^9$ (cm$^2$ year)$^{-1}$). Such a flat dependence suggests that the occurrence probability is only weakly dependent on event strength. An incorrect extrapolation of the space-based data to evaluate the occurrence probability of stronger events is still used occasionally (Gabriel and Patrick 2003; Miroshnichenko and Nymmik 2014). However, such a flat distribution cannot be extrapolated to infinity (the energy integral diverges; Newman 2005), and the distribution must roll off if it does not exhibit a cutoff (as proposed earlier by Lingenfelter and Hudson 1980). From the modern data, there is no clear indication of whether and how often very strong events can occur on long time scales. Thus while the modern data are informative, by themselves they do not permit a robust assessment of the occurrence probability of extreme SEP events. In particular, there were no years for which the annual $F_{30}$

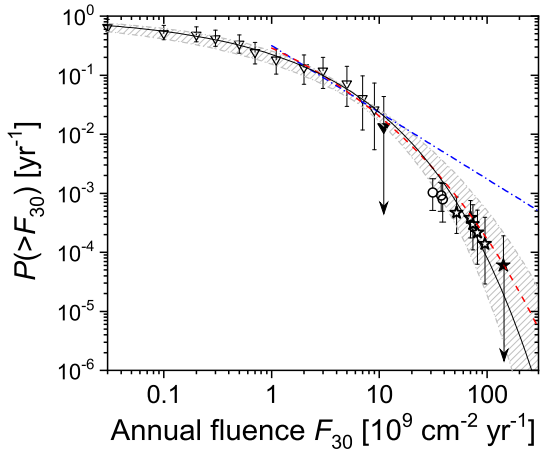

**Fig. 52** Downward cumulative frequency distribution of the occurrence of years with annual fluence of $F_{30}$ exceeding the given value (in units of $10^9$ protons/cm$^2$/year). The triangles denote the data for the space era (Shea and Smart 1990 (cycle 19); https://omniweb.gsfc.nasa.gov/ (cycles 22–24; Reedy 2006 (2001 and 2003 omitted because of large data gaps)); only years with $F_{30} > 3 \times 10^7$ protons cm$^{-2}$ year$^{-1}$ are shown, with some closely adjacent data points omitted for clarity). The circles and stars indicate historical SEP events based on terrestrial cosmogenic data, not yet confirmed and confirmed, respectively. Open symbols correspond to the measured/estimated fluences, filled symbols—a conservative upper bound. Error bars bound the 90% confidence interval, estimated using the Poisson distribution assuming that the events occur with a constant-in-time probability and independently of each other. The solid black curve corresponds to the best-fit Weibull distribution ($P = \exp(-(F_{30}/(3.66 \times 10^8))^{0.4})$. The gray-hatched area denotes the 68% confidence interval of the Weibull-shaped OFD obtained by considering the concentration of cosmogenic isotopes in lunar rocks (see Sect. 7.5). The red dashed ($Y = 0.9 * (1 - \exp(-(-X + 8.2)/1.7))$, where $X = \log(F_{30})$, $Y = \log(P)$) and blue dash-dot ($P = 4.68 \times 10^9 F_{30}^{1.13}$) curves correspond to extrapolations by Gopalswamy (2018) of the occurrence frequency of SEP events based on GOES data, using modified exponential and power-law distributions, respectively. Image modified after Poluianov et al. (2018)

fluence exceeded $10^{10}$ cm$^{-2}$ during the modern era (the black filled triangle in Fig. 52).

The confirmed cosmogenic-nuclide based SEP events are represented in Fig. 52 by open stars with error bars. The right-most open star corresponds to the event of 774 AD, while the leftmost point corresponds to the five confirmed events (7176 BC, 5259 BC, 660 BC, 774 AD, and 993 AD) found over the Holocene. The three yet unconfirmed event candidates (5410 BC, 1052 AD and 1279 AD) are indicated with the open circle. Further systematic search for cosmogenic SEP events continues (e.g., Miyake et al. 2021; Brehm et al. 2021), including new annual measurements of the cosmogenic isotopes. This is a slow and laborious process; we expect more events similar to that of 993 AD to be found and analyzed. However, we do not expect to find any events larger than that of 774 AD over the past ten millennia. This event was strong enough to be clearly observed in the decadal INTCAL series of $\Delta^{14}C$ (Usoskin and Kovaltsov 2012), and it is unlikely that stronger events were left unidentified in the earlier analyses (Usoskin and Kovaltsov 2012; Miyake et al. 2016). This is illustrated in Fig. 53, where the events of 774 AD and 993 AD are shown, along with two hypothetical events, representing two-fold

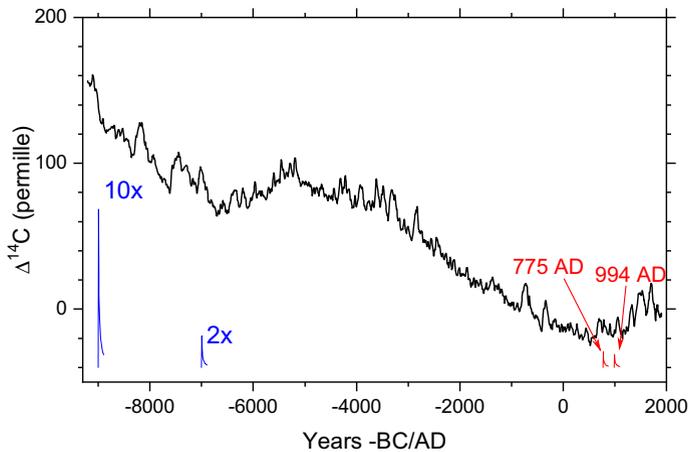

**Fig. 53** The radiocarbon $\Delta^{14}$C IntCal09 record (Reimer et al. 2009) over the Holocene (black curve) along with up-to-scale responses to the known events of 774–775 AD and 993–994 AD (red spikes). Also shown are two hypothetical spikes corresponding to events two- and ten-times larger than that of 774–775 AD. The time resolution is five years for all curves. Image adapted from Miyake et al. (2020b)

and ten-fold increases over the 774 AD event. One can see that even an event twice as large as 774 AD is unlikely to be missed, considering the well-defined time profile of such events, viz. a very fast (1–2-year) rise followed by a gradual few-decades decay. Most of the seeming spikes in Fig. 53 are caused by grand minima of solar activity and have much longer rise times of several decades and thus cannot be confused with SEP events. A ten-fold event cannot be missed regardless of the exact time profile. Thus, it can be conservatively said that no events with $F_{30} > 10^{11}$ cm$^{-2}$ occurred during the Holocene (Usoskin 2017). This is shown as the black star in Fig. 52. Accordingly, the event of 774 AD can serve as the strongest $F_{30}$ SEP event over eleven millennia. We note, however, that additional soft-spectrum events of this order may exist, awaiting discovery via analysis of $^{36}$Cl content in ice cores of events in which $^{10}$Be and $^{14}$C enhancements are not detectable (Mekhaldi et al. 2021). Mekhaldi et al. point out that the majority of GLEs observed since 1956 have soft spectra.

In order to fit the observational data, we assume that the OFD of SEP annual fluence occurrence has the shape of a Weibull (1951) distribution. The integral Weibull distribution for the probability $P(> F)$ of the annual fluence exceeding a given value F to can be written as

$$P(> F) = \exp\left(-(F/F_0)^k\right), \quad (12)$$

where k and $F_0$ are fitted parameters. In this case the mean fluence $(F)$ can be defined as

$$\langle F \rangle \equiv \int F \frac{dP}{dF} dF. \tag{13}$$

One can constrain this distribution by observational facts, as described below:

- The mean $F_{30}$ annual fluence reconstructed from lunar samples (Sect. 7.5) is between 1.0 and $1.3 \times 10^9$ (cm$^2$ year)$^{-1}$ which limits the range of the $\langle F \rangle$ value of the distribution.
- The fact that no events with $F_{30} > 10^{11}$ (cm$^2$ year)$^{-1}$ have been found, and are unlikely to be found, over the Holocene, poses an upper 90% confidence interval limit $P(> 10^{11}$ (cm$^2$ year)$^{-1}) \leq 1.9 \times 10^{-4}$ year$^{-1}$.
- Five historical events (7176 BC, 5259 BC, 660 BC, 774 AD, 993 AD) with $F_{30} > 5 \times 10^{10}$ (cm$^2$ year)$^{-1}$ (Mekhaldi et al. 2015; O'Hare et al. 2019; Paleari et al. 2022; Brehm et al. 2022) have been so far discovered and confirmed during the Holocene, but additional similar events may be found as the search continues. Accordingly, a lower limit of $P(> 5 \times 10^{10}$ (cm$^2$ year)$^{-1})$- $\geq 1.2 \times 10^{-4}$ year$^{-1}$ has been set.

The corresponding distribution function, shown as the gray hatched area in Fig. 52, encompasses all the data points and is consistent with the data from lunar rocks. In particular, a fast roll-off of the distribution for $F_{30}$ exceeding $10^{10}$ (cm$^2$ year)$^{-1}$ is observed implying that the probability of extreme event occurrence decreases dramatically with its severity.

Interestingly, the extrapolation of the space-era SEP event data by Gopalswamy (2018; shown in Fig. 47) using the modified exponential distribution (the red dashed curve in Fig. 52) agrees with the full-dataset reconstruction for intervals $\gtrsim 10$ years, while the power-law (blue dash-dot) extrapolation over-predicts $F_{30}$ for 1000-year and 10,000-year SEP events. Figure 52 indicates that, roughly, events with the $F_{30}$ fluences of $10^{10}$, $5 \times 10^{10}$ and $10^{11}$ cm$^{-2}$ are expected to occur once per roughly a century, a millennium and $10^4$ years, respectively.

### 7.7 Effective energy of different detectors: OFD for $F_{200}$

Energetic particles are recorded by different "detectors", either specifically designed and built space-borne or ground-based instruments, or naturally existing detectors such as terrestrial or lunar/meteoritic archives. While space-borne detectors can directly measure the energy spectrum of particles, other detectors are integral spectrometers providing only "count rates". The energy response of such a detector is quantified via the yield function (the response of the detector to the unit flux of particles at a given energy), or by the "effective energy" (e.g., Asvestari et al. 2017b; Koldobskiy et al. 2018) of the detector for a typical energy spectrum of incoming particles. Figure 54 illustrates the differential response function of neutron monitors, cosmogenic isotopes in terrestrial archives, and lunar rocks, for a typical hard-spectrum SEP event. The response function is a product of the detector's yield function and the energy spectrum of incoming particles. For example, the response function of a polar sea-level NM (blue dashed line) is quite

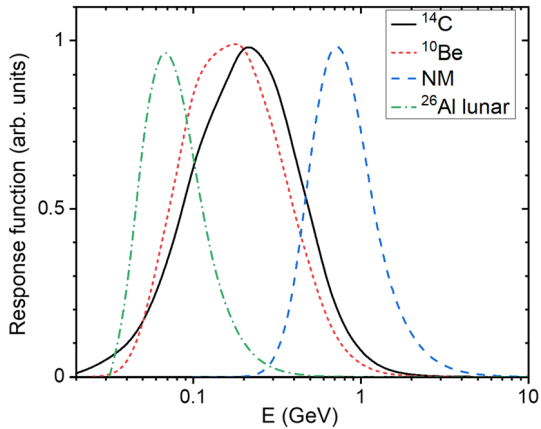

**Fig. 54** Normalized (to unity) response functions of different particle detectors for the hard-spectrum GLE (No. 69) on 20 January 2005 (event-integrated spectrum from Raukunen et al. 2018). The yield functions were taken from: Poluianov et al. (2018) for $^{26}$Al in lunar rocks (at the depth of 1 g/cm$^2$); Poluianov et al. (2016) for both $^{10}$Be (with atmospheric mixing following the parameterization of Heikkilä et al. 2009) and global $^{14}$C; and Mishev et al. (2013) for a polar sea-level neutron monitor

narrow peaking around 700 MeV which is close to the effective energy of a NM to SEPs (e.g., Asvestari et al. 2017b). This is a relatively high energy since the products of a primary particle's full atmospheric nucleonic cascade need to reach the ground with energy $\gtrsim$ 100 MeV to be detected by a NM. This can be clearly seen in Table 8 where for the strongest $F_{30}$ fluences, SEP events may have only a moderate high-energy, i.e., GLE, response (e.g., August 1972), while the strongest GLEs (e.g., February 1956) may lack a commensurate $F_{30}$ response. The effective energy of $^{26}$Al production by a SEP event in a shallow layer of lunar rock is about 50 MeV, much lower than the $\sim$ 800 MeV effective energy of a neutron monitor (Poluianov et al. 2018), since no cascade is needed and primary particles can produce the isotope directly. Thus, different proxy data provide probes for different energy ranges of primary SEP particles.

Differential response functions of terrestrial cosmogenic isotopes are shifted to lower energies (peaking at about 200 MeV) with respect to NMs since they are produced mostly in the stratosphere and do not require a full cascade to develop. The effective energy of both $^{10}$Be and $^{14}$C is roughly 200 MeV (Kovaltsov et al. 2014). Accordingly, the effective fluence of SEPs with energy above 200 MeV, $F_{200}$, can be directly assessed from the terrestrial cosmogenic proxy data, more-or-less independently of the exact particle spectrum. The cumulative OFD of $F_{200}$ is shown in Fig. 55. The distribution also rolls off at higher fluences but not as sharply as that for $F_{30}$ (Fig. 52). The gap between the space-era (triangles) and proxy-based data points (circles) is noteworthy. This is likely caused by the observational threshold for a SEP event in the low-resolution cosmogenic data. This gap is shorter in the $F_{30}$ OFD (Fig. 52), possibly due to the streaming effect or the uncertainty of $F_{30}$ for 1956 which is based on ionospheric measurements (Shea and Smart 1990; Webber et al. 2007; Usoskin et al. 2020b). At present, some combination of these effects (observational threshold, uncertainty of $F_{30}$ for 1956, streaming limit) is the most likely explanation for the gap. But the fluence values below the 68% confidence interval for the strongest events of the modern era leave open the possibility that the gap might also indicate two separate distributions, with the higher fluence branch arising from an as yet unrecognized source or mechanism.

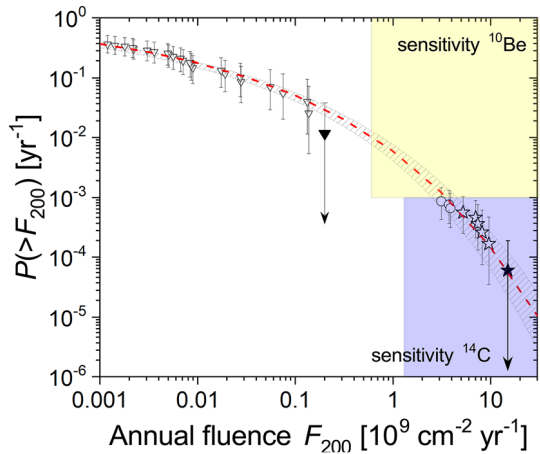

**Fig. 55** Downward cumulative frequency distribution of the occurrence of years with an annual $F_{200}$ value exceeding a given value (in units of $10^9$ protons/cm$^2$/year). The triangles denote the data for the space era (Koldobskiy et al. 2021) and the stars indicate extreme SEP events confirmed in terrestrial cosmogenic data. The open circles correspond to yet-unconfirmed events of 5410 BC, 1052 AD and 1279 AD. Open symbols correspond to the measured/estimated fluences, filled symbols denote a conservative upper bound. Error bars bound the 90% confidence interval. The red-dashed line and gray-hatched area denote the CFD for the best-fit Weibull distribution ($P = \exp(-((x/9.75 \times 10^{-4})^{0.236}))$) and its 90% confidence interval. The range of sensitivity of $^{14}$C and $^{10}$Be-based reconstructions is shown by blue and yellow boxes, respectively. Image updated after Usoskin et al. (2020b)

### 7.8 Does the 774 AD SEP event imply an ~ X3000 flare?

Early estimates of the bolometric energy of the 774 AD flare ranged from $10^{34}$ to $10^{36}$ erg [e.g., Miyake et al. (2012), taking Emslie et al. (2012) into account (Sect. 7.4.1); Shibata et al. 2013; Maehara et al. 2015]. From Eq. (5) in Sect. 3.1.6, the lowest of these values would imply a ~ X3000 SXR flare. For comparison, the corresponding estimates for the Carrington flare are ~ X45 and ~ $5 \times 10^{32}$ erg. On the basis of log–log scatterplots in Cliver and Dietrich (2013) of: (a) $F_{30}$ versus 1–8 Å SXR fluence and (b) flare bolometric energy versus SXR fluence, Cliver et al. (2014) obtained an estimate of ~ X230 for the 774 AD event, with an estimated radiative flare energy of ~ $2 \times 10^{33}$ erg. Similar scaling relationships between > 10 MeV peak proton flux, CME mass and speed, and SXR class have been developed by Gopalswamy et al. (2003), Takahashi et al. (2016), and Papaioannou et al. (2022). Because of its relative insensitivity to SEP event spectra (Kovaltsov et al. 2014) and closer relationship to $^{14}$C and $^{10}$Be production (Poluianov et al. 2019; Fig. 54), Cliver et al. (2020b) considered $F_{200}$ in their

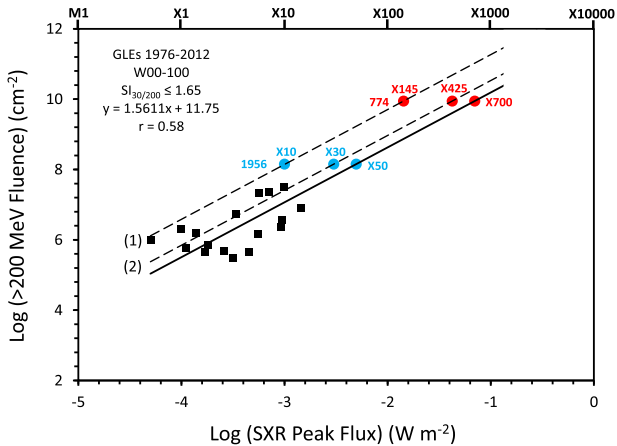

**Fig. 56** Scatter plot of the log of the > 200 MeV proton fluence versus the log of the SXR intensity of associated flares for hard-spectrum GLEs (Cliver et al. 2020a) observed from 1976 to 2012 that originated in solar longitudes from W00-100. The solid line is a reduced major axis fit, with the equation of the line given. Positions of the points of the 1956 February 23 and the 774 AD SEP events on the fit line are determined by their > 200 MeV fluence values. Dashed lines, labeled (1) and (2), parallel to the fit line have been drawn through the light blue points determined by the range of inferred SXR classes from X10 (line No. 1) to X30 (No. 2) for the 1956 February 23 GLE-associated flare and the > 200 MeV fluence of that event. From these dashed lines, the corresponding values of the SXR class for the 774 AD SEP event can be inferred from the > 200 MeV fluence for that event (red data points). Image reproduced with permission from Cliver et al. (2020b), copyright by AAS

determination of the intensity of the SXR flare associated with the 774 AD proton event.

Using the 23 February 1956 GLE (with estimated SXR class ∼ X10-30) as a bridge, and a $F_{200}$ value of $8.8 \times 10^9$ cm$^{-2}$ ($Q_{774}/Q_{1956}$ ratio of 62; no magnetic field correction), Cliver et al. (2020b) estimated a SXR classification of ∼ X285 ± 140 for 774 AD ($1.9 \pm 0.7 \times 10^{33}$ erg from Eq. (5)) from correlations

$$\log(F_{200}) = 1.5611 \times \log(\text{SXR}_{\text{int}}) + 12.48 \quad (14)$$

$$\log(F_{200}) = 1.5611 \times \log(\text{SXR}_{\text{int}}) + 12.09 \quad (15)$$

between the logs of flare SXR intensity (SXR$_{\text{int}}$; in W m$^{-2}$) and $F_{200}$ for hard spectrum GLEs occurring after 1976 (with known SXR flare intensities) corresponding to lines (1) (Eq. (14); 1956 = X10) and (2) (Eq. (15); X30) in Fig. 56. A bolometric energy of ∼ $1.9 \times 10^{33}$ erg meets the $10^{33}$ erg superflare criterion and is approximately four (five) times larger than that inferred for the Carrington (4 November 2003) flare. The nominal ∼ $1.9 \times 10^{33}$ erg and X285 estimates for 774 lie within our working estimates (X180 (− 100, + 300); 1.4 (− 0.6, + 1.4) × $10^{33}$ erg) for the largest possible solar flare based on the April 1947 active region. An X285 class SXR flare also falls within the X200-310 range encompassed by the power-law and modified exponential fits in to the SXR peak flux distribution (Fig. 8) for a 10,000-year flare.

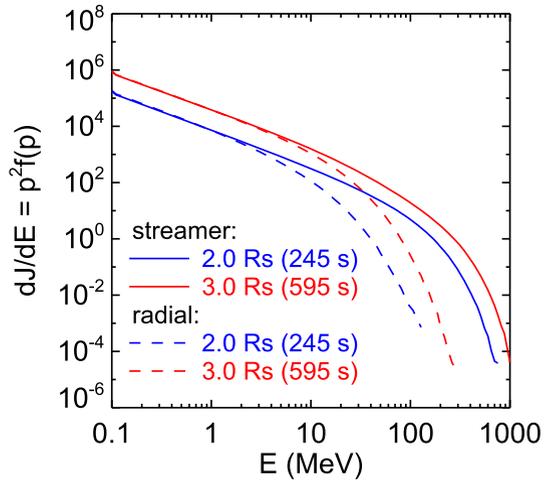

Fig. 57 Simulated differential intensity spectra versus proton energy for a streamer-like (solid lines) and radial (dashed lines) coronal magnetic field at $\sim$ W55 for shock altitudes of 2 $R_\odot$ and 3 $R_\odot$. Image adapted with permission from Kong et al. (2017), copyright by AAS

Because of its strong SEP association, the 774 AD flare is taken to have been eruptive implying a total (bolometric plus CME (6 x $\sim$ 1.9 x $10^{33}$ erg = $\sim$ 1.2 x $10^{34}$ erg)) energy of $\sim$ 1.4 × $10^{34}$ erg and an active region potential magnetic energy $\gtrsim 10^{35}$ erg. These estimates assume that the 774 AD proton event originated in a single eruption. Two equally intense SEP events occurring in close succession (within a few months) to yield the $F_{200}$ value inferred for 774 AD (based on a $Q_{774}/Q_{1956}$ ratio of 62) would reduce the estimate for a single flare to X180 ± 90 (radiative energy = 1.4 ± $\sim$ 0.6 × $10^{33}$ erg), and three such closely-spaced SEP events would further reduce the estimate to $\sim$ X140 ± 70 (1.2 ± $\sim$ 0.5 × $10^{33}$ erg), within the range of 1000-year estimates for an extreme flare, still implying a threshold level $10^{33}$ erg solar superflare.

Recent theoretical work by Kong et al. (2017) calls attention to the potential importance for high-energy proton acceleration of the magnetic field topology at the solar footpoint of the field line connected to Earth. Specifically, the simulation by Kong et al. (2017) indicates that a streamer at the nominal W55 solar footpoint of the magnetic fieldline to Earth can enhance the acceleration efficiency of an impinging quasi-perpendicular shock for protons with energy > 100 MeV by about two orders of magnitude above that for the scenario of quiet field at this longitude, with a much smaller effect at 10 MeV (Fig. 57).

In a Fraunhofer map of the Sun's disk,[19] Cliver et al. (2020b) noted a filament—the low coronal manifestation of a streamer—located $\sim$ 25° east of the active region (Greenwich 17351; with a maximum field strength of 4300 G (Livingston et al. 2006)) that produced the W80 flare associated with the 23 February GLE. The filament, embedded in a string of northern hemisphere active regions that included the GLE-parent flare (Cliver et al. 2014), was thus located at the nominal W55 footpoint of the magnetic fieldline to Earth. Given that the February 1956 SEP event was six times larger at > 1 GV than the next largest individual event (12 November

---

[19] https://www.ngdc.noaa.gov/stp/space-weather/solar-data/solar-imagery/composites/full-sun-drawings/fraunhofer/

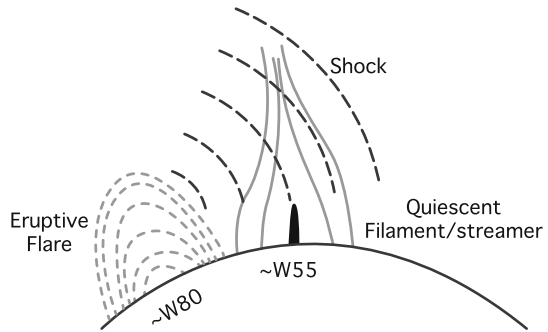

**Fig. 58** Cartoon showing a shock from an eruptive flare at ∼ W80 impinging on a streamer at ∼ W55, a favorable situation for acceleration of high-energy SEPs directed toward Earth that existed for the 23 February 1956 event. Image reproduced with permission from Cliver et al. (2020b), who adapted from Wild (1969)

1960; Raukunen et al. 2018; Usoskin et al. 2020a, b) and that the 774 AD proton event is estimated to have been ∼ 70 times larger than the 1956 event, Cliver et al. speculated that the scenario for out-sized hard-spectrum SEP events proposed by Kong et al. might have applied to these two proton events to account for the gap between the directly observed and cosmogenic-based SEP events in both the scatter plot in Fig. 56 and the OFD in Fig. 55. (Because of the hard spectrum of the 774 AD SEP event, we can assume (after Cliver et al. 2020a) that the associated eruptive flare also originated near the west limb of the Sun). A cartoon of this well-placed streamer scenario, patterned and repurposed after a similar schematic by Wild (1969), is given in Fig. 58. In the Kong et al. simulation, proton trapping in the lower-lying closed field loops of a helmet-streamer configuration play an important role in the enhanced acceleration of high-energy protons; such protons face the same escape challenge (Hudson 2018) as those accelerated in a flare-resident process.

# 8 Conclusion

## 8.1 Summary

This review of extreme solar events, geomagnetic storms, and flares on Sun-like stars has covered peak size parameters of sunspot groups, flares (both on the Sun and Sun-like stars), radio bursts, CMEs, historical fast transit ICMEs, proton events, geomagnetic storms, and aurorae with a goal of documenting/estimating worst-case or limiting values for each of these phenomena.

For convenience, a summation of these parameters for key space weather phenomena is given here:

(a)  Flare bolometric energy and SXR class:

   (1) Most intense directly observed flares: bolometric energy of $4.3 \times 10^{32}$ erg (∼X35) on 4 November 2003; $\sim 5 \times 10^{32}$ erg (∼X45) on 1 September 1859; (2) strongest inferred flare: $1.9 \pm 0.7 \times 10^{33}$ erg (X285 ± 140) for the 774–775 AD cosmogenic-nuclide-based SEP event (reducible to $1.2 \pm \sim 0.5 \times 10^{33}$ erg (X140 ± 70), per flare) for a cluster of three closely-spaced, equal-fluence SEP events); (3) theoretical estimate for the largest possible flare based on the largest observed spot group in the modern era

(April 1947; 6132 μsh): X180 (− 100, +300) (1.4 (− 0.6, +1.4)×$10^{33}$ erg);

(b) Coronal mass ejections (CMEs):

(1) Fastest directly observed CME: 3387 km s$^{-1}$ on 10 November 2004; (2) most energetic directly observed CME: 4.2×$10^{33}$ erg on 9 September 2005; (3) most energetic inferred CME: ∼1.2×$10^{34}$ erg, for a single 774 flare; (iv) theoretical estimate for the most energetic possible CME; 8.4×$10^{33}$ erg, (based on largest possible flare estimate for April 1947 active region);

(c) Solar energetic proton (SEP) events:

(1) Largest fluence event directly observed at >30 MeV: 8.4×$10^9$ cm$^{-2}$, composite event in August 1972; at 200 MeV: 1.4×$10^8$ cm$^{-2}$, 23 February 1956; (2) largest inferred fluence event at >30 MeV: ∼$10^{11}$ cm$^{-2}$, 774 AD; at >200 MeV: ∼$10^{10}$ cm$^{-2}$ 774 AD;

(d) Geomagnetic storms:

(1) Most intense modern event: Dst = − 589 nT, March 1989; (2) strongest reconstructed/inferred events: September 1859, − 949 (±∼30) nT; − 907 (±132) nT, May 1921; ∼−1000 nT, February 1872; (3) strongest possible events based on direct observations: − 1200 nT, July 2012, − 1400 nT, August 1972; (4) strongest theoretically possible event: ∼− 2000 to − 2500 nT.

The above observed and theoretical peak values are given in the comprehensive Table 10 along with estimates for 100- and 1000-year events where available. The first column in Table 10 lists the observed/inferred extreme values, the central columns list the sizes of 100-year and 1000-year events from frequency distributions fitted to modified exponential and power-law functions (taken from Gopalswamy et al. 2018, unless otherwise noted; with preference given to the exponential function fits in the discussion below), and the final column lists extreme values based on theoretical models or worst-case calculations.

## 8.2 On the reliability of the 100- and 1000-year event estimates in Table 10

In general, with SEP events being a notable exception, the exponential-based 100-year estimates in Table 10 are close to their directly observed values − sunspot group area (5800 μsh vs. 6132 μsh), SXR flare class (X44 vs. X35 ± 5), CME speed (3800 km s$^{-1}$ vs. 3387 km s$^{-1}$), and geomagnetic storms (− 603 nT vs. − 589 nT). This gives some confidence that the 1000-year estimates have validity, but an

Table 10 Observed, statistically expected, and modelled extreme solar and solar-terrestrial events

| Parameter | Observed/ inferred maximum | 100-year Event(a) Exponential | 100-year Power-law | 1000-year Event(a) Exponential | 1000-year Power-law | Modelled maximum |
|---|---|---|---|---|---|---|
| Sunspot group area (μSH) | 6132(b) | 5800 | 7100 | 8200 | 13,600 | – |
| GOES flare SXR class | X35 ± 5(c)/ ∼ X285 ± 140(e) ∼ X140 ± 70(e) | X44 | X42 | X100 | X115 | X180(d) |
| Flare TSI Fluence ($10^{32}$ erg) | 4.3(f) 4.3(f)/ ∼ 20(e) ∼ 10(e) | 4.4(g) 5.0(h) | 4.2(g) 4.9(h) | 10(g) 9(h) | 12(g) 10(h) | 14(d) |
| ∼ 1.5 GHz radio emission ($10^6$ sfu) | 1(i) | – | 3.2–12(j) | – | 61 – 200(j) | – |
| CME speed (km/s) | 3387(k) | 3800 | 4500 | 4700 | 6600 | (l) |
| CME kinetic E ($10^{32}$ erg) | 42(m)/ ∼ 120(e) | 44 | 70 | 100 | 300 | 84(d) |
| ICME transit time | 14.6(n) | – | – | – | – | (l) |
| > 30 MeV pr fluence ($10^{10}$ cm$^{-2}$) | 0.84(o)/ ∼ 10(p) | 1.6 | 2.1 | 5 | 16 | (l) |
| > 200 MeV pr fluence ($10^{9}$ cm$^{-2}$) | 0.14(q)/ ∼ 10(p) | 0.6(r) | – | 3.5(r) | – | – |
| Dst (nT) | – 589(s) – 907(u) – 949(v) ≲ – 1000(w) | – 603 | – 774 | – 845 | – 1470 | – 2000 to – 2500(t) |
| Lowest lat of O.H. aurora (°) | 24.2(x) | – | – | – | – | – |

**Table 10** continued

Listed are sunspot group area, GOES SXR flare class, flare bolometric energy, decimetric (∼ 1.5 GHZ) radio emission, CME speed and kinetic energy, ICME transit time, SEP event fluences at > 30 MeV and > 200 MeV, magnetic storm Dst values, and equatorward extent of overhead aurora

(a) Unless otherwise noted, the (rounded) values in these columns are from Gopalswamy (2018)

(b) 8 April 1947 (Greenwich)

(c) Estimate for 4 November 2003 event from Cliver and Dietrich (2013) based on Kiplinger and Garcia (2004), Thomson et al. (2004, 2005), Brodrick et al. (2005) and Tranquille et al. (2009)

(d) Based on estimates of Aulanier et al. (2013), Kazachenko et al. (2017), and Tschernitz et al. (2018) for Greenwich 14886 in Table 1 (and 6:1 CME:flare energy apportionment)

(e) Inferred, based on the 774 AD SEP event for a single flare (top) and for the case of three SEP-equivalent flares (bottom) (Cliver et al. 2020b); CME kinetic energy estimate from Sect. 7.8 is for a single eruption

(f) 4 November 2003 (Emslie et al. 2012; Woods et al. 2006)

(g) Assuming a linear scaling between total radiated energy and GOES flare class (Gopalswamy et al. 2018)

(h) Based on Eq. (5) in Sect. 3.1.6 for corresponding GOES flare SXR class entry

(i) 6 December 2006 (Gary 2008)

(j) (Giersch et al. 2017; Nita et al. 2002)

(k) 10 November 2004, Gopalswamy (2018)

(l) CME speed (Sect. 5.1), transit time (Sect. 5.2)

(m) 9 September 2005, Gopalswamy (2018)

(n) 4 August 1972, Cliver et al. (1990a, b)

(o) Composite August 1972 event, Jiggens et al. (2014)

(p) 774 AD (based on a 70-fold multiple for 23 February 1956 (Usoskin et al. 2020b)

(q) 23 February 1956 (Usoskin et al. 2020b)

(r) Estimates from Fig. 55

(s) 14 Mar 1989 (Allen et al. 1989)

(t) Liu et al. (2020); Vasyliūnas (2011)

(u) 15 May 1921, Love et al. (2019c), with uncertainty of ± 132

(v) 2 September 1859 (Hayakawa et al. 2022)

(w) 4 February 1872, Hayakawa et al. (2018a; Sect. 6.2)

(x) 4 February 1872 (Hayakawa et al. 2018a)

important caveat applies for geomagnetic storms. Three geomagnetic storms (in September 1859, February 1872, and May 1921) appear to have surpassed the $-845$ nT minimum Dst value of the 1000-year event within the last $\sim 160$ years. This might mean that the exponential function is not a good choice for storms but it may also be a reflection of the short Dst data base (1957–2016).

The largest deviations of the sizes of directly observed events from their exponential-based 100-year are seen for proton event fluences, viz., (i) the $> 30$ MeV fluence where the largest observed modern event ($8.4 \times 10^9$ cm$^{-2}$; composite event in August 1972) is a factor of $\sim 2$ below the (annual) 100-year estimate; and (ii) the $> 200$ MeV fluence where the February 1956 event was $\sim 4$ times smaller than the 100-year estimate. In regard to (i), we note that Gopalswamy (2018) obtained reduced 100- and 1000-year estimates for Ellison-Ramaty and Band function fits to the $F_{30}$ size distribution. Item (ii) for $F_{200}$ could imply a disconnect between the low and high fluence branches of the occurrence frequency distribution in Fig. 55, although the uncertainties are large. The recent discoveries of smaller candidate historical SEP events reduces the gap between the modern and historical data points but the ultimate confirmation of the SEP hypothesis awaits direct observation of a significantly ($\gtrsim 3$ times) stronger $F_{200}$ event than that of February 1956.

The $> 200$ MeV fluence for the 774 AD SEP event ($\sim 10^{10}$ cm$^{-2}$; Cliver et al. 2020b) in Fig. 55 is notable in that it exceeds the largest such space age event (23 February 1956; $\sim 1.4 \times 10^8$ cm$^{-2}$; Usoskin et al. 2020b) by a factor of $\sim 70$ ($\pm 30$) and the 1000-year estimate by a factor of $\sim 3$. Even if the lower limit multiple of 40 were applied and the 774 AD event consisted of three equal-sized $F_{200}$ events, each of these would still be an order of magnitude larger than the 1956 event.

For the solar flares for which the bolometric energy has been directly observed (Woods et al. 2006), the largest TSI fluence value is $\sim 4 \times 10^{32}$ erg, corresponding to an observed/estimated (off-scale at X18.4) SXR class of $\sim$ X35 for an event on 4 November 2003. Such flares are expected to occur once per century (Fig. 8). Superflares with SXR class $\sim$ X100 (bolometric energy $\sim 10^{33}$ erg) may occur once per millennium. A once-per-10,000-year event may have been observed based on the estimate of $\sim 1.9 \times 10^{33}$ erg (SXR class $\sim$ X285) for an assumed single source for the 774 AD SEP event (Cliver et al. 2020b).

### 8.2.1 An important note on GOES SXR event recalibration

The transition of GOES 16 to operations at the end of 2017 triggered a recalibration of earlier GOES SXR data from 1976–2017 and an ongoing update of the NOAA SXR data base that will be completed in 2022 (Hudson et al. 2022).[20] As a result, all of the pre-GOES 16 1–8 Å SXR peak fluxes given in this review will need to be increased by a factor of 1/0.7 (1.43). For example, the peak SXR classes for the 1 September 1859 (X45 ± 5) and 774 AD ($\sim$ X285 ± 140) events will increase to X64.4 ± 7.2 and ∼ X410 ± 200, respectively, while that for the largest possible solar flare changes from X180 (− 100, + 300) to X260 (− 140, + 430). In addition, relationships such as Eq. (5) for $\mathcal{F}_{TSI}$ and Eq. (6) for $\Phi_r$, both with $\left(\frac{\mathcal{C}_{GOES}}{\mathcal{C}_{GOES,X1}}\right)$ as the independent variable, will need to have $\mathcal{C}_{GOES,X1}$ set to $1.43 \times 10^{-4}$ W m$^{-2}$ when $\mathcal{C}_{GOES}$ is the recalibrated SXR intensity. The data sets for recent studies bearing on SXR class considered in this review (e.g., Tschernitz et al. 2018; Cliver et al. 2020a, b) are homogeneous in that they include no events occurring after 18 December 2017 when GOES 16 became operational.

---

[20] Hudson et al. also obtained revised values for saturated events such as 4 November 2003, giving an additional correction for such events.

### 8.3 Discussion

#### 8.3.1 Is the Sun a Sun-like star?

The principal reasons for thinking that the Sun is not "Sun-like" (in the same sense as *Kepler* Sun-like superflare stars) are that: (1) the Sun-like superflare stars have strongly imbalanced spot coverage with a large filling factor—they appear to be spot-dominated, while the Sun is faculae dominated (Willson and Hudson 1988; Foukal and Lean 1988); (2) as of yet there are no spectroscopically confirmed Sun-like superflare stars with rotation periods > 20 days. Items (1) and (2) are related. From the long-term study of 72 Sun-like stars begun at Lowell Observatory in 1984, (Lockwood et al. 1997), Radick et al. (1998) reported "a clear distinction between the younger and older stars in the correlation pattern between long-term photometric and chromospheric emission variations: the young, active stars become fainter as their chromospheric emission increases, whereas the older stars, including the Sun, tend to become brighter as their chromospheric emission increases. A simple interpretation of this behavior is that the long-term variability of the young stars is spot-dominated, whereas for the older stars like the Sun [it] is faculae-dominated." This finding has been strengthened by continued observation (Lockwood et al. 2007; Radick et al. 2018). Radick et al. noted that the dividing line between spot-dominated and facula-dominated Sun-like stars (defined as "spectral type late-F through G, unevolved or, at most, slightly evolved") in a scatter plot of chromospheric emission (log $R'_{HK}$; Noyes et al. 1984) versus temperature (B-V) occurs near the location of the Vaughan-Preston (VP; Vaughan and Preston 1980) gap (the relative absence of G-K main sequence stars with intermediate magnetic activity levels; − 4.85 ≲ log $R'_{HK}$ ≲ − 4.65; e.g., Reinhold et al. 2019). For a sample of 18 F-K stars, Metcalfe et al. (2016) found that G-type stars on the low activity side of the VP-gap have rotation periods ∼ 15 days, comparable to the rotation periods for the slowest rotating spectroscopically-validated superflare stars in Table 2, but well below that of the Sun.

In a sample of ∼ 4000 Sun-like stars (F7-G4; $T_{eff}$ within 150 K of the Sun, log(g) > 4.2) with known rotation periods, Montet et al. (2017) identified 463 that exhibited year-to-year brightness variability at the 3σ level. They then used the photometrically-based $S_{ph}$ index of magnetic activity (Mathur et al. 2017) to separate the 463 stars into those which were spot-dominated and those which were facula-dominated. As shown in Fig. 59, they found "a transition between anticorrelated [with long-term brightness variations] (star-spot-dominated) variability and correlated (facula-dominated) variability between rotation periods of 15–25 days, suggesting the transition between the two modes is complete for stars at the age of the Sun." The relative absence of Sun-like stars with rotation periods near 25 days implies that they move rapidly through this marker (van Saders et al. 2016; Metcalfe et al. 2016; Metcalfe and van Saders 2017) or that they do so "quietly" (with a near balance of spots and faculae; e.g., Radick et al. 2018), or both, making detection of a rotation period difficult.

While the rotation period is a prime factor in determining a star's spot coverage and capability to produce superflares, a recent study by Reinhold et al. (2020) suggests that there can be substantial deviations from this rule for a given star. The Sun exhibited an example of such anomalous behavior, for a constant (or nearly constant, e.g., Obridko and Shelting 2016) rotation period during the Maunder Minimum (Eddy 1976; Usoskin et al. 2013). The work by Reinhold et al. raises the possibility that the Sun is in an uncharacteristically subdued state of activity compared to the peers to which it is commonly compared, even though it is not currently in a state like the Maunder Minimum.

Reinhold et al. (2020) combined 4 years of *Kepler* photometry with astrometric data from the *Gaia* spacecraft on 369 solar-like stars, with rotation periods between 20 and 30 days, effective temperatures between 5500 and 6000 K, with log $g > 4.2$, for a range of metallicity ([Fe/H] from $-0.8$ to 0.3 dex) about that ($\sim 0.0$) of the Sun and (based on *Gaia* measurements) on isochrones that put their ages between 4 and 5 Gyr. They also identified 2529 "pseudosolar" stars, with similar characteristics for which a rotation period could not be determined. Reinhold et al. find that for those stars for which the rotational modulation is sufficient to reliably determine

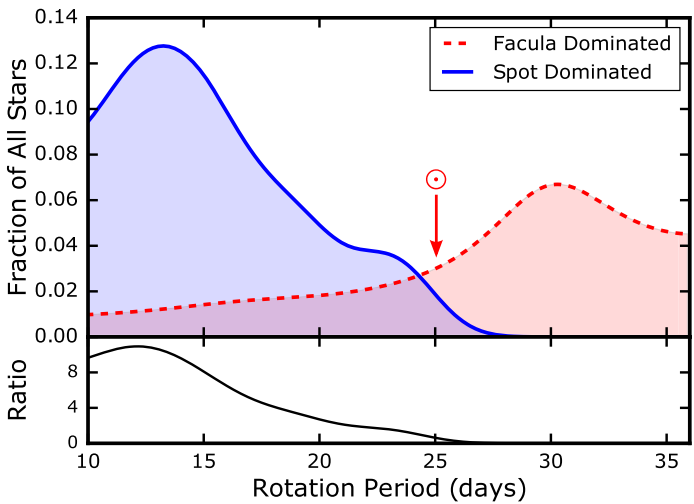

**Fig. 59** Occurrence frequencies (and their ratio) of 463 spot- and facula-dominated *Kepler* Sun-like stars as a function of rotation period. The arrow indicates the solar rotation period. Image reproduced with permission from Montet et al. (2017), copyright by AAS

a rotation period, their variability exceeds that of the Sun. Specifically, the characteristic variability (which they define as the range between the 5th and 95th percentile of 4 years of photometric brightness measurements relative to the mean brightness) of the solar-like stars with measurable rotation periods (0.38%) was found to be approximately five times higher than the average value (of 0.07%) of the solar data, and 1.8 times higher than even the most variable periods observed for the Sun. They conclude that "These stars appear nearly identical to the Sun except for their higher variability. Therefore, we speculate that the Sun could potentially also go through epochs of such high variability." Thus while Maehara et al. (2017) suggested that a presumably rare "super-active" phase of solar activity is required for solar super flares, the work of Reinhold et al. (2020) hints that such phases are the norm for Sun-like stars.

In the context of the present review, the result of Reinhold et al. (2020) raises the following questions: Can the Sun's activity level become more like that of its Sun-like peers (a set defined in terms of fundamental stellar properties) in the future? In its present evolutionary state, is the Sun capable of changing its spots? Can the Sun's dynamo change in an opposite manner from that during the Maunder Minimum to produce more spots rather than fewer? The discovery of superflares on solar-type stars that lack "hot Jupiters" (Sect. 3.2.2)—and potentially on slower rotating Sun-like stars (Okamoto et al. 2021)—makes the answers to these questions of more than academic interest. They raise further questions such as: Did any of the 369 solar-like stars with rotation periods of $\sim$ 25 days in the Reinhold et al. sample produce superflares? Or, perhaps more revealingly, did any of the "2529" "pseudosolar" stars that lacked detectable rotation periods, and therefore may hold a greater claim to Sun-like status (Reinhold et al. 2019, 2020) than the 369 "periodic" stars they analyzed, give rise to a superflare? *Kepler-* and TESS-based investigations that bear on the impact of superflares on the habitability of exoplanets and the Earth will require a strong ground-based spectroscopic component (e.g., Notsu et al. 2015a, b) to establish the Sun-like nature of the stars considered.

### 8.3.2 Black swans and dragon-kings

Within the past twenty years two new concepts have been introduced to describe certain types of extreme events: Black Swans (Taleb 2007) and Dragon-Kings (Sornette 2009). We describe each of them in turn as they relate to extreme solar and solar-terrestrial events.

In regard to Black Swans, Riley et al. (2018) write: "Given their rarity, it is reasonable to ask whether extreme solar events are merely 'black swans' such as crashes in the stock market (Taleb 2007). Such phenomena are defined as highly improbable events with the following characteristics: (1) unpredictable; (2) massive impact; and (3) an inherently random event for which we retrospectively concoct an explanation to explain it, making it apparently more predictable. Extreme space weather events are not unpredictable, at least once their signatures have been observed at the Sun. They undoubtedly have a massive impact, ranging from failures in the power grid, navigation assets, and communication. However, they do not need us to invent reasons for their occurrence. Given the substantial knowledge we have about them, preparing for them is within our grasp; that is, this is a solvable policy and investment problem. Extreme solar events are not 'black swans'."

Had the huge inferred SEP event in 774 AD occurred in say 2002, rather than being discovered in 2012, it would have qualified as a black swan. As it stands, we have the awareness of what the Sun is capable of doing in terms of energetic protons without having experienced the consequences. While a 774 AD class SEP event (estimated to have a $> 200$ MeV fluence $\sim 70$ times larger than that of the 1956 GLE; Usoskin et al. 2020b; Cliver et al. 2020b) does not pose the systemic threat of an extreme geomagnetic storm, the radiation impact on space instrumentation and operations related to communications, navigation, weather monitoring, etc. would be significant (see links in Footnote 1). Dyer et al. (2018) note that a 774 AD class SEP event could give air crews (and passengers) their lifetime radiation dose in a single high latitude flight.

Extreme events, worst-case scenarios, black swans—all of these terms indicate something well beyond the norm. Should we be surprised that such events hint at physics different from that underlying the bulk of the events in a distribution function? The idea that the physics of some extreme events might be different in nature from other large events of the same type (e.g., SEP events or radio bursts) has been developed by Sornette and colleagues who term such events "dragon-kings" (Sornette 2009; Sornette and Quillon 2012; Pisarenko and Sornette 2012; Wheatley et al. 2017; Aschwanden 2019; 2021). Here "king" refers to great event size while "dragon" denotes an unusual nature or aspect. From Sornette and Quillon (2012), "Dragon-kings are defined as extreme events that do not belong to the same population as the other events … [they] appear as a result of amplifying mechanisms that are not necessary fully active for the rest of the population." A clear example of a dragon-king can be seen for the intense solar decimetric radio bursts reviewed in Sect. 4 that are attributed to coherent radio emission (either electron cyclotron maser emission or plasma emission) versus the incoherent gyrosynchroton emission that applies to merely large bursts in this frequency range. The largest geomagnetic storms also present evidence of a dragon-king aspect, viz., the impact of field-aligned/auroral currents on low-latitude magnetograms (Sect. 6.3). The distribution function for geomagnetic storms is dominated by the smaller events, in this case storms which are not strong enough to have associated aurora reaching to the low-latitudes of the Dst magnetic stations—stations that have been selected for this reason. This geographic effect contributes to the apparent excess of storms stronger than the exponential-function-based estimate (Gopalswamy 2018) for a 1000-year event. In addition, there are hints in the gap in the $> 200$ MeV fluence distribution function in Sect. 7.7 and in the solar circumstances for the 1956 GLE (Sect. 7.8) that solar SEP events might be harboring a dragon-king. At present, however, the most likely (Occam's razor) conclusion is that they do not (Usoskin and Kovaltsov 2021).

The parameters in Table 10 lend themselves to an engineering strategy (e.g., stock-piling transformers; radiation hardening; communication redundancy) based on worst-case scenarios for coping with black swans and dragon-kings. At bottom, proficiency in the parallel forecasting/warning mitigation approach, as well as the validity/utility of the values in Table 10, will depend on how well we understand the far reaches of solar activity.

**Acknowledgements** E.W.C. and I.U. benefited from participation in the Institute for Space-Earth Environmental Research (ISEE) workshop on Extreme Solar Events organized by Fusa Miyake and Ilya Usoskin held at Nagoya University from 02–06 October 2018, and from membership on the International Space Science Institute (ISSI) International Teams led by Athanasios Papaioannou on high-energy solar particle events. E.W.C. thanks Nat Gopalswamy, Hisashi Hayakawa, Hugh Hudson, Eduard Kontar, Jeffrey Love, Florian Mekhaldi, Raimund Muscheler, Kosuke Namekata, Yuta Notsu, Dean Pesnell, Art Richmond, and Astrid Veronig for helpful discussions, and Lyndsay Fletcher for sponsoring his stay in Glasgow. C.J.S. is grateful for the support of the National Solar Observatory for the completion of this project. K.S. was supported by JSPS KAKENHI Grant Number 21H01131 and thanks Y. Notsu, K. Namekata, H. Maehara, D. Nogami, S. Okamoto, H. Hayakawa for helpful discussions. I.U. benefited from membership on the ISSI International Team led by Fusa Miyake and himself on extreme solar events and acknowledges support from the academy of Finland (project No. 321882 ESPERA). We thank Owen Giersch and Gelu Nita for kindly sharing data, Astrid Veronig for assistance with Fig. 16, and Peter Gallagher and Brian Dennis for providing Fig. 32. We acknowledge constructive comments from the assigned editor and two reviewers which significantly improved the paper.

Extreme solar events　　　Page 119 of 143　　　2